\documentclass[aip,pof]{revtex4-1}

\usepackage{amsfonts,amssymb}
\usepackage{graphicx}
\usepackage{amsmath}
\usepackage{bm}

\begin{document}

\title{Marangoni effects on a thin liquid film coating a sphere with axial or radial thermal gradients}

\author{Di Kang}
\email[]{di.kang@cgu.edu}
\affiliation{Claremont Graduate University}

\author{Ali Nadim}
\email[]{ali.nadim@cgu.edu}
\affiliation{Claremont Graduate University}

\author{Marina Chugunova}
\email[]{marina.chugunova@cgu.edu}
\affiliation{Claremont Graduate University}


\date{\today}


\begin{abstract} 
We study the time evolution of a thin liquid film coating the outer surface of a sphere in the presence of gravity, surface tension and thermal gradients. We derive the fourth-order nonlinear partial differential equation that models the thin film dynamics, including Marangoni terms arising from the dependence of surface tension on temperature. We consider two different imposed temperature distributions with axial or radial thermal gradients. We analyze the stability of a uniform coating under small perturbations and carry out numerical simulations in COMSOL for a range of parameter values. In the case of an axial temperature gradient, we find steady states with either uniform film thickness, or with the fluid accumulating at the bottom or near the top of the sphere, depending on the total volume of liquid in the film, dictating whether gravity or Marangoni effects dominate. In the case of a radial temperature gradient, a stability analysis reveals the most unstable non-axisymmetric modes on an initially uniform coating film.



\end{abstract}

\pacs{TBD} 

\maketitle 

\section{Introduction}
\label{sec:1}

The effect of heating on the dynamics of thin liquid films has been studied by many authors in planar geometry on both smooth and non-smooth substrates. For example an experimental study applying a local heat source to a liquid film falling down a vertical plate was carried out by \cite{kabov2002heat} They observed the deformation of the film surface into a horizontal bump shape, which became unstable above some critical heating value and eventually gave rise to vertical downstream rivulets. \cite{kalliadasis2003marangoni} used the integral boundary-layer approximation of the Navier-Stokes system and free-surface boundary conditions to model thin film coating flow on an inclined heated plane and analyzed the linear stability of steady states for different governing dimensionless groups. The stability analysis of a thin liquid film flowing down an inclined heated plane was extended to account for the effects of evaporation or condensation by \cite{bankoff}. It was also shown that the characteristic time for the development of thin spots decreases as the fourth power of the evaporation rate. Results on the stability of dry patches forming in thin liquid films flowing over a heated planar surface were described by \cite{zuber1966stability}. It was shown that for liquid metals (high wettability materials), the thermal effects were dominant in determining the stability of dry patches. The three-dimensional evolution of the long-wave Marangoni instability of thin liquid films on a planar surface was studied by \cite{oron2000nonlinear}. It was seen that the evolving two-dimensional film is unstable to small three-dimensional random perturbations.  Large scale surface deformations of a liquid film were studied by \cite{bestehorn20033d}. Based on a linear stability analysis, they concluded that the shape deformation is unstable due to the Marangoni effect caused by external heating on a smooth and solid substrate. The stability of a heated liquid film on a plane in the case of a temperature dependent fluid viscosity was analyzed by \cite{reisfeld1990nonlinear}. Through numerical computations, they discovered that when the fluid viscosity is temperature dependent, the rupture time of the film is reduced relative to the constant-viscosity case. \cite{jensen1993spreading} used lubrication theory to derive evolution equations for a surfactant contaminated liquid film height and cross-sectionally averaged bulk surfactant concentration. They also studied the special case of the governing equations corresponding to the Marangoni flow induced by a locally hot region in the liquid layer. Dynamics of thin liquid films on a heated horizontal wall with parallel grooves on its upper surface were studied experimentally and numerically by \cite{alexeev2005marangoni}. They found that the heat transfer in liquid thin film depends on the topography of the wall surface.

Fewer results are available for heated thin liquid films in non-planar geometries.  Experimental results on film breakdown and heat flux for sub-cooled water films flowing downwards on the outside of a vertical uniformly-heated tube can be found in the work of \cite{fujita1978heat}. Some numerical simulations of the flow field and associated heat transfer coefficient for a thin liquid film on a horizontal rotating disk were presented by \cite{rahman1992numerical}. \cite{wilson1994onset} studied the onset of steady Marangoni convection for a thin liquid film coating a rigid sphere for cases where the free surface was either deformable or non-deformable. It was seen that the liquid layer is always stable when heated from the outside and may be unstable when heated from the inside if the magnitude of the non-dimensional Marangoni number is sufficiently large. In this work, we consider the same geometry as \cite{wilson1994onset} but rather than solve for the temperature distribution as part of the solution after linearization, we simply assume that the temperature field is externally imposed. The lubrication approximation in the spherical geometry, as used in our own recent work \cite{kang2016dynamics} but without rotation and with added Marangoni effects, allows us to derive a nonlinear initial-boundary-value problem in the form of a single fourth-order degenerate parabolic partial differential equation for the film thickness. We use this as a basis to study the case where the temperature varies in the vertical direction in the presence of gravity. We then examine the case of radial temperature gradients as well and compare the results from numerical simulations of the nonlinear system with the predictions from a linearized stability analysis.   

Other related areas of research where Marangoni effects at a free surface play an important role include thermo-capillary migration of attached droplets along a surface with an imposed temperature gradient \cite{ford1994, sui2014moving}; evaporative instabilities in thin liquid films and climbing films \cite{hosoi2001evaporative, sultan2005evaporation}; modulating surface stresses for microfluidic actuation \cite{darhuber2005principles}; and non-isothermal spreading of liquid drops or thin fluid ridges on solid substrates \cite{ehrhard1991non, dunn2009quasi}. The mathematical and computational challenges of solving higher order partial differential equations such as the thin-film equation on complex curved surfaces\cite{greer2006fourth, azencot2015functional} are also topics of active research.

The article has the following structure: in Section~\ref{sec:2}, we describe the physical model and derive the governing nonlinear partial differential equation for the evolution of the film profile under the lubrication approximation. The driving forces in the model include gravity, surface tension and Marangoni effects. In Section~\ref{sec:3}, we consider the case of an axial temperature variation and show that it is possible to find steady states where the downward draining flow due to gravity is exactly balanced by an upward Marangoni flow. Depending on the amount of fluid present, one can obtain a uniform film, or have the accumulation of fluid near the top or bottom of the sphere. The streamfunction for the internal flow is obtained for the case of uniform film thickness, illustrating that while the profile is steady, there is continual flow occurring within the film. In Section~\ref{sec:4}, we consider the case with a radially varying temperature field. In the absence of gravity, we complete a linear stability analysis starting with a uniform film and identify the set of unstable modes that include both axisymmetric and non-axisymmetric shapes. We also carry out numerical simulations of the evolution of a slightly perturbed initial profile and compare the numerical growth rates of the unstable modes with the predictions from the linearized analysis. For the axisymmetric case, even in the presence of gravity and with Marangoni effects, we find an energy functional for the system with a dissipation property, i.e., with the energy monotonically decreasing during the time evolution.

\section{Model formulation and governing equations}
\label{sec:2}

In this section we derive the model for the time evolution of a thin viscous liquid film on a solid spherical object of radius $R$ in the presence of gravity $\bm{g}$ and with an imposed steady temperature field which will be described later. The derivation parallels the one in our recent paper\cite{kang2016dynamics} which focused on the effect of centrifugal forces but without Marangoni effects. Here, we do not consider rotational effects but do allow for surface tension variation with temperature and non-axisymmetric shapes. Let $r$ denote the radial distance from the origin, $\theta$ the polar angle ($\theta=0$ corresponds to the positive $z$-direction) and $\phi$ the azimuthal angle 
in spherical coordinates.  The imposed temperature field is a function of position given by $T(r,\theta,\phi)$. The function $h(\theta,\phi,t)$ denotes the thickness of the thin film coating at time $t$ at each point on the spherical substrate. We assume that the model satisfies the standard lubrication approximation constraints. The interface of the thin film $r=R+h(\theta,\phi,t)$ is the zero-level set of the function
\begin{equation}
\mathcal{F}(r,\theta,\phi,t)=r-R-h(\theta,\phi,t)=0\,.\label{calF}
\end{equation}
The kinematic boundary condition ${D\mathcal{F}}/{Dt}=0$ results in
\begin{equation}
\frac{\partial h}{\partial t}=v_{r}-\frac{v_{\theta}}{r}\frac{\partial h}{\partial\theta}-\frac{v_{\phi}}{r\sin\theta}\frac{\partial h}{\partial\phi}\,, \label{kinematic BC}
\end{equation}
at $r=R+h(\theta,\phi,t).$

The continuity equation $\bm{\nabla \cdot v} = 0$ takes the form
\begin{equation}
\frac{1}{r^{2}}\frac{\partial}{\partial r}(r^{2}v_{r})+\frac{1}{r\sin\theta}\frac{\partial}{\partial\theta}\left(\sin\theta\,v_{\theta}\right)+\frac{1}{r\sin\theta}\frac{\partial}{\partial\phi}v_{\phi}=0\,.\label{continuity}
\end{equation}
Multiplying this equation by $r^2 \sin\theta$, integrating along the $r$-direction from $r=R$ to $r=R+h$, and applying the Leibniz rule to bring the derivatives outside the integrals, we obtain
\begin{equation}
\begin{aligned}\sin\theta\,(R+h)^{2}v_{r}|_{R+h} & +\frac{\partial}{\partial\theta}\left(\sin\theta\int_{R}^{R+h}rv_{\theta}dr\right)+\frac{\partial}{\partial\phi}\left(\int_{R}^{R+h}rv_{\phi}dr\right)\\
 & -\sin\theta(R+h)v_{\theta}|_{R+h}\frac{\partial h}{\partial\theta}-(R+h)v_{\phi}|_{R+h}\frac{\partial h}{\partial\phi}=0\,,
\end{aligned}
\end{equation}
where the no-penetration boundary condition $v_r=0$ at the solid surface $r=R$ has also been applied. When combined with the kinematic boundary condition, this yields the time evolution equation for the thin film thickness $h(\theta,\phi,t)$ as
\begin{equation}
\sin\theta\,(R+h)^{2}\,\frac{\partial h}{\partial t}+\frac{\partial}{\partial\theta}\left(\sin\theta\int_{R}^{R+h}rv_{\theta}dr\right)+\frac{\partial}{\partial\phi}\left(\int_{R}^{R+h}rv_{\phi}dr\right)=0\,.\label{eq:evolution}
\end{equation}
In order to get the partial differential equation which models the time evolution of the coating thickness  $h(\theta,\phi,t)$, we need to relate the velocity components $v_{\theta}$ and $v_{\phi}$ to $h(\theta,\phi,t)$ and its derivatives. This can be done by integrating the components of the momentum equation under the Reynolds lubrication approximation.  

Let us introduce the modified pressure $P$ as follows:
\begin{equation}
P=p-\rho\,\bm{g\cdot x}\,.
\end{equation}
Here $\bm{x}$ is the position vector and $\bm{g}$ is the gravitational acceleration which is taken to act in the negative $z$-direction. In terms of the modified pressure, under the lubrication approximation, the $r$-, $\theta$- and $\phi$-components of the momentum equation simplify to\cite{kang2016dynamics}
\begin{eqnarray}
\frac{\partial P}{\partial r}&=&0\,,\label{rmom} \\
\frac{1}{r}\frac{\partial P}{\partial\theta}&=&\frac{\mu}{r^{2}}\frac{\partial}{\partial r}\left(r^{2}\frac{\partial v_{\theta}}{\partial r}\right)\,,\label{thetamom} \\
\frac{1}{r\sin\theta}\frac{\partial P}{\partial\phi}&=&\frac{\mu}{r^{2}}\frac{\partial}{\partial r}\left(r^{2}\frac{\partial v_{\phi}}{\partial r}\right)\,.\label{phimom}
\end{eqnarray}
Eq.~(\ref{rmom}) implies that the modified pressure is independent of $r$ and may only depend on $\theta$, $\phi$ and $t$. The normal stress balance at $r=R+h$, requires the pressure in the film to be equal to the uniform pressure $p_{o}$ in the external air plus the capillary contribution associated with the surface tension $\sigma$ of the interface and its curvature $\nabla\cdot\hat{{\bf n}}$, i.e., $p|_{r=R+h}=p_{o}+\sigma\nabla\cdot\hat{{\bf n}}$. Here, $\hat{{\bf n}}=\nabla{\cal F}/|\nabla{\cal F}|$ (with ${\cal F}$ given by Eq.~(\ref{calF})) is the unit normal at the interface pointing toward the exterior. When the film thickness $h(\theta,\phi,t)$ is much smaller than the sphere radius $R$, the curvature can be approximated in a series in $h$, with the first two terms given by:
\begin{equation}
\nabla\cdot\hat{\bf{n}}\approx\frac{2}{R}-\frac{1}{R^{2}}\left(2h+\frac{1}{\sin\theta}\frac{\partial}{\partial\theta}\left(\sin\theta\frac{\partial h}{\partial\theta}\right)+\frac{1}{\sin^{2}\theta}\frac{\partial^{2}h}{\partial\phi^{2}}\right).\label{curvature}
\end{equation}
In terms of the modified pressure, the normal stress balance becomes:
\begin{equation}
P(\theta,\phi,t)=p_{o}+\sigma\nabla\cdot\hat{{\bf n}}+\rho g(R+h)\cos(\theta)\,.\label{fullmodP}
\end{equation}

To incorporate Marangoni effects into the analysis, we need to consider the tangential stress condition at the interface. The Marangoni effect can be thought of as an effective tangential stress along an interface due to presence of surface tension gradients. When the dependence of surface tension $\sigma$ on the temperature gives rise to such stresses resulting in a flow, this phenomenon is called thermo-capillary convection. The tangential stress balance at the interface is
\begin{equation}
\mathbf{\hat{n}}\cdot(\bm{\tau}_{2}-\bm{\tau}_{1})\cdot\mathbf{\hat{t}}=\nabla_{s}\sigma,
\end{equation}
where $\mathbf{\hat{n}}$ and $\mathbf{\hat{t}}$ are the normal and tangential unit vectors at the interface and $\bm{\tau}_1$ and $\bm{\tau}_2$ are the viscous stresses in the liquid and air phases respectively, the latter being negligible. For small variations in temperature, the dependence of surface tension on temperature can be linearized about a base state in the form
\[
\sigma(T)=\sigma(T_{0})+\left.\frac{\partial\sigma}{\partial T}\right|_{T=T_{0}}(T-T_{0})+\cdots=\sigma_{0}+\sigma_{1}(T-T_{0})+\cdots
\]
where $\sigma_{0}\equiv\sigma(T_{0})$ and $\sigma_{1}\equiv\left.\frac{\partial\sigma}{\partial T}\right|_{T_{0}}$. Typically, surface tension is a decreasing function of temperature, meaning that $\sigma_1<0$.

Under the lubrication approximation, the $\theta$- and $\phi$-components of the tangential stress balance at the interface take the forms:
\begin{equation}
\mu\,r\frac{\partial}{\partial r}\left(\frac{v_{\theta}}{r}\right)=\frac{1}{r}\frac{\partial\sigma}{\partial\theta}\quad\mbox{at}\quad r=R+h,\label{eq:tangential BC1}
\end{equation}
\begin{equation}
\mu\,r\frac{\partial}{\partial r}\left(\frac{v_{\phi}}{r}\right)=\frac{1}{r\sin\theta}\frac{\partial\sigma}{\partial\phi}\quad\mbox{at}\quad r=R+h.\label{eq:tangential BC2}
\end{equation}

Noting that the modified pressure $P$ is independent of $r$, the momentum equations (\ref{thetamom}) and (\ref{phimom}) can be integrated to obtain the following general solutions for $v_{\theta}$ and $v_{\phi}$:
\[
v_{\theta}=\frac{1}{2\mu}\frac{\partial P}{\partial\theta}r-\frac{C_{1}}{r}+C_{2},
\]
\[
v_{\phi}=\frac{1}{2\mu\sin\theta}\frac{\partial P}{\partial\phi}r-\frac{C_{3}}{r}+C_{4}.
\]
To solve for the integration ``constants'' (these are independent of $r$ but may depend on the other variables), we apply the no-slip boundary condition on the solid surface and the tangential stress balances (\ref{eq:tangential BC1}) and (\ref{eq:tangential BC2}), which include the Marangoni effects, on the liquid-air interface. These result in
\begin{eqnarray*}
C_{1}&=&-\frac{R^{2}(R+h)}{2\mu(R-h)}\frac{\partial P}{\partial\theta}+\frac{R(R+h)}{\mu(R-h)}\frac{\partial\sigma}{\partial\theta}, \\
C_{2}&=&-\frac{R^{2}}{\mu(R-h)}\frac{\partial P}{\partial\theta}+\frac{(R+h)}{\mu(R-h)}\frac{\partial\sigma}{\partial\theta}.\\
C_{3}&=&-\frac{R^{2}(R+h)}{2\mu(R-h)\sin\theta}\frac{\partial P}{\partial\phi}+\frac{R(R+h)}{\mu(R-h)\sin\theta}\frac{\partial\sigma}{\partial\phi},\\
C_{4}&=&-\frac{R^{2}}{\mu(R-h)\sin\theta}\frac{\partial P}{\partial\phi}+\frac{(R+h)}{\mu(R-h)\sin\theta}\frac{\partial\sigma}{\partial\phi}.
\end{eqnarray*}

Having found the relationships between $v_{\theta}$, $v_{\phi}$ and the film thickness $h(\theta,\phi,t)$, the integrals within the evolution equation (\ref{eq:evolution}) can be evaluated and simplified. In the lubrication limit, the film thickness $h$ is everywhere small compared to the sphere radius $R$, so that all factors $R+h$ and $R-h$ that appear in the resulting equation can be approximated by $R$ to leading order. This results in the following self-contained fourth-order partial differential equation for the evolution of the film thickness:
\begin{equation}
\frac{\partial h}{\partial t}-\frac{1}{R^{2}\sin\theta}\frac{\partial}{\partial\theta}\left(\frac{h^{3}\sin\theta}{3\mu}\frac{\partial P}{\partial\theta}-\frac{h^{2}\sin\theta}{2\mu}\frac{\partial\sigma}{\partial\theta}\right)-\frac{1}{R^{2}\sin^2\theta}\frac{\partial}{\partial\phi}\left(\frac{h^{3}}{3\mu}\frac{\partial P}{\partial\phi}-\frac{h^{2}}{2\mu}\frac{\partial\sigma}{\partial\phi}\right)=0\,,
\label{eq:evolution h eqn}
\end{equation}
where
\begin{equation}
P(\theta,\phi,t)=\rho g R \,\cos\theta-\frac{\sigma}{R^{2}}\left(2h+\frac{1}{\sin\theta}\frac{\partial}{\partial\theta}\left(\sin\theta\frac{\partial h}{\partial\theta}\right)+\frac{1}{\sin^{2}\theta}\frac{\partial^{2}h}{\partial\phi^{2}}\right).
\label{eq:pressurefield}
\end{equation}

The Marangoni terms in the evolution equation (\ref{eq:evolution h eqn}) are the contributions to the flux components proportional to ${\partial\sigma}/{\partial\theta}$ and ${\partial\sigma}/{\partial\phi}$. However, there is a subtle issue with this derivation. In particular, since surface tension $\sigma$ appears in the expression for the pressure $P$, it may seem that differentiating $P$ with respect to $\theta$ or $\phi$, as needed in the above evolution equation, would introduce additional Marangoni terms involving the derivatives of $\sigma$. However, scaling arguments given below will show that in the normal stress balance which leads to the dependence of $P$ on $\sigma$, one can simply use the leading constant term $\sigma_0$ for the surface tension, whereas in the actual Marangoni terms, it will be the next term $\sigma_1$ that plays the main role.

For the purpose of scaling the equation in the following sections, we can define the characteristic film thickness $H$ to be the average thickness of the thin film based on its initial profile. One can then define the dimensionless film thickness to be $\hat{h}=h/H$ . The lubrication approximation requires $\epsilon={H}/{R}\ll1$. Also, one can choose an arbitrary time scale $\tau$ whose value can be determined later by scaling arguments and define a dimensionless time $\hat{t}=t/\tau$. Within the thin film equation (\ref{eq:evolution h eqn}), we have driving forces associated with gravity, surface tension and Marangoni effects. The scaling we shall choose for each of the cases below will be one in which all three of these effects are of similar orders of magnitude, balancing one another. This will be done for two different imposed  temperature fields (vertical and radial distributions) in the next two sections.

\section{Vertical temperature field}
\label{sec:3}

\subsection{The leading order equation}

Let us suppose that an externally imposed temperature field exists which varies linearly in the vertical coordinate $z$ with a vertical temperature gradient $k_1$; that is,
\[
T=T_{0}+k_{1}\,z = T_0+ k_1 \,r\, \cos\theta \,.
\]
From the assumed linear dependence of surface tension on temperature, we thus find
\[
\sigma=\sigma_{0}+\sigma_{1}\,k_{1}\,(R+h(\theta,\phi,t))\,\cos\theta \approx \sigma_{0}+\sigma_{1}\,k_{1}\,R\,\cos\theta\,,
\]
at the liquid-air interface.

Using this result in the evolution equation (\ref{eq:evolution h eqn}), scaling $h$ with $H$ and $t$ with $\tau$ as suggested earlier, and simplifying to get the equation for $\partial\hat{h}/\partial\hat{t}$, results in a dimensionless equation in which the following three dimensionless groups appear respectively in front of the contributions to the flux due to gravity, surface tension and Marangoni effects (in the $\theta$-derivative):
\begin{equation}
\label{acd}
{\cal G}\equiv\frac{\rho gH^{2}\tau}{3\mu R}\,, \quad {\cal S}\equiv \frac{\sigma_{0}H^{3}\tau}{3\mu R^{4}}\,,\quad \mbox{and}\quad {\cal M}\equiv\frac{\sigma_{1}k_{1}H\tau}{2\mu R}\,.
\end{equation}
Note that the Marangoni parameter $\cal M$ may be negative since $\sigma_1$ and $k_1$ can have either sign. We seek a model in which all the above mentioned effects are present and of the same orders of magnitude. To balance these three terms, we thus need
\[
\sigma_{1}k_{1}\sim\epsilon^{2}\frac{\sigma_{0}}{R}\sim\epsilon\rho gR\,.
\]
Equivalently, the constant part of surface tension $\sigma_0$ must be fairly large compared to gravitational effects ($\sigma_0\sim \rho g R^2 /\epsilon$) in order for capillarity and gravity to both be important. Also the typical variation in surface tension across the length of the film ($\sigma_1 k_1 R$) due to temperature gradients, must be quite small compared to the constant part of the surface tension $\sigma_0$, i.e., $\sigma_1 k_1 R\sim \epsilon^2 \sigma_0$. In this limit, all three effects (gravity, capillarity and Marangoni) contribute equally to the film evolution and the dimensionless groups $\cal G$, $\cal S$ and $\cal M$ introduced above are all of order unity. Time scale $\tau$ can be chosen to make any one of the three parameters equal to one exactly, if desired. Also, in the normal stress balance at the interface, $p|_{r=R+h}=p_{o}+\sigma\nabla\cdot\hat{{\bf n}}$, one can replace $\sigma$ with $\sigma_0$ without affecting the subsequent evolution equation at leading order.

Under this scaling, in the $\phi$-derivative term of the evolution equation, only the capillarity term due to the normal stress balance contributes at leading order. The scaled evolution equation (from which the hats have been dropped for clarity), takes the form:
\begin{multline*}
\frac{\partial h}{\partial t}+\frac{1}{\sin\theta}\frac{\partial}{\partial\theta}\left\{\sin\theta h^{3}\left[{\cal G} \sin\theta+{\cal S}\frac{\partial}{\partial\theta}\left(2h+\frac{1}{\sin\theta}\frac{\partial}{\partial\theta}\left(\sin\theta\frac{\partial h}{\partial\theta}\right)+\frac{1}{\sin^{2}\theta}\frac{\partial^{2}h}{\partial\phi^{2}}\right)\right]-{\cal M}\sin^{2}\theta h^{2}\right\}\\
+\frac{1}{\sin^{2}\theta}\frac{\partial}{\partial\phi}\left\{{\cal S} h^{3}\frac{\partial}{\partial\phi}\left(2h+\frac{1}{\sin\theta}\frac{\partial}{\partial\theta}\left(\sin\theta\frac{\partial h}{\partial\theta}\right)+\frac{1}{\sin^{2}\theta}\frac{\partial^{2}h}{\partial\phi^{2}}\right)\right\}=0.
\end{multline*}
where parameters $\cal G$, $\cal S$ and $\cal M$ are given by equation (\ref{acd}).

With the change of variable $x=-\cos\theta$, the equation becomes
\begin{multline}
\label{EvolEqVert}
\frac{\partial h}{\partial t}+\frac{\partial}{\partial x}\left\{ (1-x^{2}) h^{3}\left[{\cal G}+{\cal S}\frac{\partial}{\partial x}\left(2h+\frac{\partial}{\partial x}\left((1-x^{2})\frac{\partial h}{\partial x}\right)+\frac{1}{1-x^{2}}\frac{\partial^{2}h}{\partial\phi^{2}}\right)\right]-{\cal M}(1-x^{2})h^{2}\right\}\\
+\frac{1}{1-x^{2}}\frac{\partial}{\partial\phi}\left\{ {\cal S} h^{3}\frac{\partial}{\partial\phi}\left[2h+\frac{\partial}{\partial x}\left((1-x^{2})\frac{\partial h}{\partial x}\right)+\frac{1}{1-x^{2}}\frac{\partial^{2}h}{\partial\phi^{2}}\right]\right\} =0\,,
\end{multline}
over the range $-1<x<1$ and $t>0$. It is this form of the equation that is used in the numerical simulations presented below.

\subsection{Uniform films and their stability}

\begin{figure}
\begin{centering}
\includegraphics[scale=0.15]{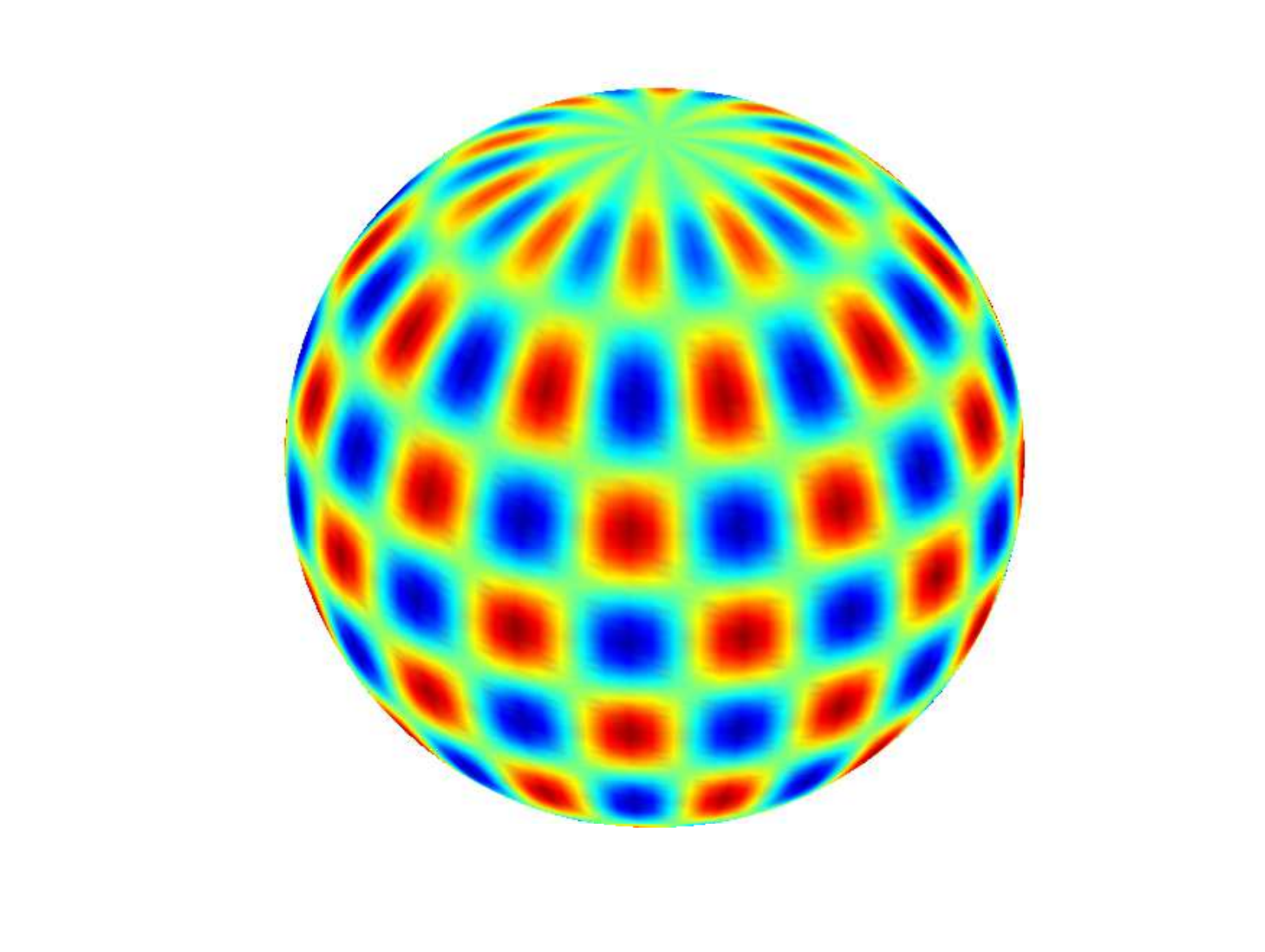}\includegraphics[scale=0.15]{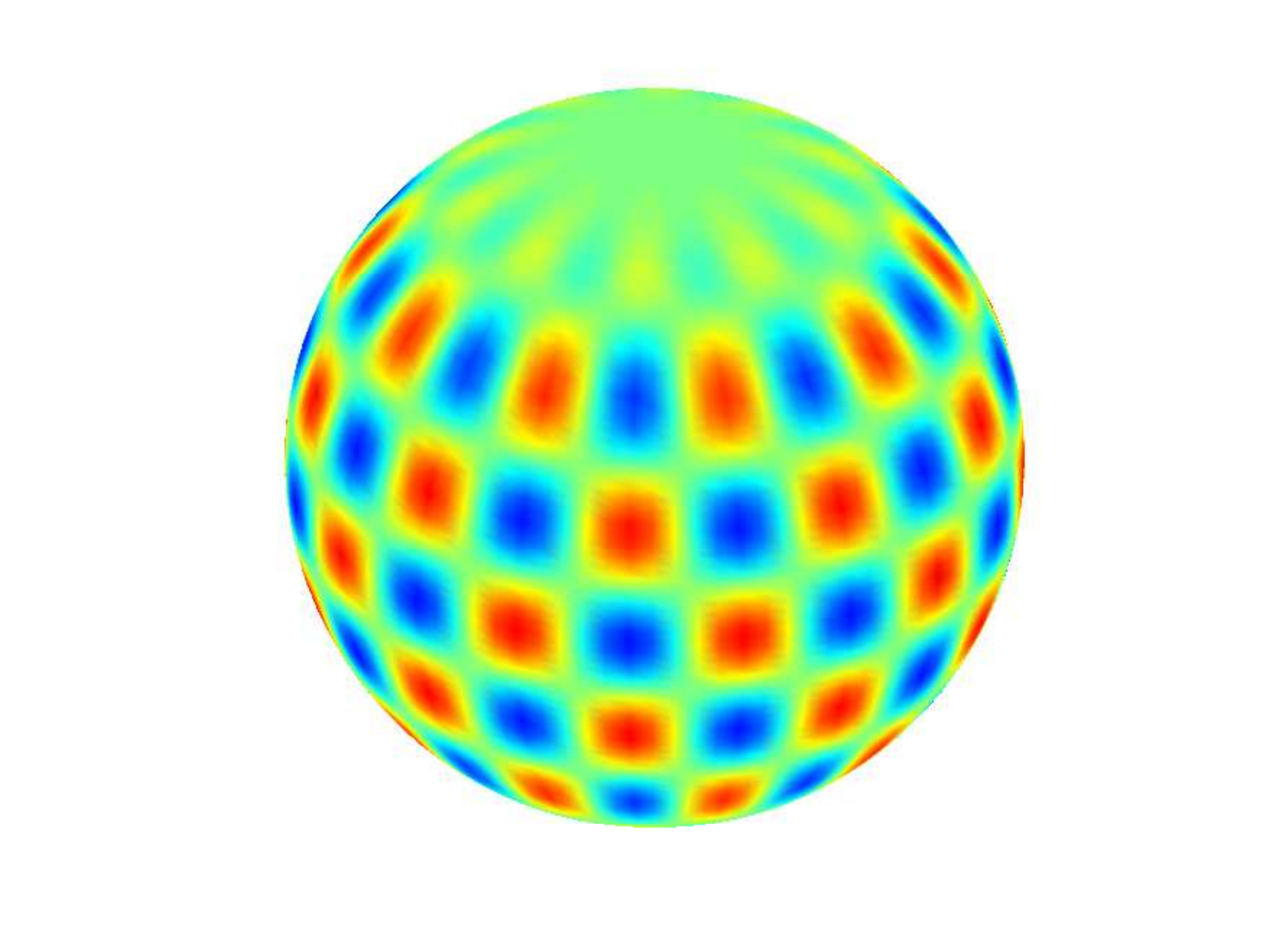}\includegraphics[scale=0.15]{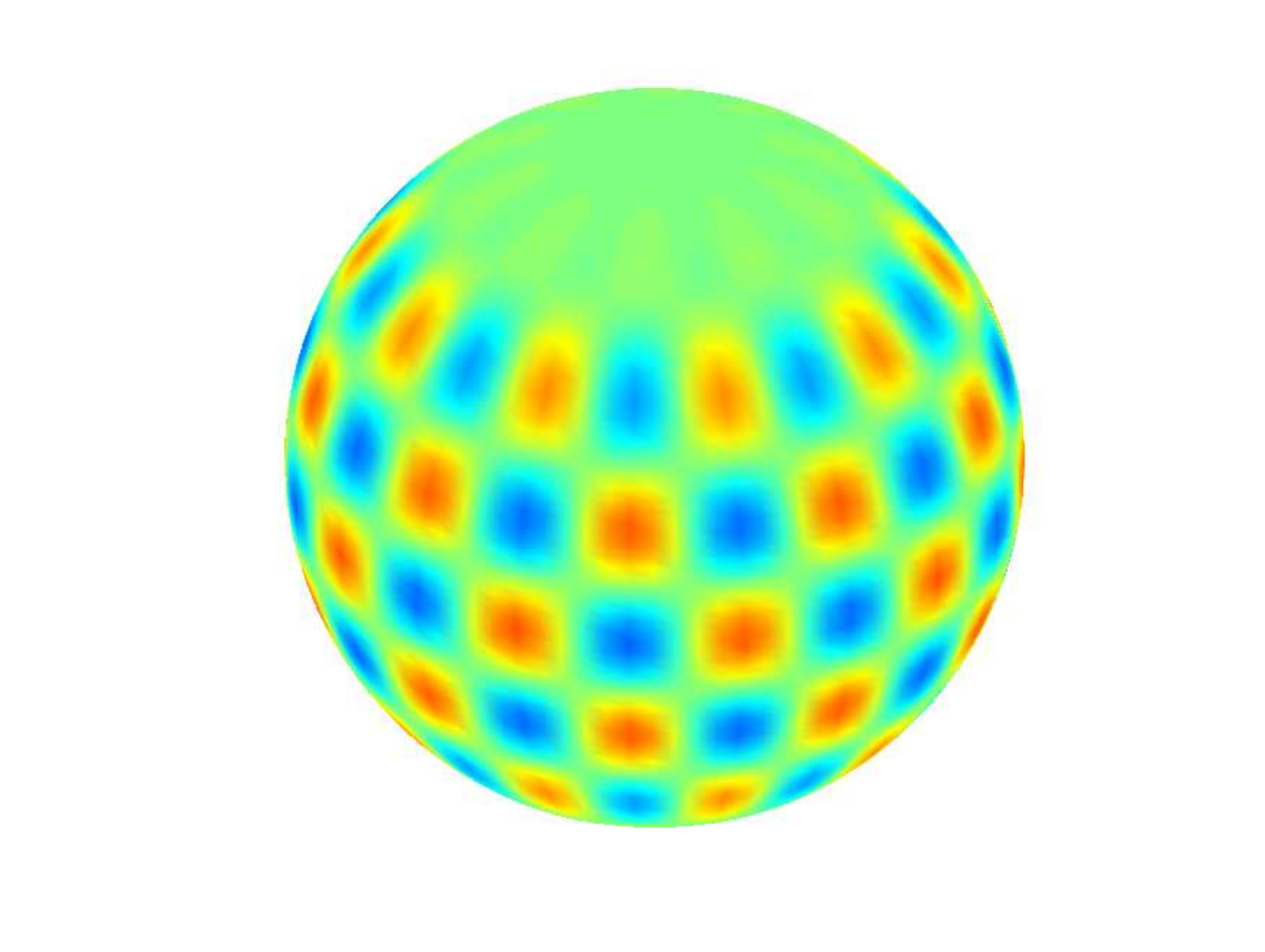}\includegraphics[scale=0.15]{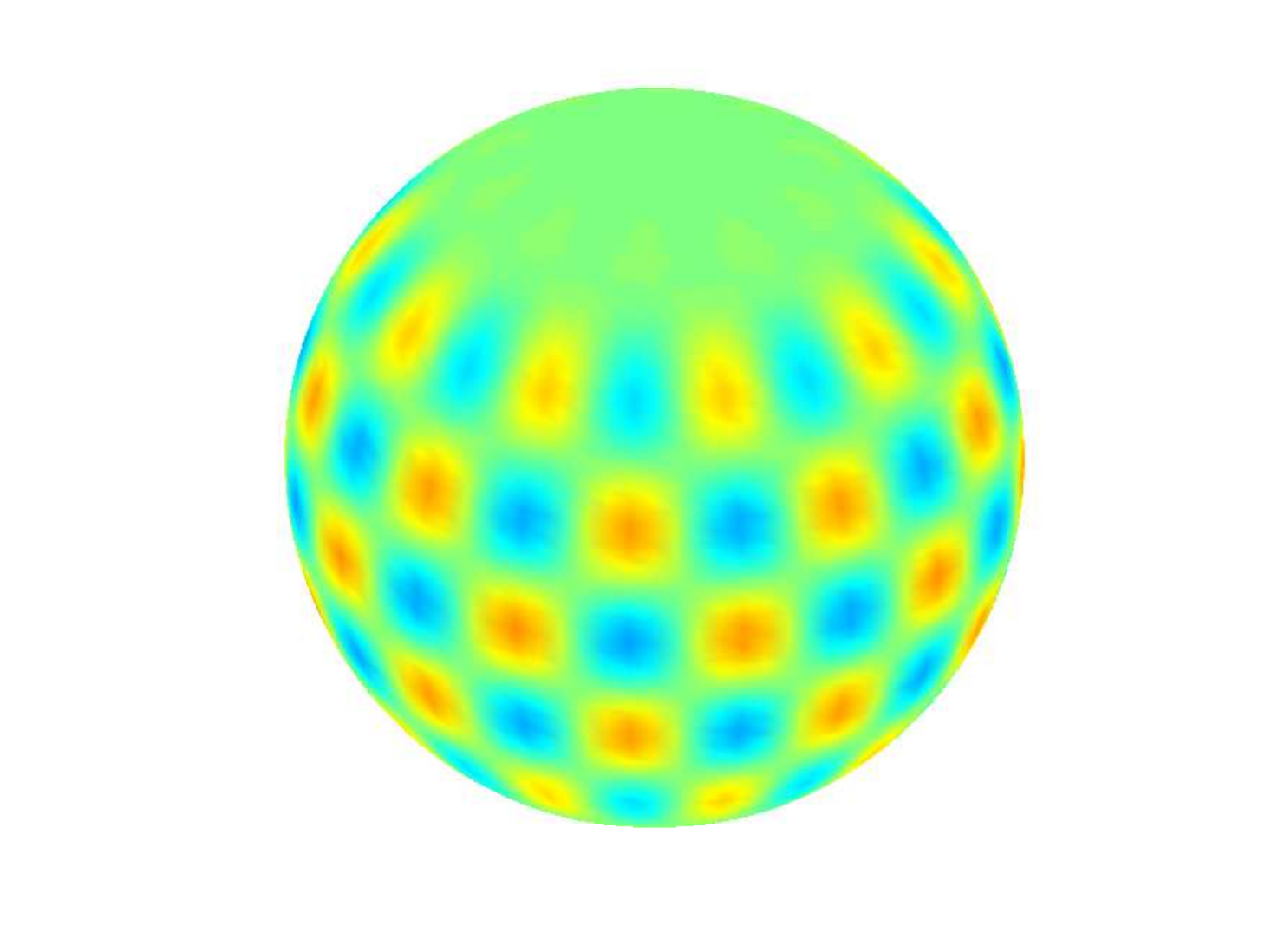}\includegraphics[scale=0.15]{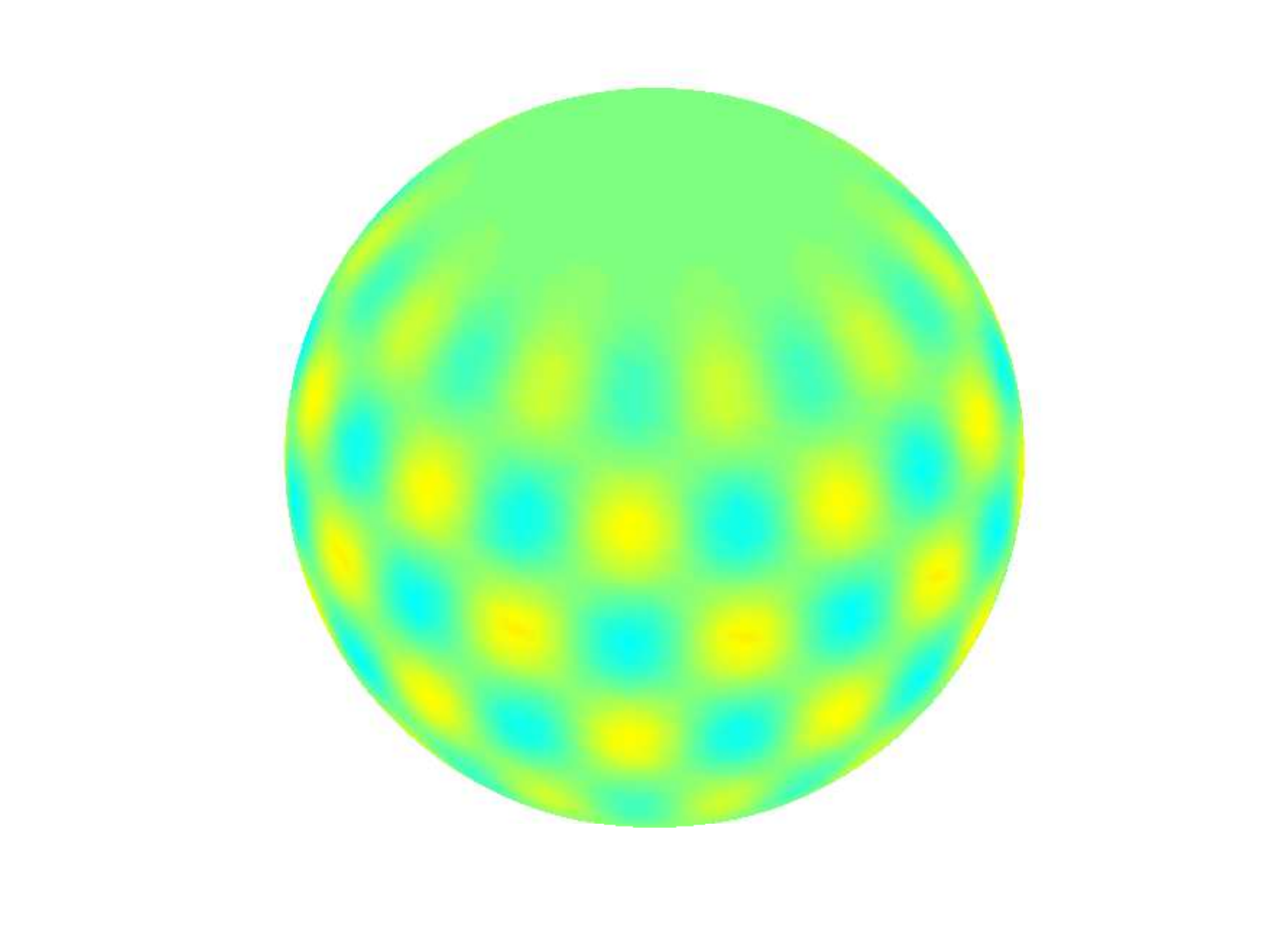}
\par\end{centering}
\protect\caption{When ${\cal G}=1$, ${\cal S}=1$ and ${\cal M}=-1$, the slightly perturbed initial state $h=1+10^{-5}\cos(10x)\cos(10\phi)$ will become uniform eventually. The time values from left to right are $t=0$, $t=10^{-5}$, $t=2\times10^{-5}$, $t=3\times10^{-5}$ and $t=5\times10^{-5}$.
\label{fig:vertical a=00003D1 d=00003D-1 to uniform}}
\end{figure}

\begin{figure}
\begin{centering}
\includegraphics[scale=0.15]{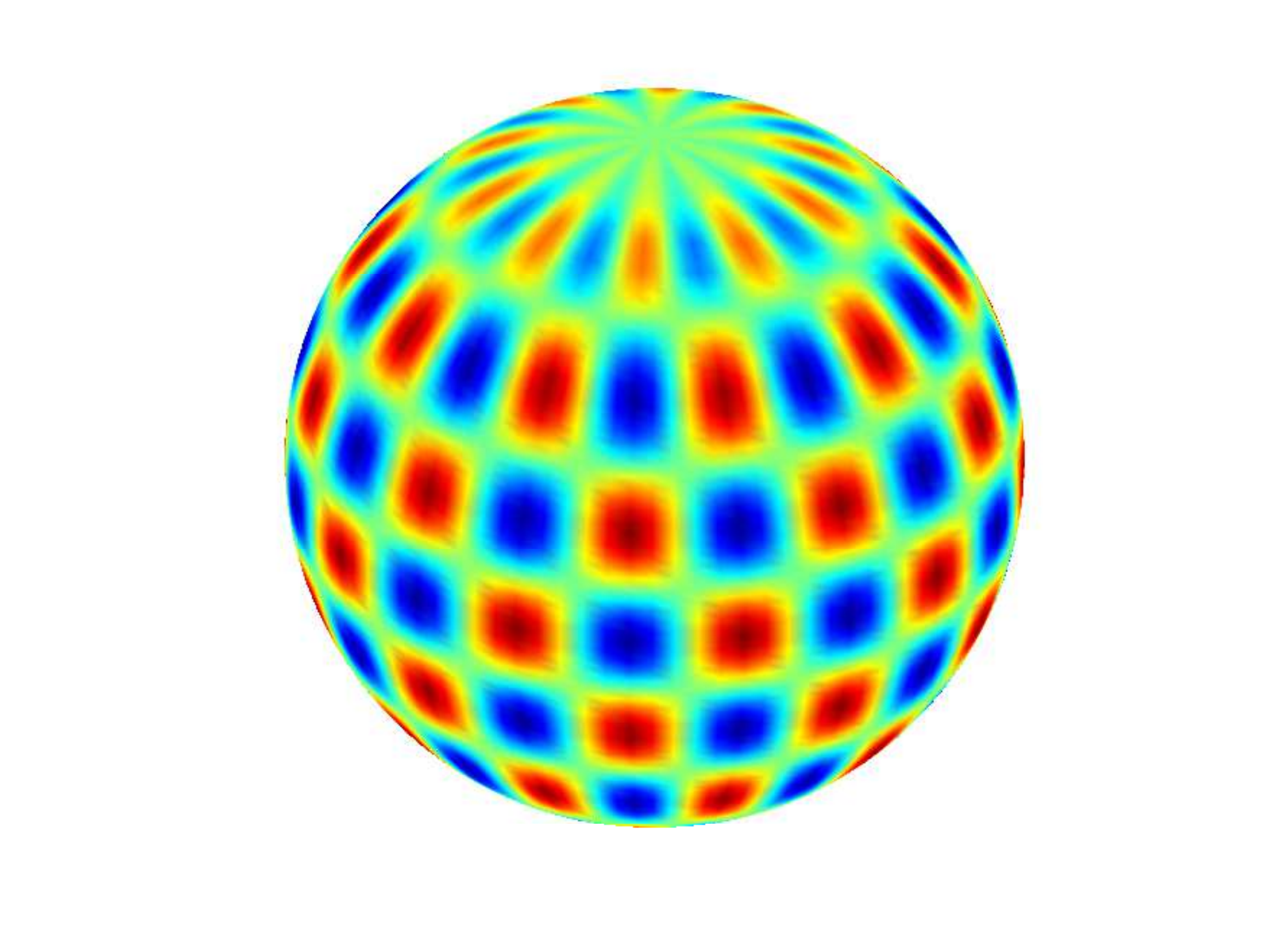}\includegraphics[scale=0.15]{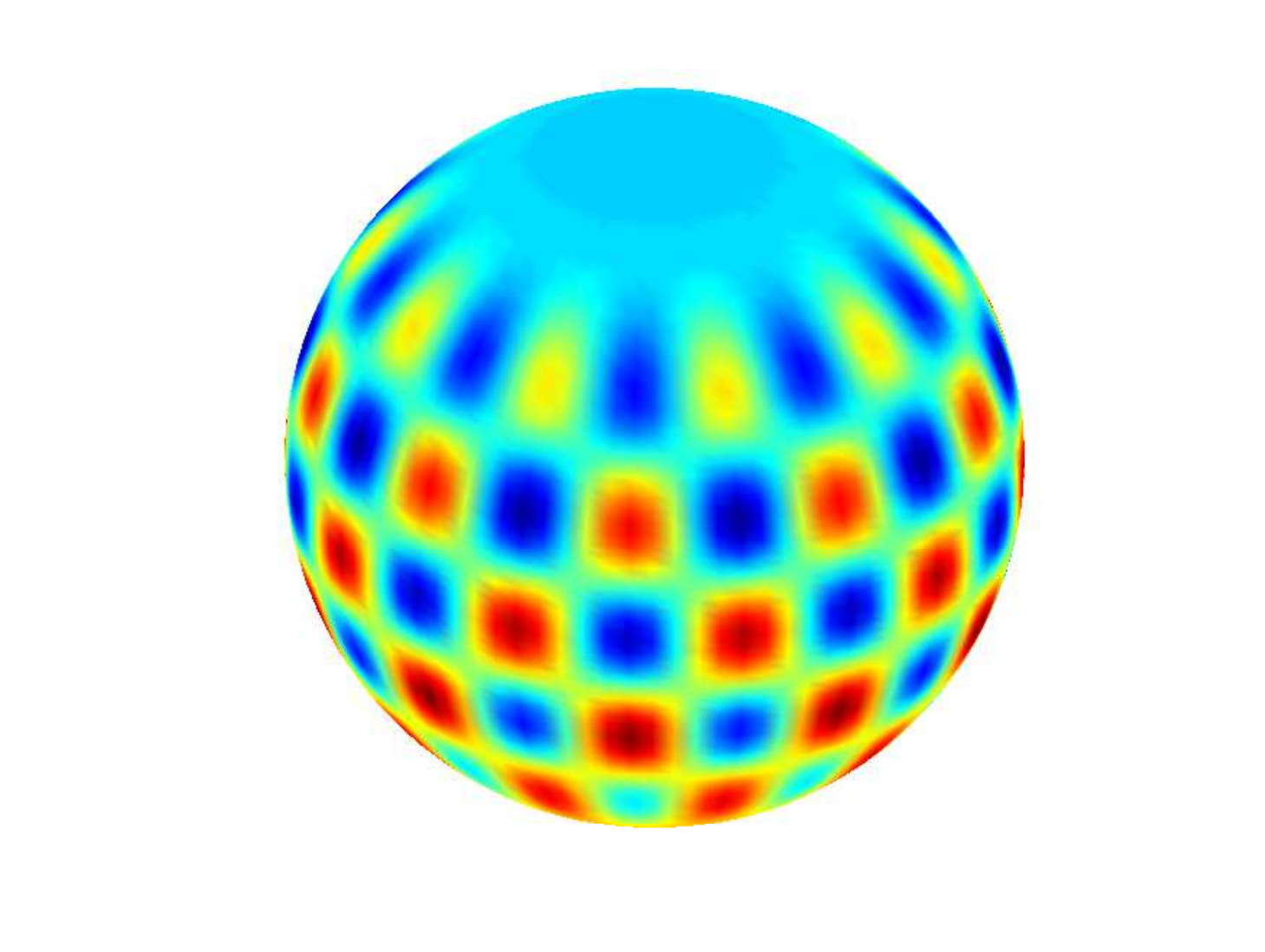}\includegraphics[scale=0.15]{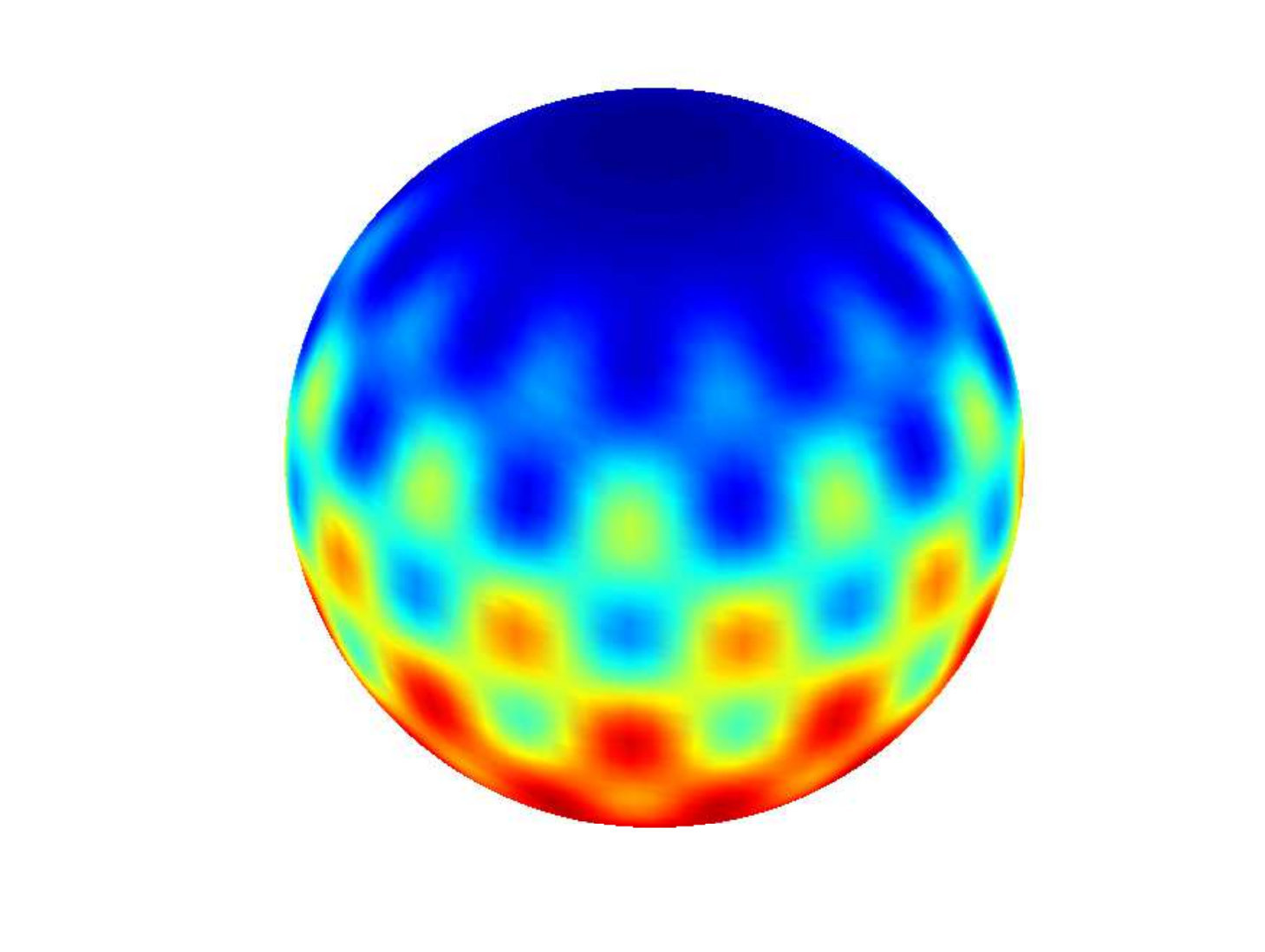}\includegraphics[scale=0.15]{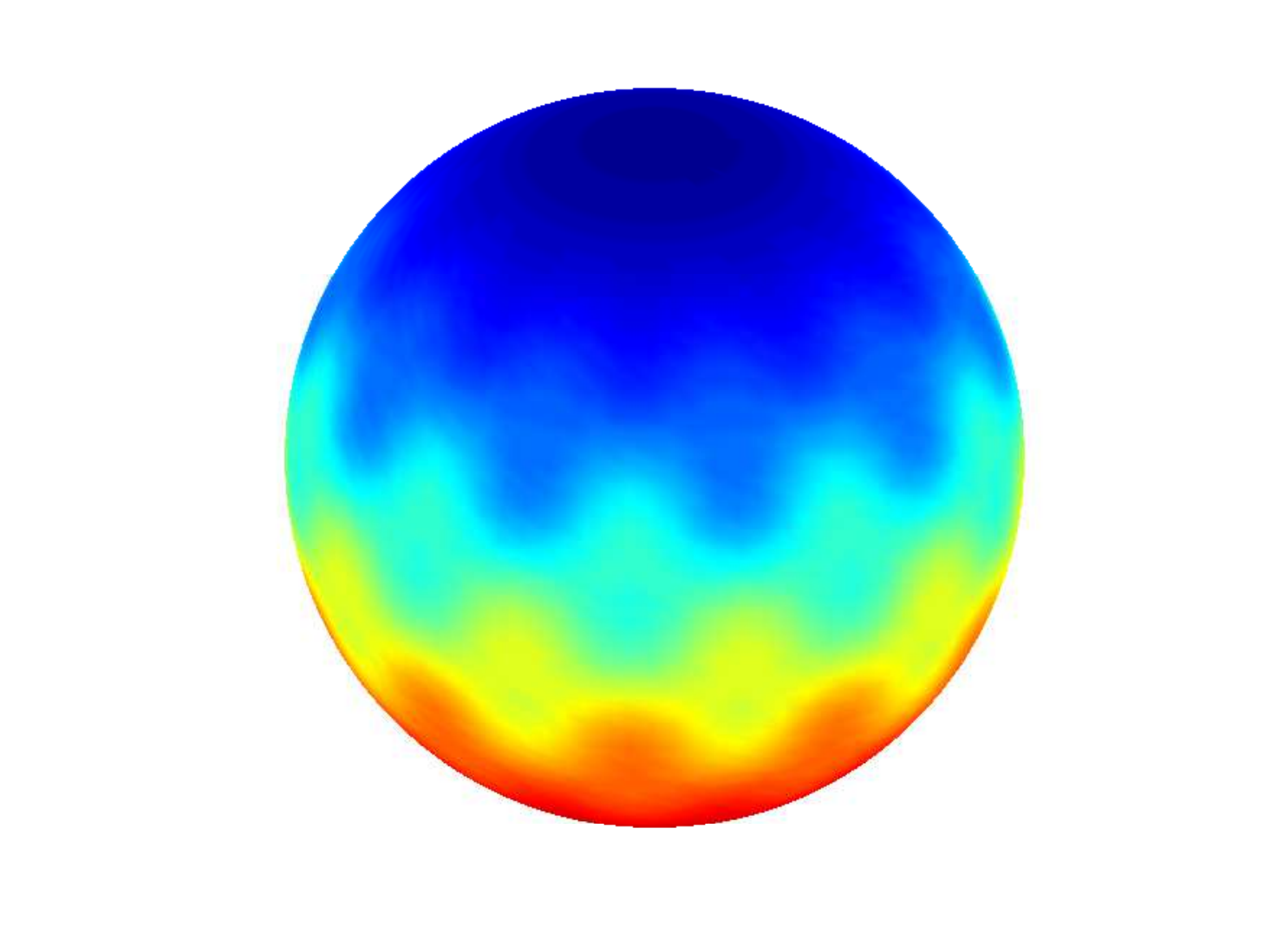}\includegraphics[scale=0.15]{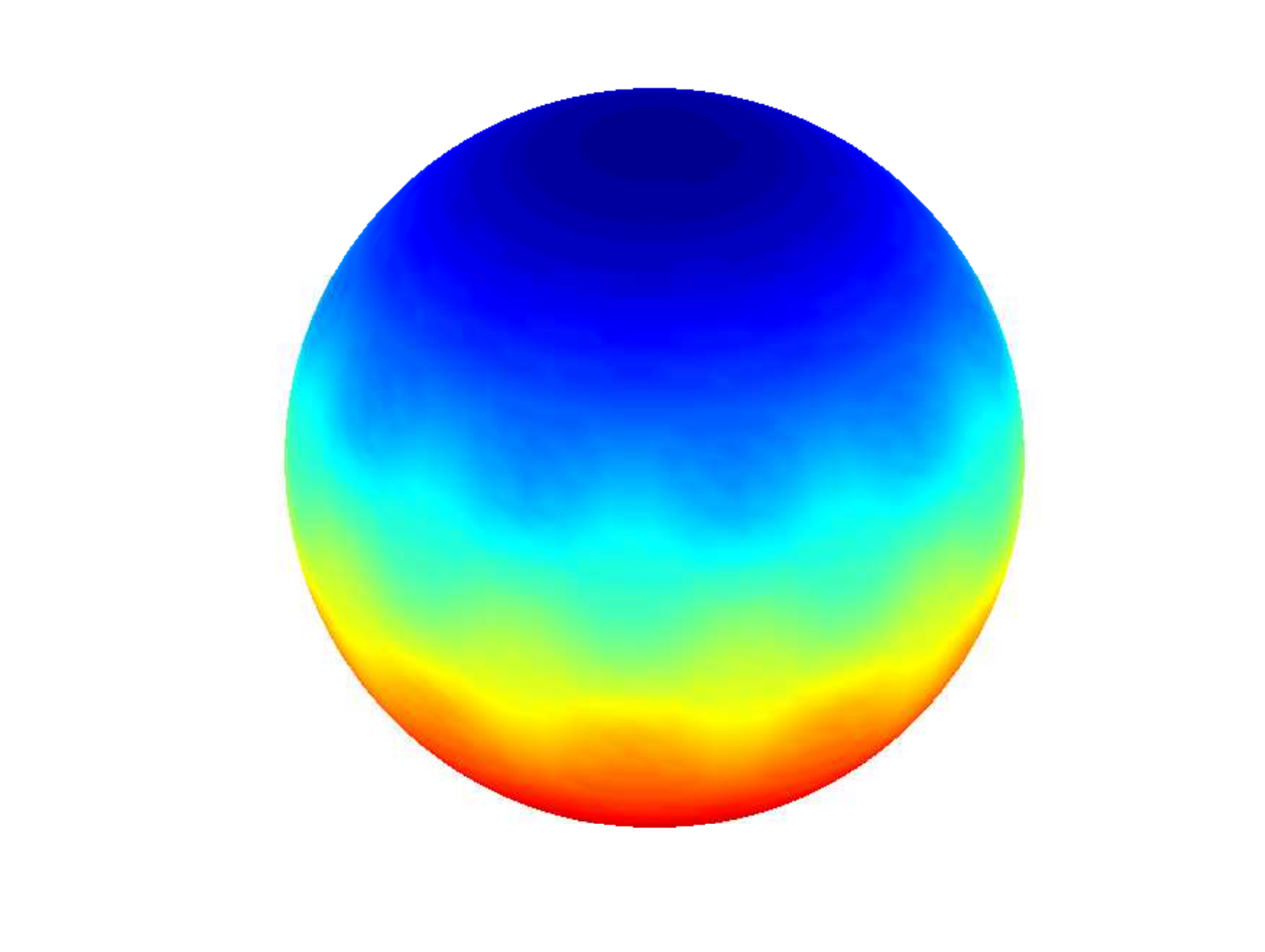}
\par\end{centering}
\protect\caption{When ${\cal G}=1$, ${\cal S}=1$ and ${\cal M}=1$, with the initial thickness given by $h=1.01+10^{-5}\cos(10x)\cos(10\phi)$, the solution shown at times $t=0$, $t=5\times10^{-5}$, $t=10^{-4}$, $t=1.5\times10^{-4}$, $t=2\times10^{-4}$ develops a hanging drop profile near the bottom of the sphere, being thicker at the bottom than on top. 
\label{fig:vertical-down}}
\end{figure}

\begin{figure}
\begin{centering}
\includegraphics[scale=0.15]{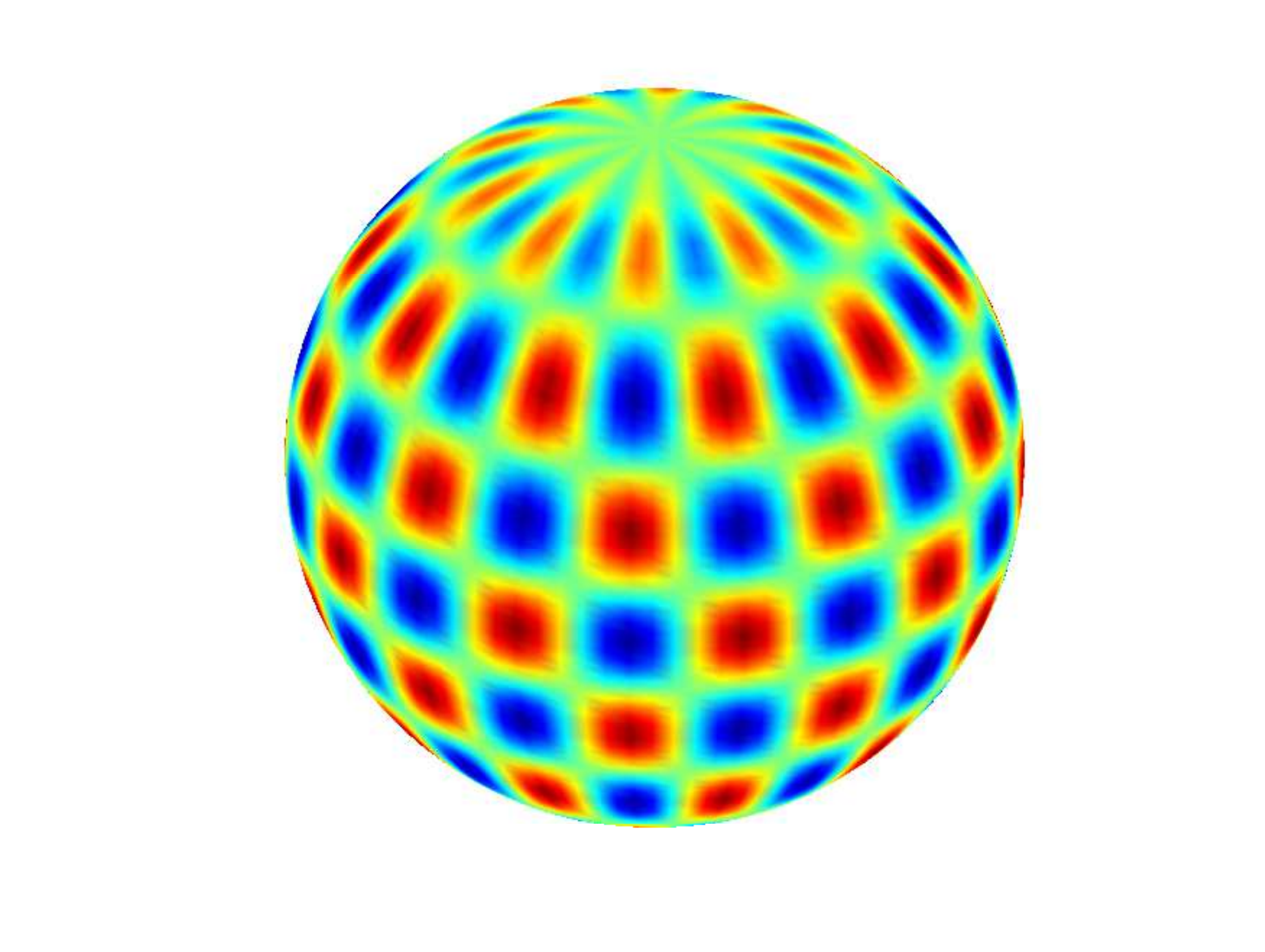}\includegraphics[scale=0.15]{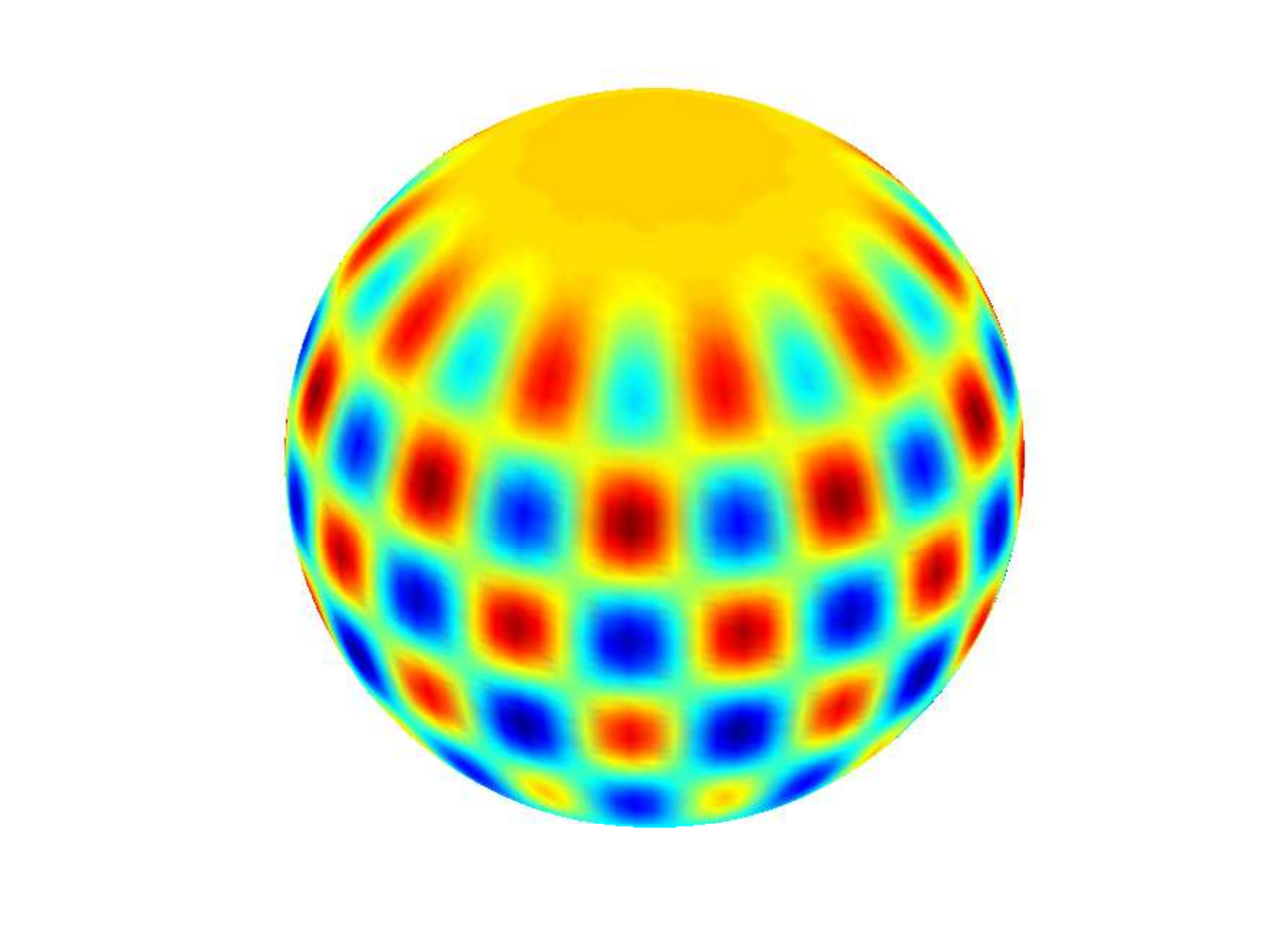}\includegraphics[scale=0.15]{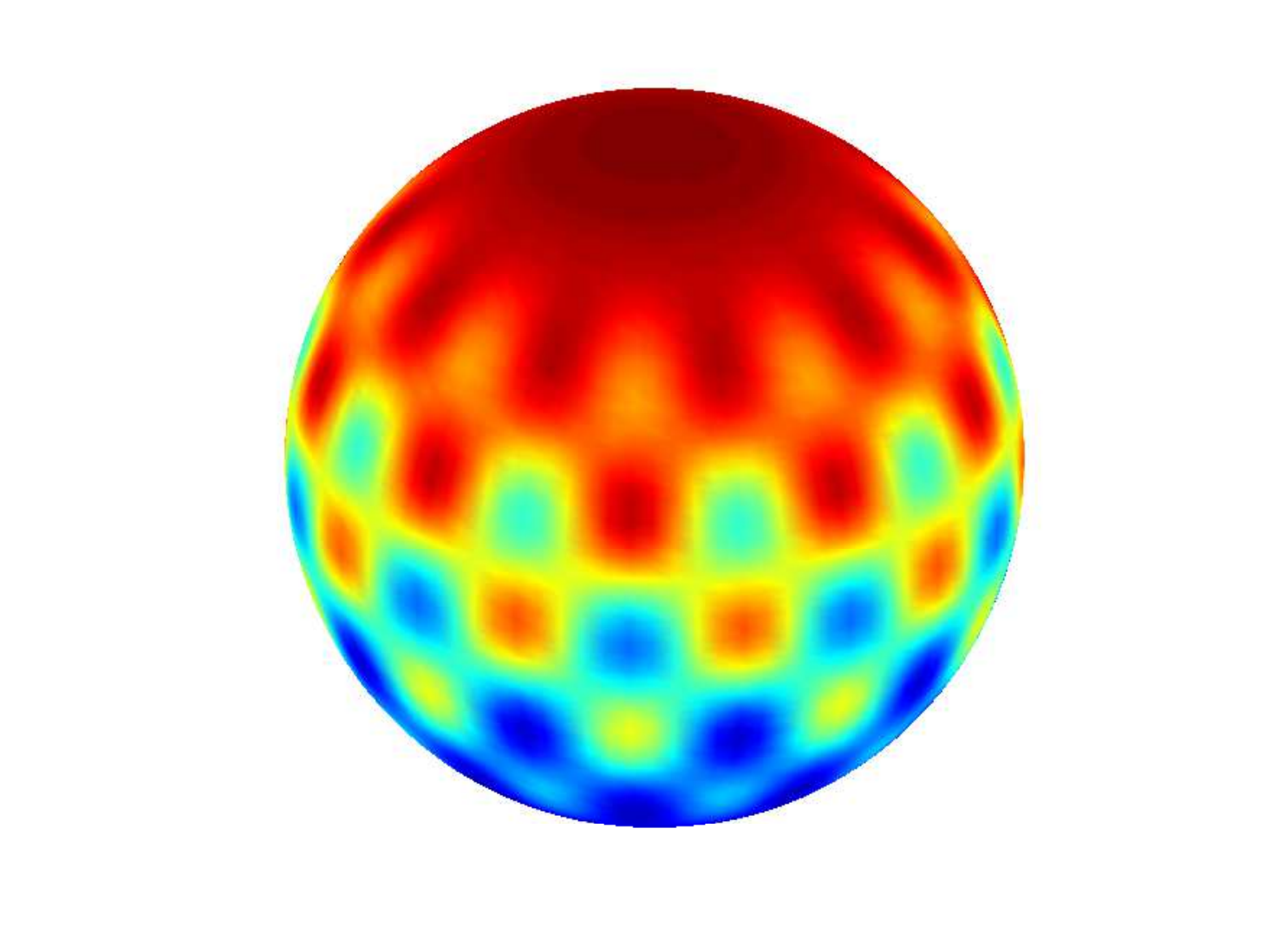}\includegraphics[scale=0.15]{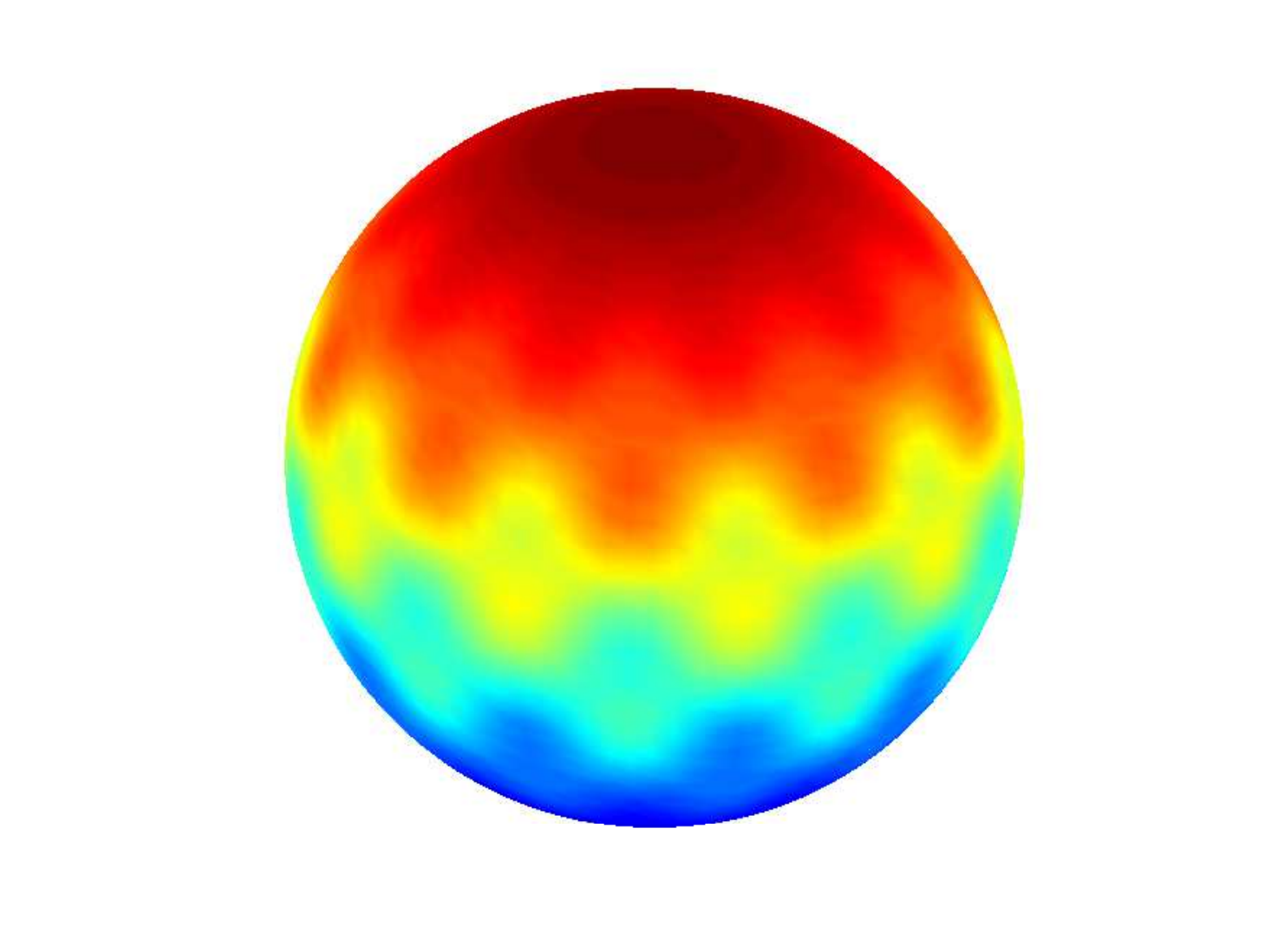}\includegraphics[scale=0.15]{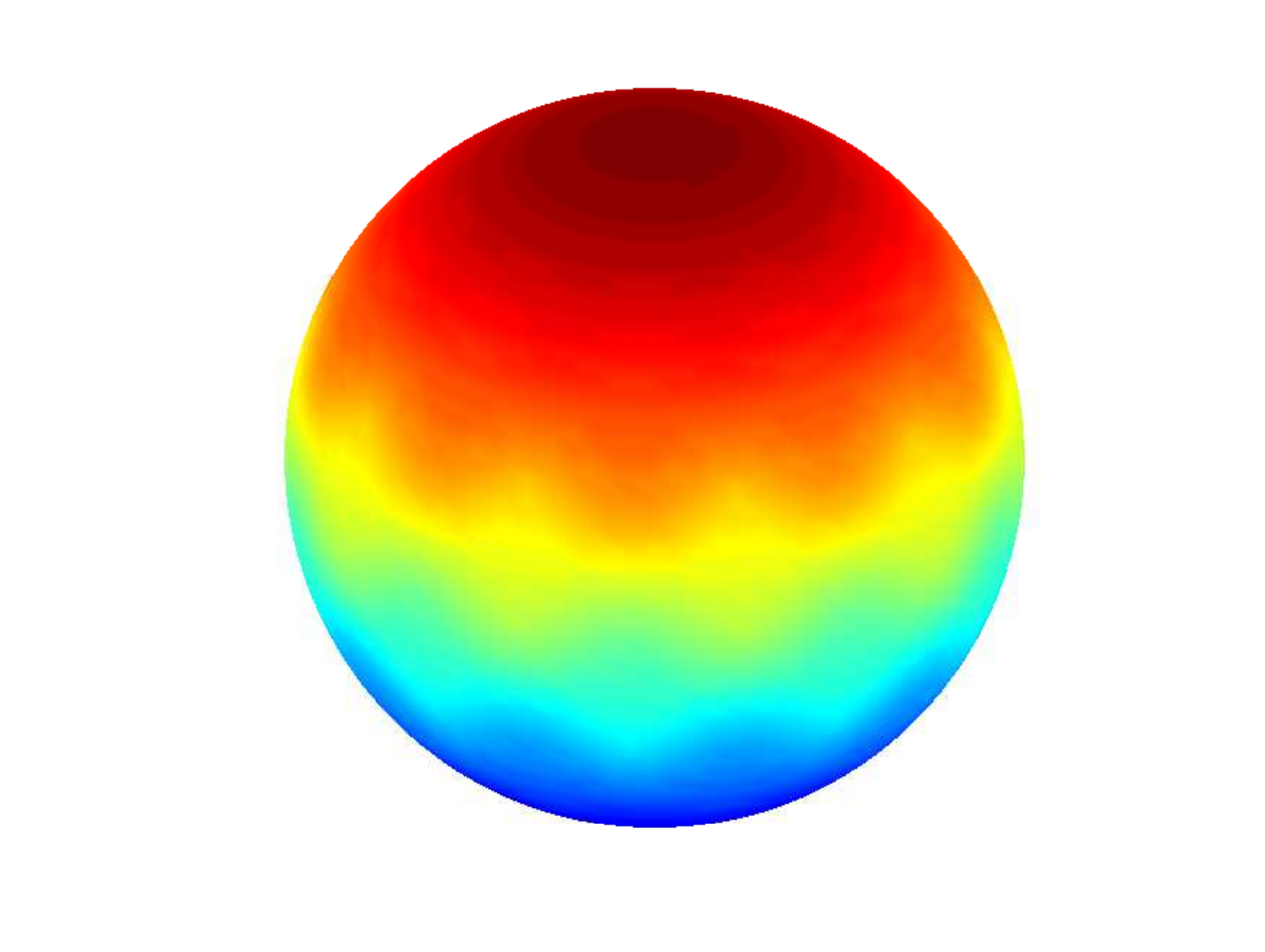}
\par\end{centering}
\protect\caption{When ${\cal G}=1$, ${\cal S}=1$ and ${\cal M}=1$, with initial thickness given by $h=0.99+10^{-5}\cos(10x)\cos(10\phi)$, the solution shown at times
$t=0$, $t=5\times10^{-5}$, $t=10^{-4}$, $t=1.5\times10^{-4}$,
$t=2\times10^{-4}$ develops a bump near the top of the sphere, being thicker at the top than at the bottom. 
\label{fig:vertical-up}}
\end{figure}

\begin{figure}
\begin{centering}
\includegraphics[scale=0.15]{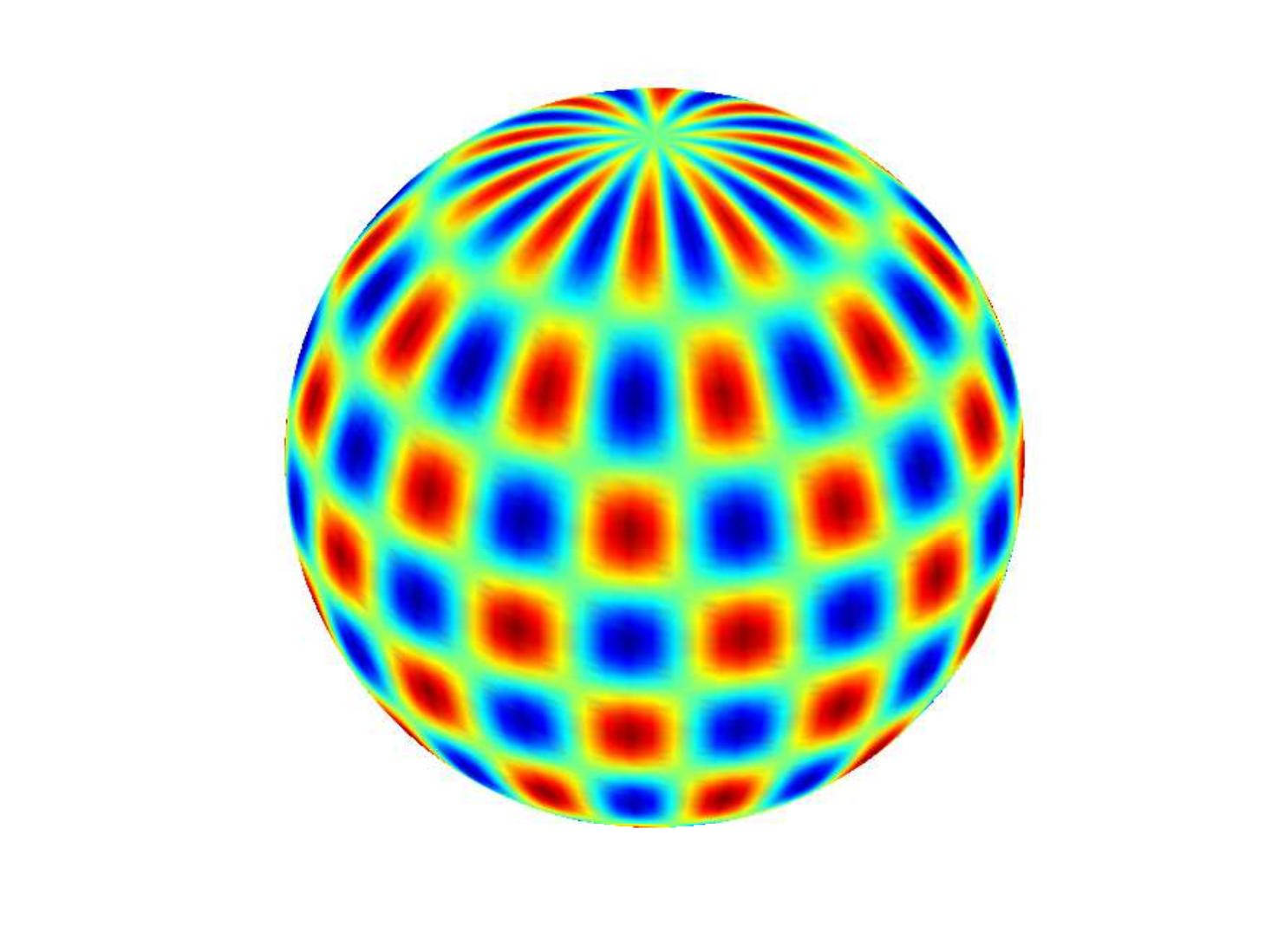}\includegraphics[scale=0.15]{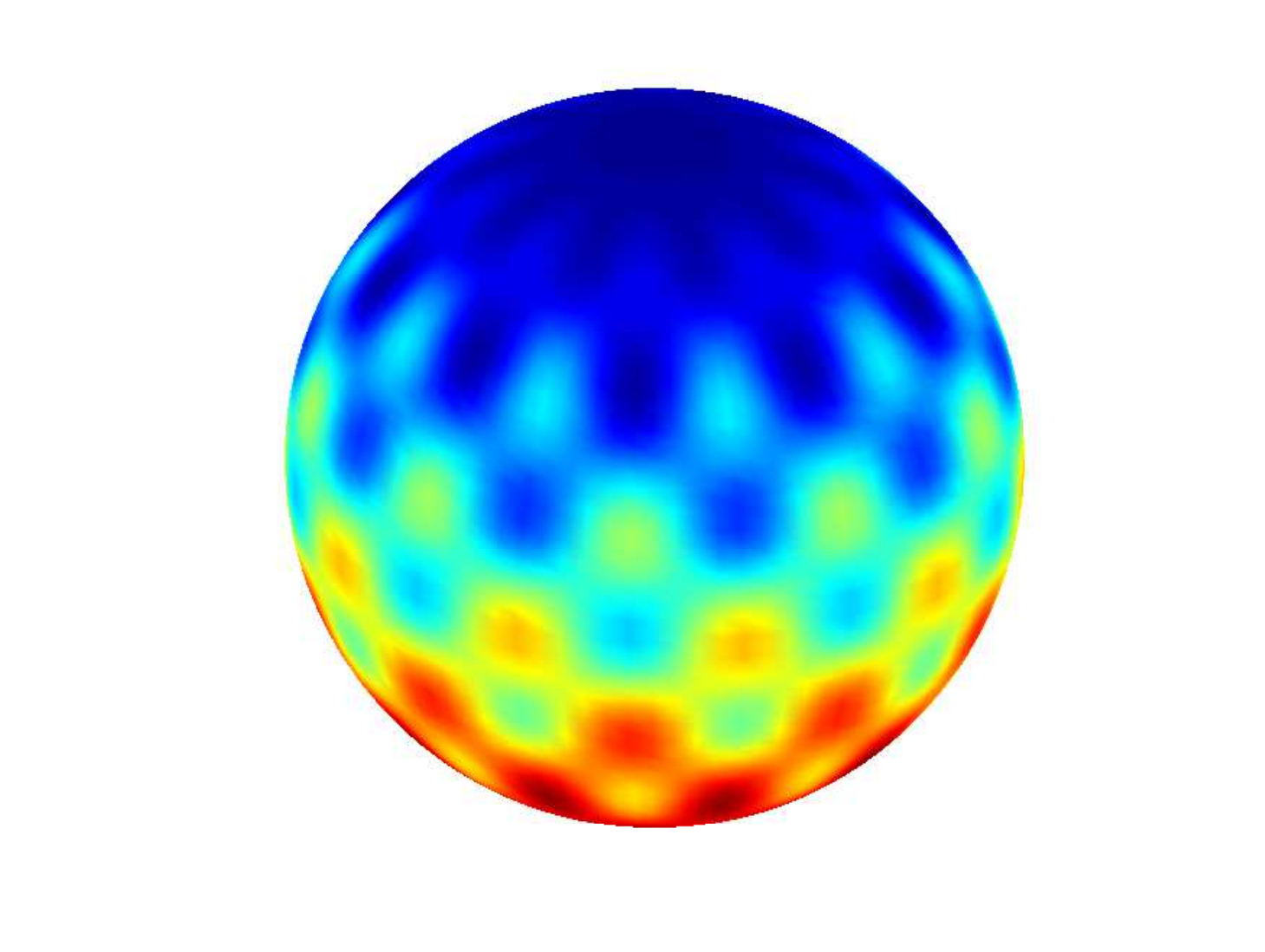}\includegraphics[scale=0.15]{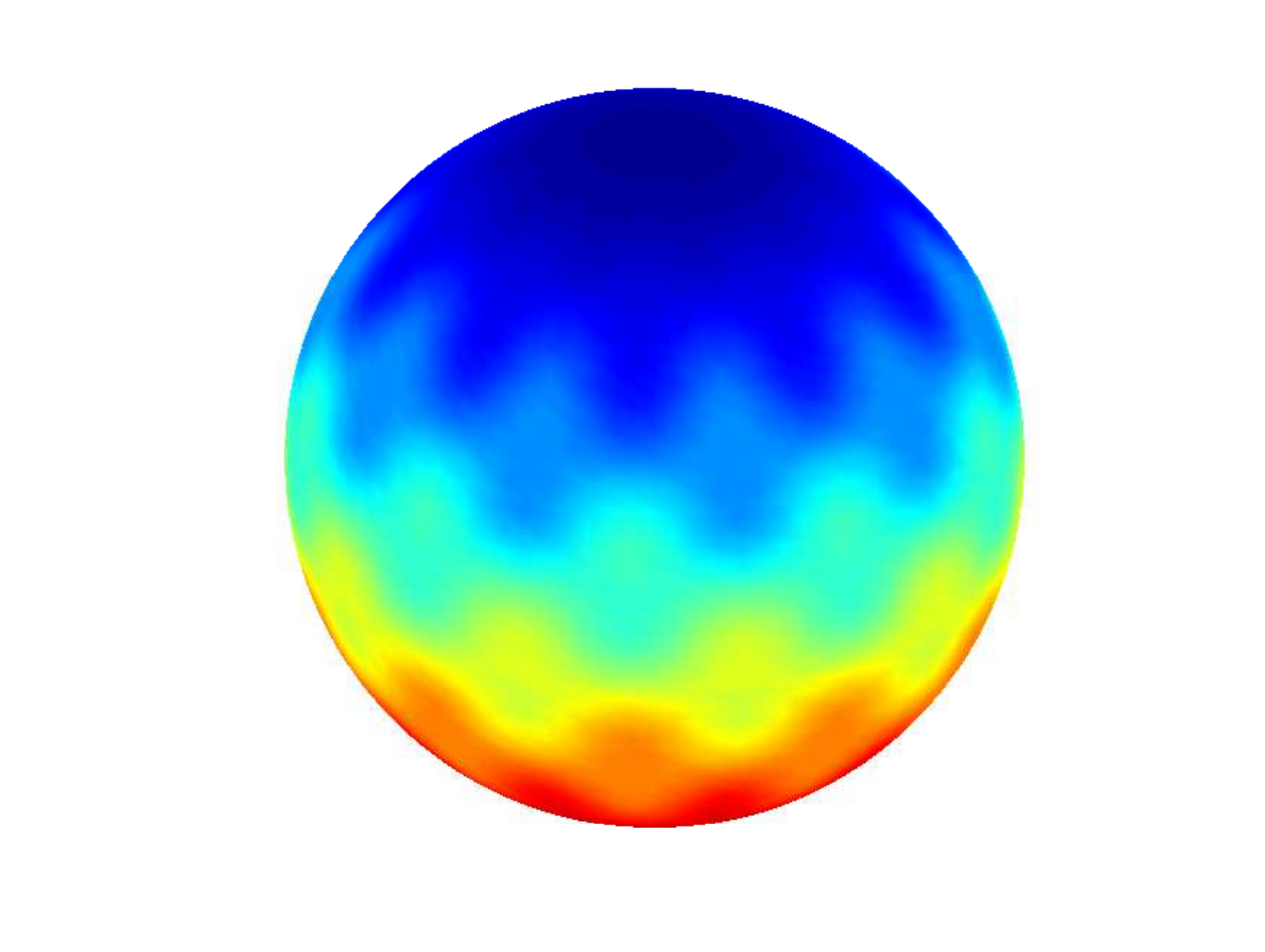}\includegraphics[scale=0.15]{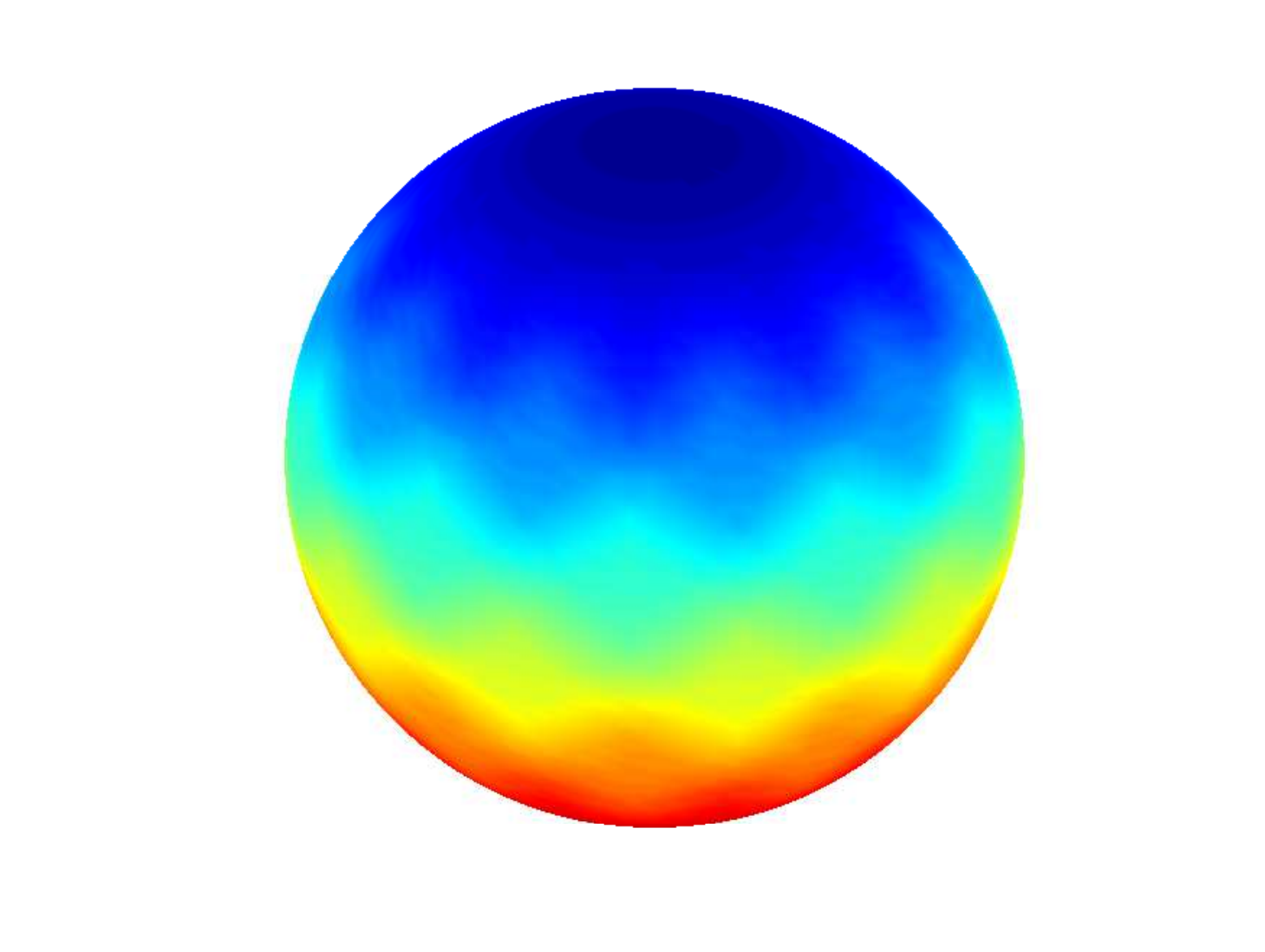}\includegraphics[scale=0.15]{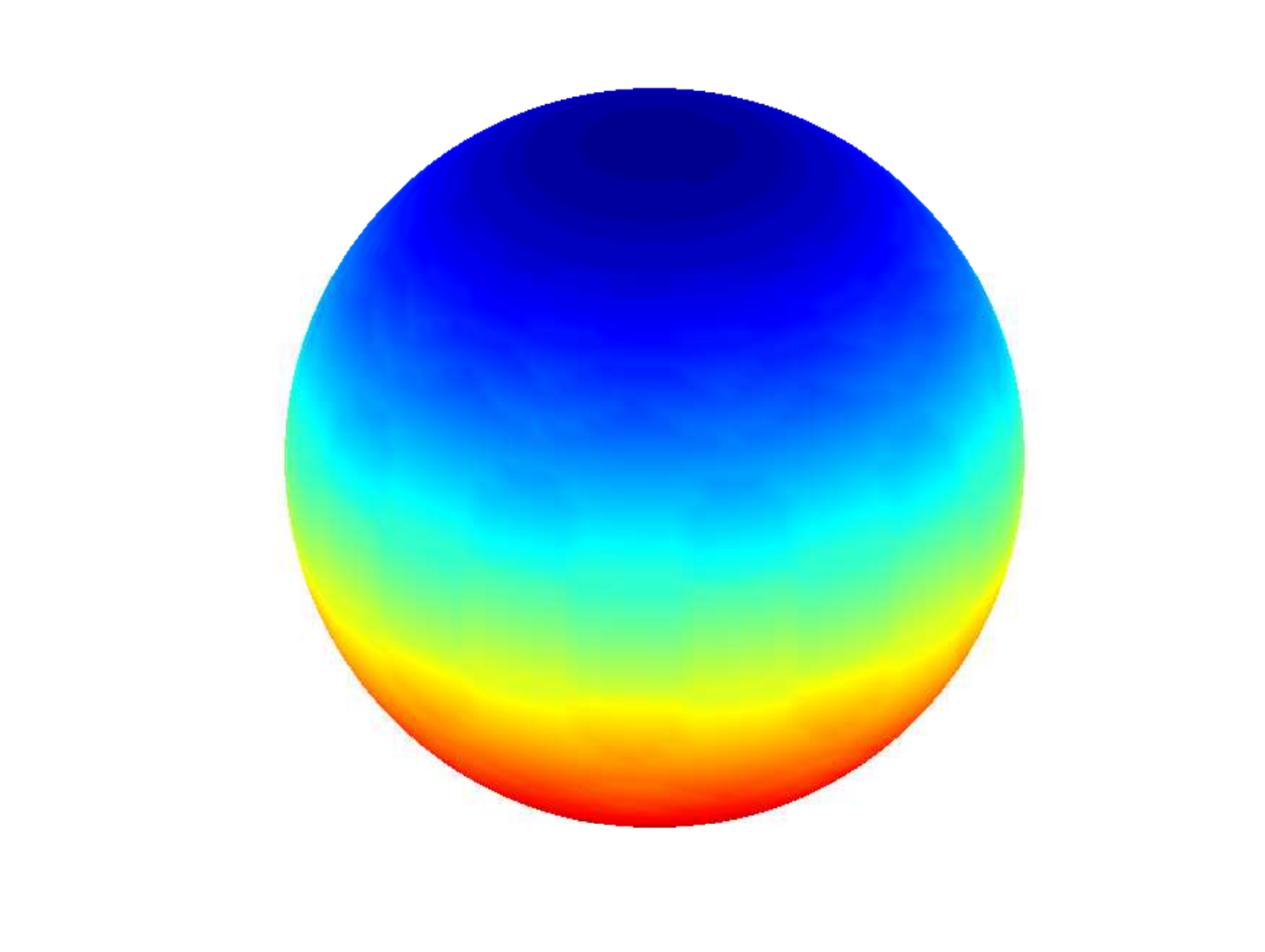}
\par\end{centering}
\protect\caption{When ${\cal G}=2$, ${\cal S}=1$ and ${\cal M}=1$, the solution shown at times $t=0$, $t=10^{-5}$, $t=2\times10^{-5}$, $t=3\times10^{-5}$ and $t=5\times10^{-5}$ develops a hanging drop shape near the bottom of the sphere. 
\label{fig:vertical-unstable}}
\end{figure}

\begin{figure}
\begin{centering}
\includegraphics[scale=0.15]{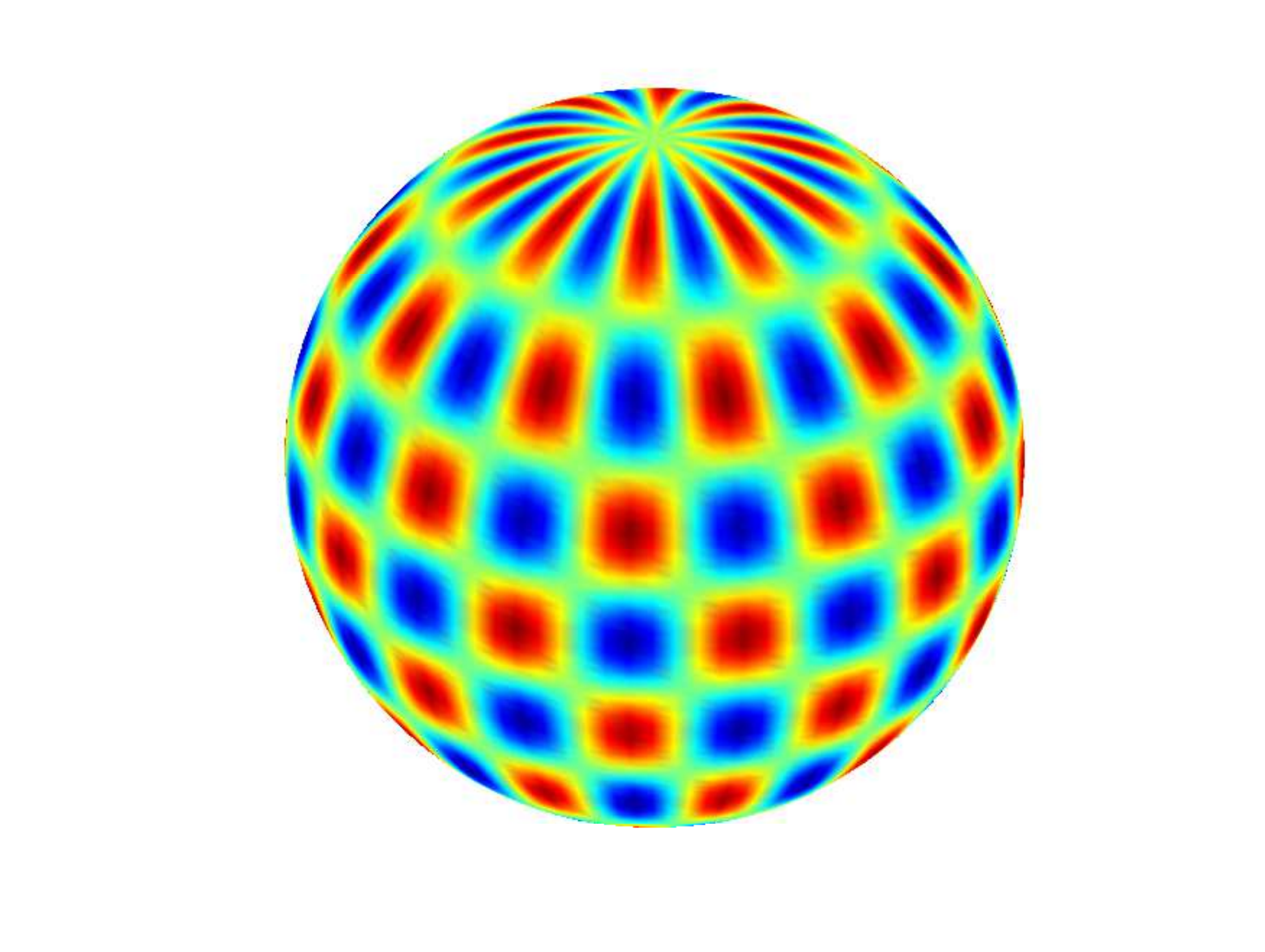}\includegraphics[scale=0.15]{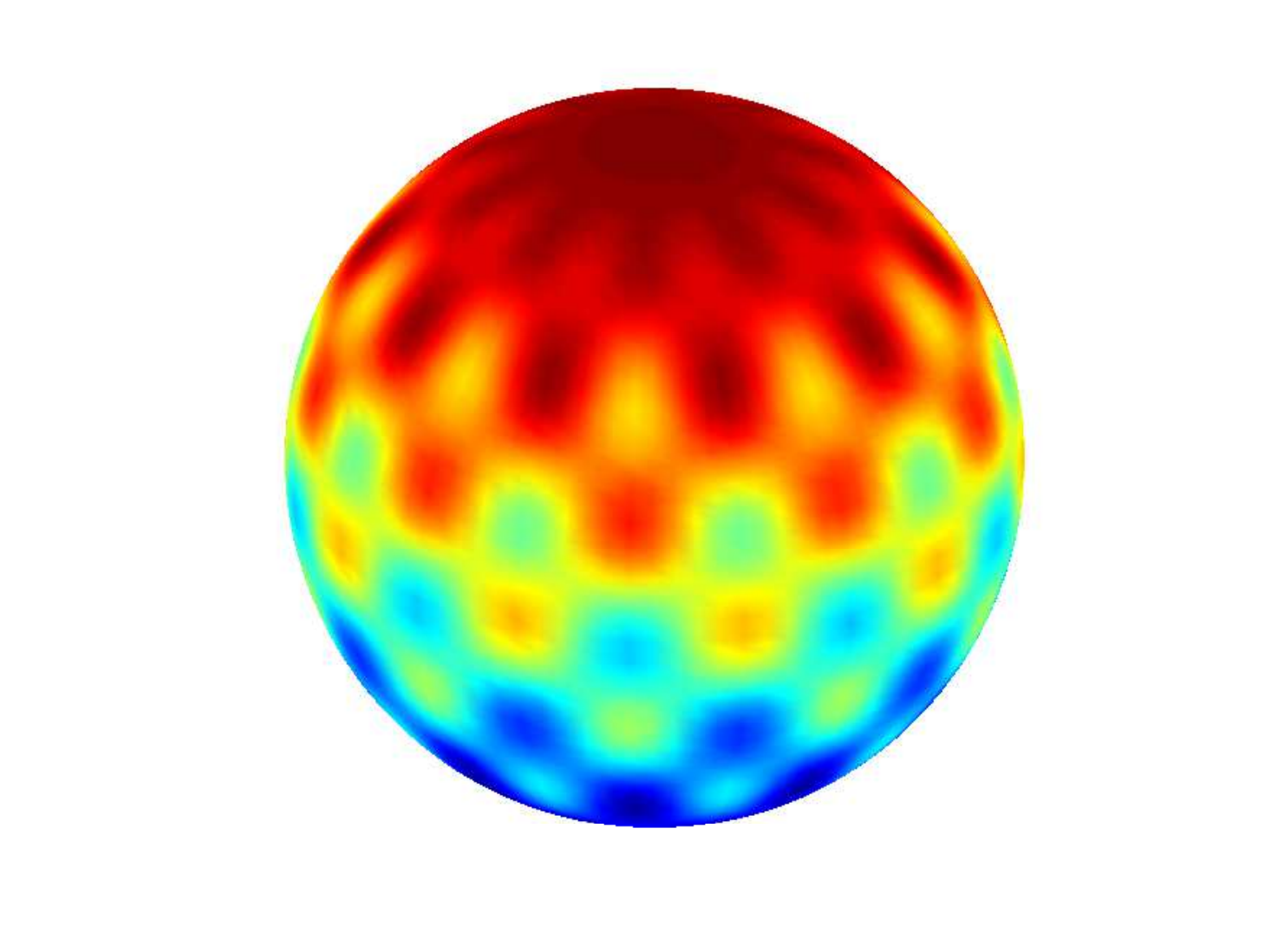}\includegraphics[scale=0.15]{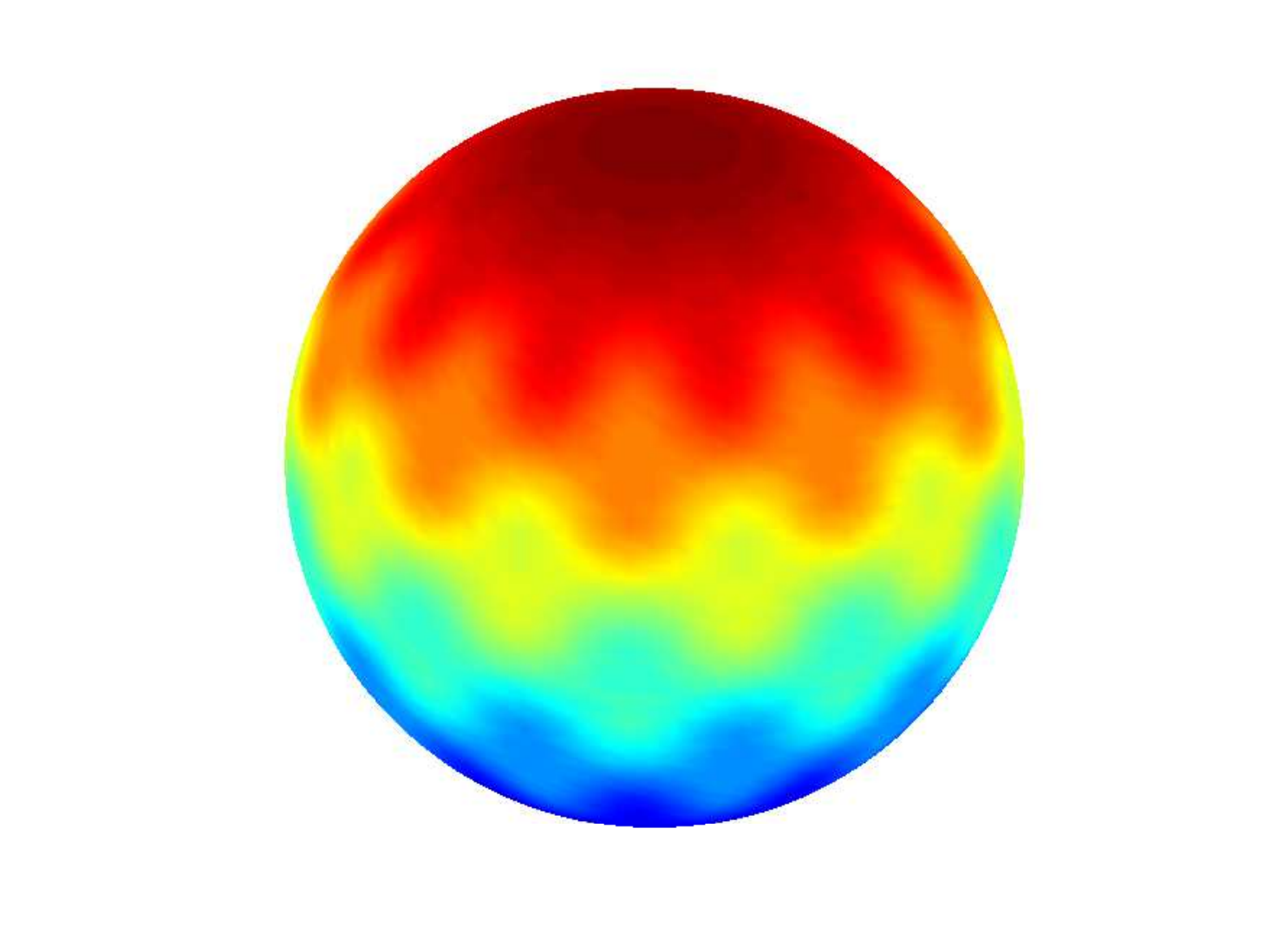}\includegraphics[scale=0.15]{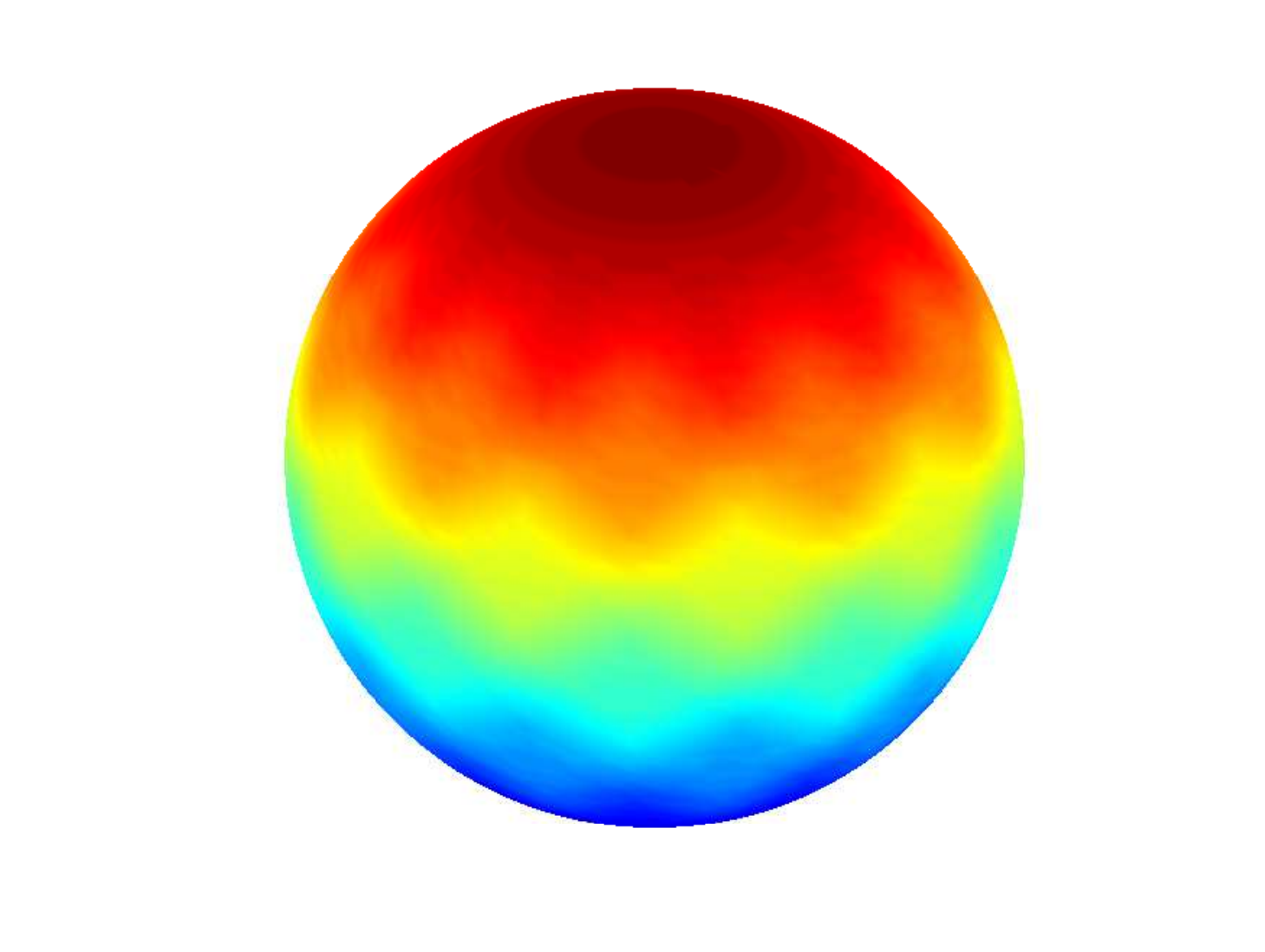}\includegraphics[scale=0.15]{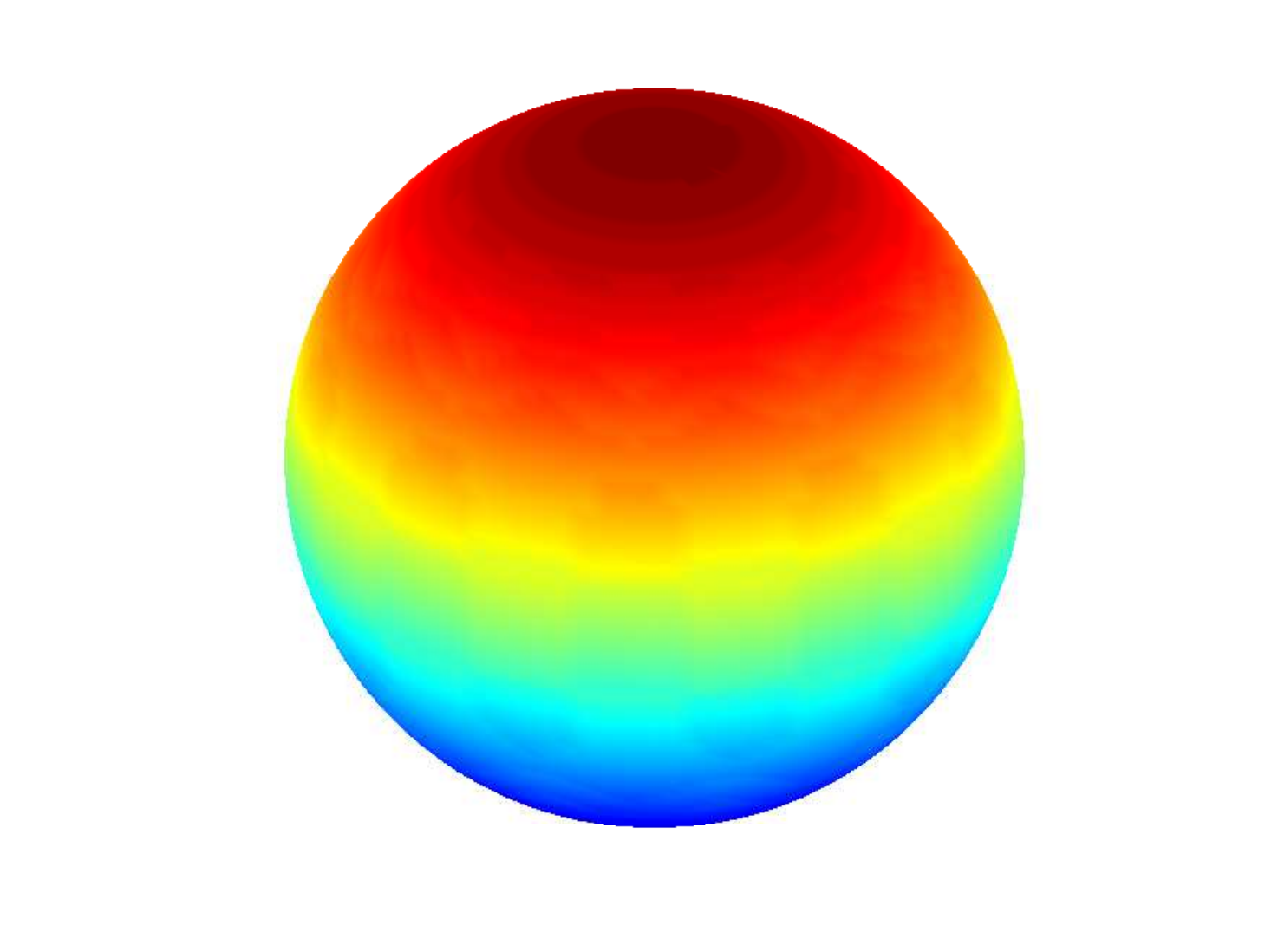}
\par\end{centering}
\protect\caption{When ${\cal G}=1$, ${\cal S}=1$ and ${\cal M}=2$, the solution shown at times $t=0$, $t=10^{-5}$, $t=2\times10^{-5}$, $t=3\times10^{-5}$ and $t=5\times10^{-5}$ develops a steady bump near the top of the sphere. 
\label{fig:vertical-unstable2}}
\end{figure}

\begin{figure}
\begin{centering}
\includegraphics[scale=0.45]{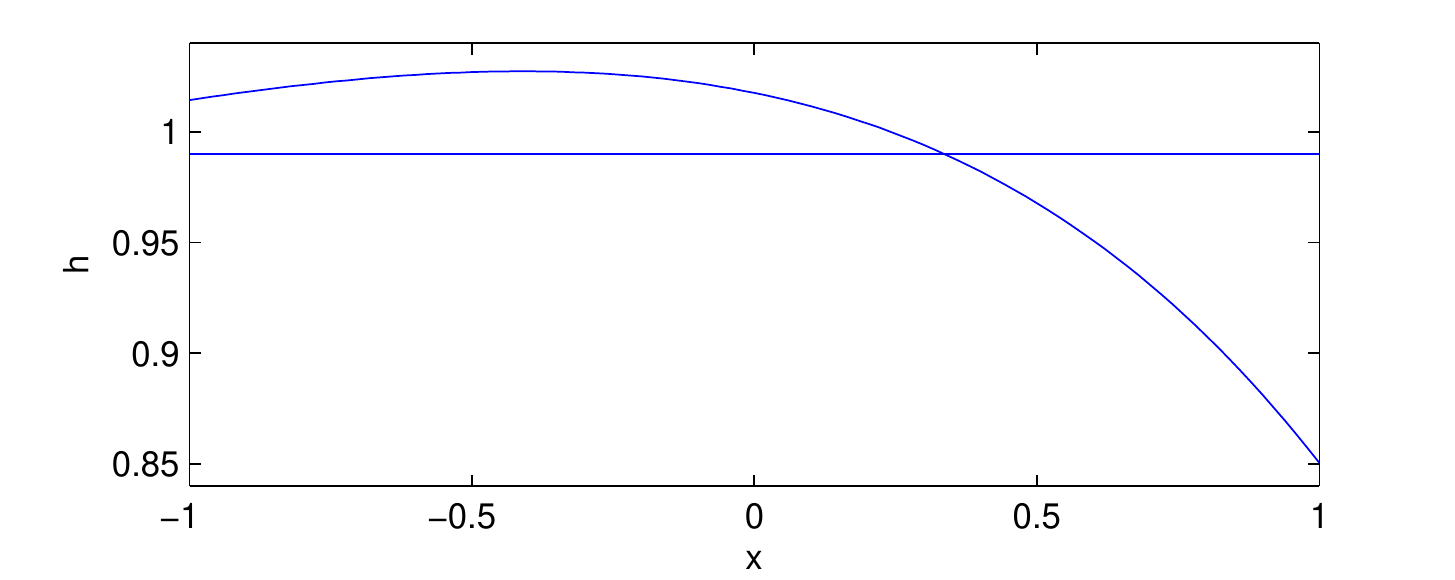}\includegraphics[scale=0.45]{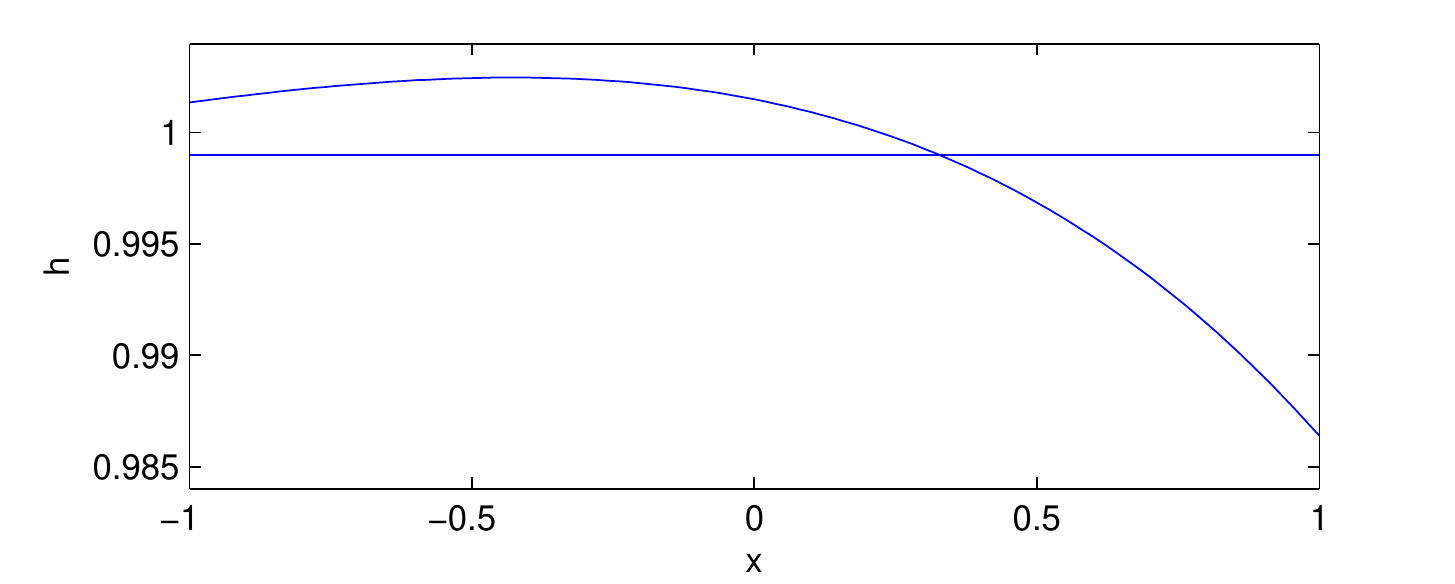}
\par\end{centering}
\begin{centering}
\includegraphics[scale=0.45]{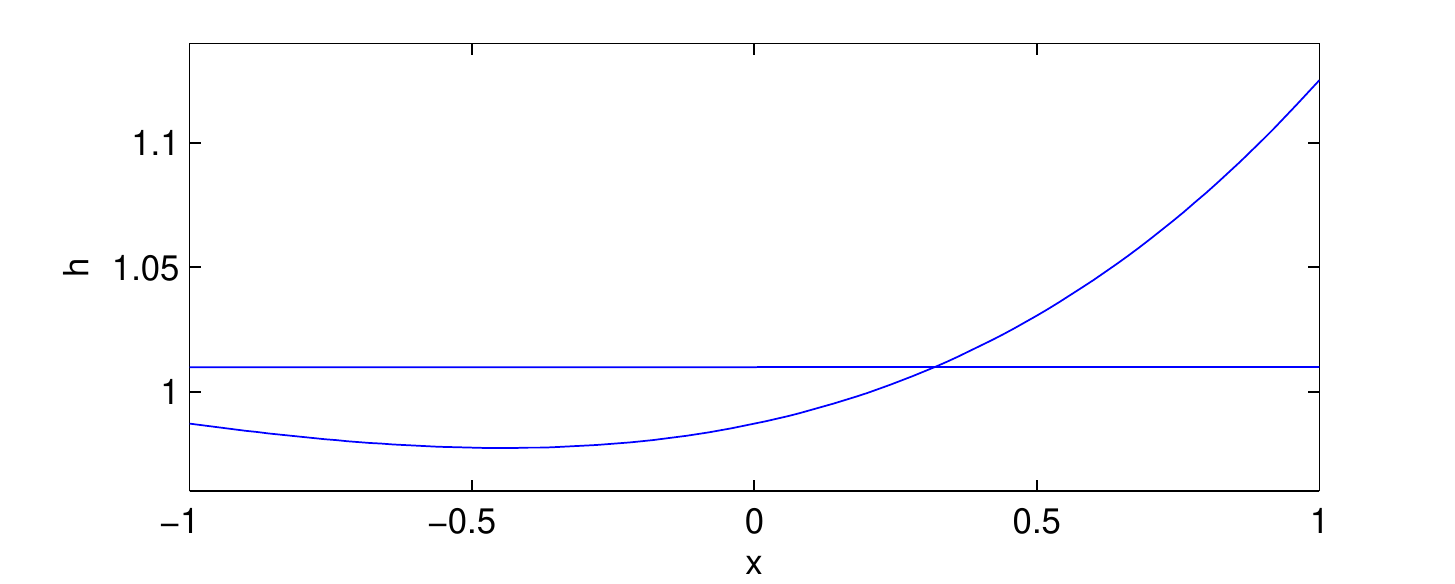}\includegraphics[scale=0.45]{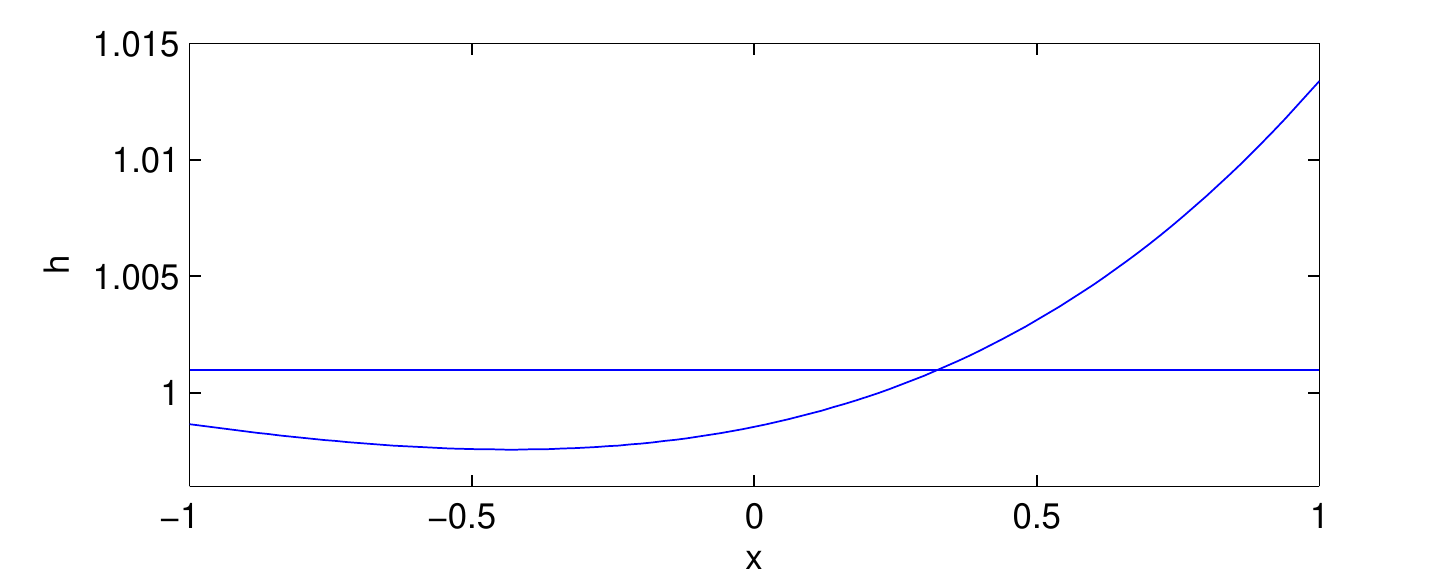}
\par\end{centering}
\protect\caption{The initially uniform and the final steady state profiles when ${\cal G}=10$, ${\cal S}=1$ and ${\cal M}=10$; the initial conditions in the top left and top right profiles are given by $h=0.99$ and $h=0.999$, respectively, and those in the bottom left and bottom right profiles are $h=1.01$ and $h=1.001$. 
\label{fig:vertical-shape-change}}
\end{figure}

Equation (\ref{EvolEqVert}) admits a steady solution with a uniform film thickness $h$ provided that ${\cal G}h^3-{\cal M}h^2=0$, or when $h={\cal M}/{\cal G}$. In terms of dimensional parameters, this corresponds to a film of thickness $3\sigma_1 k_1 / (2\rho g)$ provided that $\sigma_1 k_1 >0$. For a given slope $\sigma_1$ of the surface tension versus temperature curve and a given vertical temperature gradient $k_1$, only a film of that particular thickness can remain uniform along the entire spherical surface. In that state, the draining of the film due to gravity perfectly balances the upward flow caused by the temperature gradient (and Marangoni effect) to keep the film steady and perfectly uniform. A natural question to ask is what happens if an initial film of that thickness is perturbed slightly in $\theta$ and $\phi$, or if the volume of the film is such that its initially uniform thickness is either smaller or larger than that particular value.

For convenience, if we focus on the case ${\cal G}-{\cal M}=0$, the uniform film thickness would be $h=1$. To check the stability of that state by numerical simulations, in Figure \ref{fig:vertical a=00003D1 d=00003D-1 to uniform} we show the evolution of a slightly perturbed film computed by COMSOL and observe that, without perturbing the overall volume of the film, such a film relaxes back to its original uniform state where the gravity and Marangoni effects are in balance. The uniform steady state is thus stable with respect to small perturbations which do
not change the total mass (or volume) of the liquid.

However, that uniform state is unstable to perturbations that change the mass (or average initial thickness) of the liquid film away from its equilibrium state. If the the perturbation increases the mass slightly, the gravitational term dominates and and the uniform shape develops a droplet profile at the bottom of the sphere, as seen in Figure \ref{fig:vertical-down}, but if the perturbation decreases the mass or the average film thickness, the Marangoni effect wins and the uniform profile develops a bump near the top of the sphere, as seen in Figure \ref{fig:vertical-up}.

If ${\cal G}-{\cal M}\neq 0$, then the uniform thickness $h=1$ is not a stationary solution. Starting with a film of average initial thickness 1, if ${\cal G}-{\cal M}>0$, the combined gravity and Marangoni effects will drive the fluid downward and the steady state shape will be like a hanging drop near the bottom of the sphere, i.e., gravity wins (see Figure \ref{fig:vertical-unstable}); whereas if ${\cal G}-{\cal M}<0$, the combined gravity and Marangoni effect will drive the fluid upward to form a bump near the top of the sphere, i.e., Marangoni wins (see Figure \ref{fig:vertical-unstable2}). 

The smaller the initial perturbation in volume, the smaller would be the deviation of the final stationary shape from the uniform one. This can be seen in Figure \ref{fig:vertical-shape-change} where for the case ${\cal G}={\cal M}=10$ that still corresponds to equilibrium thickness $h=1$, the initial profile and the final steady state profile are plotted when the initial thickness has values 0.99, 0.999, 1.001 and 1.01. As expected, initial films that are thicker than the uniform equilibrium value develop maxima near the bottom of the drop  (at $x=1$), whereas the initially thinner films develop minima near the bottom. Interestingly, the steady state film thickness is not a monotonically increasing or decreasing function in either case.

\subsection{Streamlines for the steady surface flow}

\begin{figure}
\begin{centering}
\includegraphics[scale=0.45]{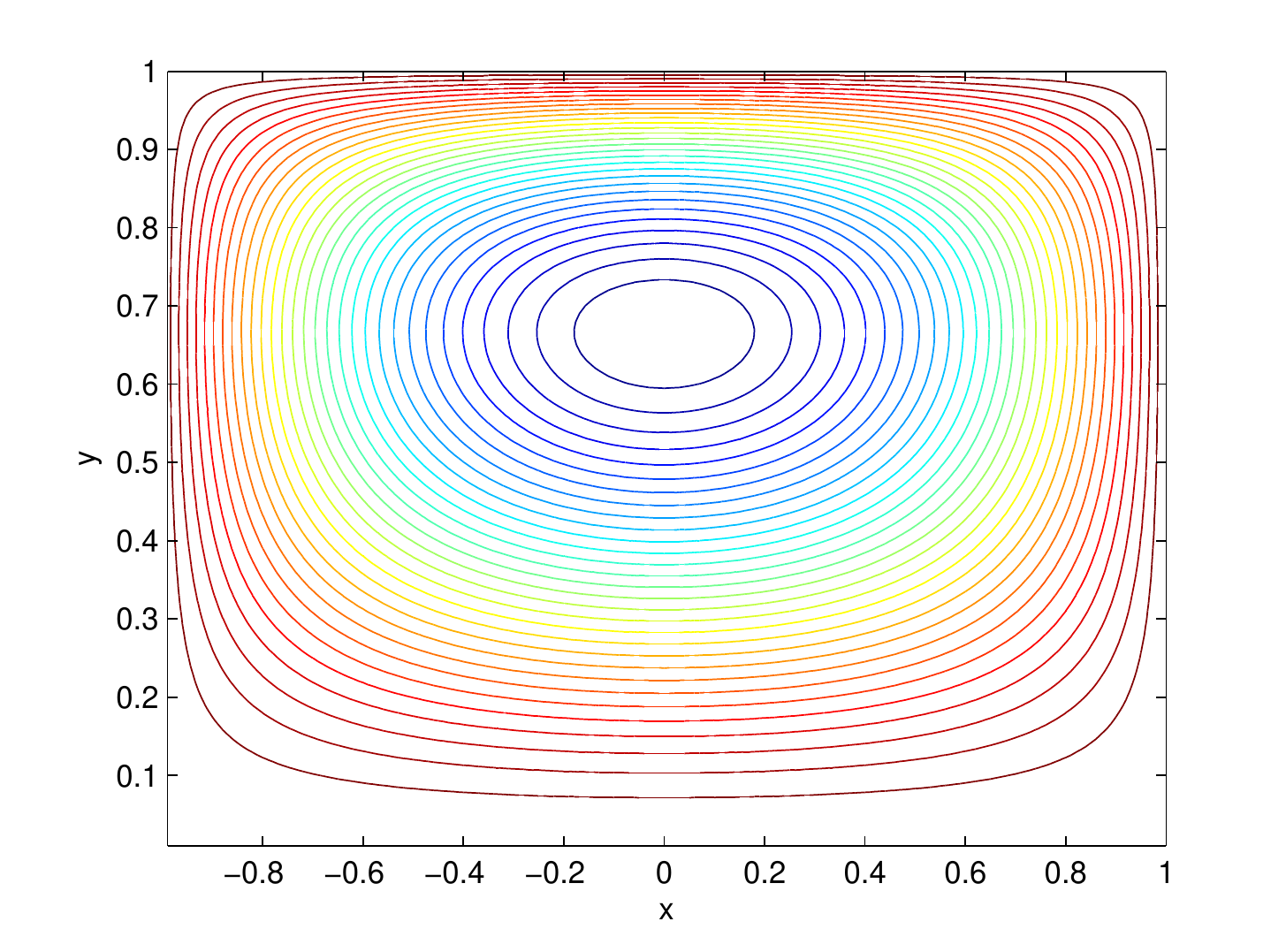}
\includegraphics[scale=0.45]{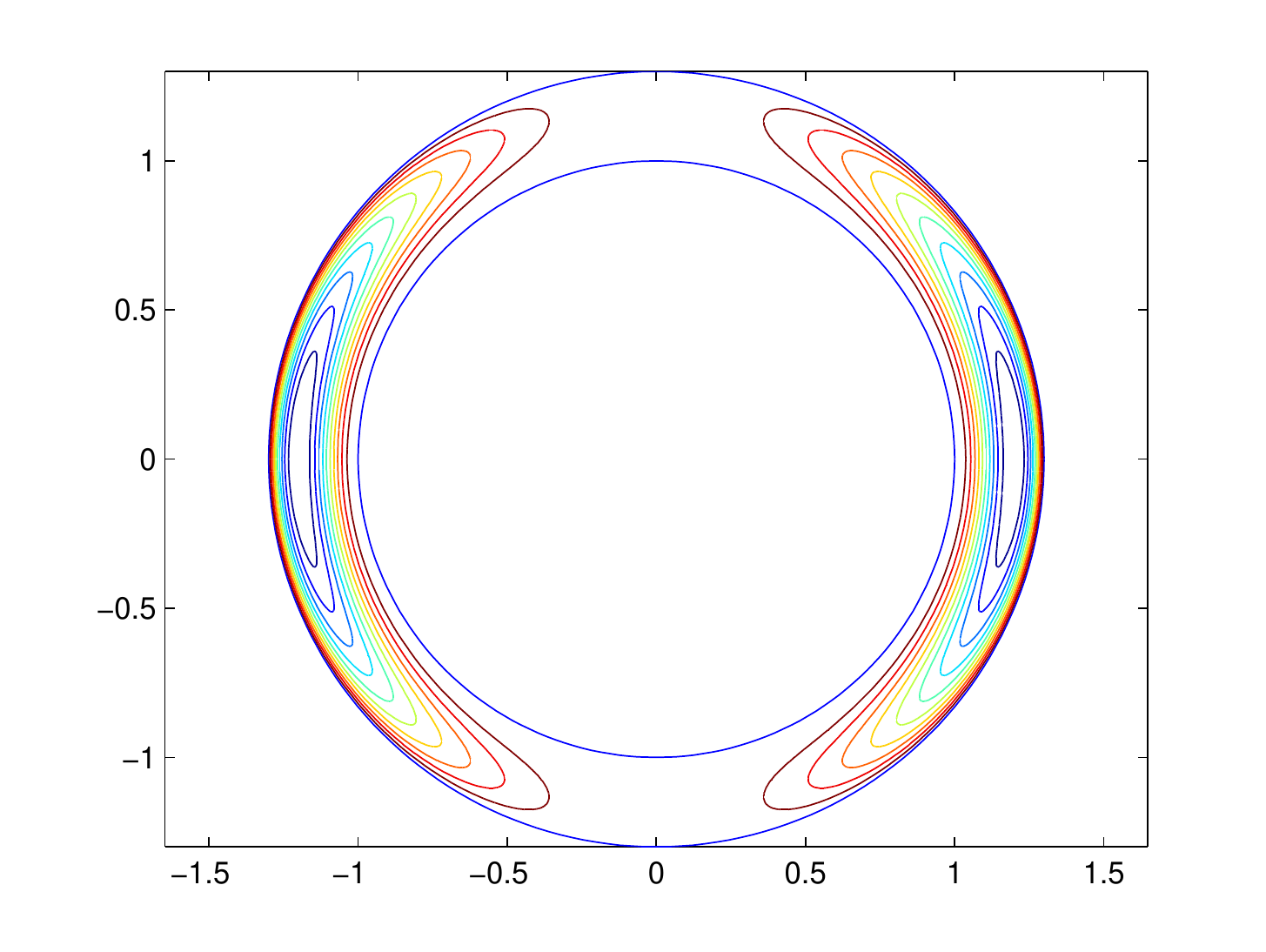}
\par\end{centering}
\protect\caption{Streamlines in Cartesian and spherical coordinates. \label{fig:stream-lines}}
\end{figure}

For the case $\mathcal{G}=\mathcal{M}$, The uniform film $h=1$ is a steady state solution. To obtain the streamlines of the flow taking place within this uniform film, it is convenient to transform the radial coordinate to a new variable $y$ (scaled and dimensionless) that ranges from 0 to 1 across the film thickness. Let us choose $y={(r-R)}/{(\epsilon R)}$. Recall that $\epsilon=H/R$ where $H$ would be the dimensional uniform film thickness. Rewrite Eq.~(\ref{thetamom}) and Eq.~(\ref{eq:tangential BC1}) in terms of variable $y$ and take the leading order to get 
\[
\frac{\partial^2 v_{\theta}}{\partial y^2}=\frac{\epsilon^2 R}{\mu}\frac{\partial P}{\partial \theta}
\]
with boundary conditions
\[
v_{\theta}=0 \quad \text{at} \quad y=0\,,
\]
\[
\frac{\partial v_{\theta}}{\partial y}=\frac{\epsilon}{\mu}\frac{\partial\sigma}{\partial\theta} \quad \text{at} \quad y=1\,.
\]
Note that only variable $y$ is dimensionless, with all other variables and parameters still being dimensional.
Solve for $v_{\theta}$ to get
\[
v_{\theta}=\frac{\epsilon^2 R}{2\mu}\frac{\partial P}{\partial\theta}(y^2-2y)+\frac{\epsilon}{\mu}\frac{\partial\sigma}{\partial\theta}y\,.
\]
While it may appear that the two terms in this equation are of different orders, because the Marangoni term is an order of $\epsilon$ smaller than the gravity term, the two terms in the equation are of the same order of magnitude, as will be seen below. When the film is of uniform thickness, from the expression (\ref{eq:pressurefield}) for the pressure field we find that 
\[
\frac{\partial P}{\partial \theta}=-\rho gR\sin\theta\,.
\]
Also, from (\ref{acd}), when $\cal G=\cal M$, we have that: 
\[
\frac{\partial\sigma}{\partial\theta}=-\sigma_1 k_1 R \sin\theta =- \frac{2}{3}\rho gRH\sin\theta.
\]
Thus the expression for $v_{\theta}$ simplifies to: 
\[
v_{\theta}=\frac{\epsilon^2 \rho g R^2 \sin\theta}{\mu}\left(\frac{y}{3}-\frac{y^2}{2}\right)\,.
\]
This is a quadratic profile which changes sign at $y=2/3$, with the fluid flowing downward due to gravity in the bulk region $0<y<2/3$, but flowing back up due to the Marangoni effect in $2/3<y<1$ near the interface, with zero net flux across the whole cross section. Substituting $v_\theta$ into the continuity equation (\ref{continuity}), we can solve for $v_r$:
\[
v_r=\frac{\epsilon^3 \rho g R^2 \cos\theta}{3\mu}(y^3-y^2)\,.
\]
The streamfunction corresponding to this flow is given by
\[
\psi=\frac{\epsilon^2 \rho g R^2}{6\mu}\,\sin^2\theta\,(y^3-y^2)=
\frac{\epsilon^2 \rho g R^2}{6\mu}\,(1-x^2)\,(y^3-y^2) \,.
\]
In this case, the velocity components are related to the streamfunction by $v_r = \frac{\epsilon}{\sin\theta}\frac{\partial \psi}{\partial \theta}$ and $v_\theta = \frac{-1}{\sin\theta} \frac{\partial \psi}{\partial y}$. 
The streamlines are illustrated in Figure~\ref{fig:stream-lines}.

\section{Radial temperature field}
\label{sec:4}

\subsection{The leading order equation}

In this section, we consider the specific case where the externally imposed temperature field is a linear function of $r$ only:
\[
T=T_{0}+k_{2}(r-R).
\]
Temperature $T_0$ is a uniform temperature along the surface of the solid sphere and depending on whether the gradient $k_2$ is positive or negative, temperature increases or decreases outward from the surface.
As such, the surface tension $\sigma$ of the liquid-air interface at coordinate $(\theta,\phi)$ depends directly on the film thickness $h(\theta,\phi,t)$. In this sense, we can treat $\sigma$ as a function of $h$:
\[
\sigma=\sigma_{0}+\sigma_{1}k_{2}h.
\]
The sign of the product $\sigma_1 k_2$ determines whether surface tension is higher or lower in regions where the film is thinner or thicker.

Using this result for the surface tension in the evolution equation (\ref{eq:evolution h eqn}), scaling $h$ with $H$ and $t$ with $\tau$ as before and simplifying results in a dimensionless evolution equation in which the following three dimensionless groups appear respectively in front of the contributions to the flux due to gravity, surface tension and Marangoni effects (in the $\theta$-derivative):
\begin{equation}
\label{eq:GSN}
{\cal G}=\frac{\rho gH^{2}\tau}{3\mu R},\quad {\cal S} = \frac{\sigma_{0}H^{3}\tau}{3\mu R^{4}},\quad \mbox{and}\quad {\cal N}=\frac{\sigma_{1}k_{2}H^{2}\tau}{2\mu R^{2}}.
\end{equation}
The first two parameter are the same as before, while the third, measuring Marangoni effects, is a little different. To balance these terms so that all effects are comparable and present at leading order we now need
\[
\sigma_{1}k_{2}\sim\epsilon\frac{\sigma_{0}}{R}\sim\rho gR\,,
\]
which is different from the scaling in the case of a vertical temperature gradient with respect to the Marangoni effect.

Under this scaling, surface tension and Marangoni effects will also appear in the $\phi$-derivative terms. The leading order evolution equation assumes the form
\begin{multline*}
\frac{\partial h}{\partial t}+\frac{1}{\sin\theta}\frac{\partial}{\partial\theta}\left(\sin\theta h^{3}\left({\cal G}\sin\theta+{\cal S}\frac{\partial}{\partial\theta}\left[2h+\frac{1}{\sin\theta}\frac{\partial}{\partial\theta}\left(\sin\theta\frac{\partial h}{\partial\theta}\right)+\frac{1}{\sin^{2}\theta}\frac{\partial^{2}h}{\partial\phi^{2}}\right]\right)+{\cal N}\sin\theta h^{2}\frac{\partial h}{\partial\theta}\right) \\
+\frac{1}{\sin^{2}\theta}\frac{\partial}{\partial\phi} \left[ {\cal S}h^{3}\frac{\partial}{\partial\phi}\left( 2h+\frac{1}{\sin\theta}\frac{\partial}{\partial\theta} \left( \sin\theta\frac{\partial h}{\partial\theta} \right)+\frac{1}{\sin^{2}\theta}\frac{\partial^{2}h}{\partial\phi^{2}} \right) +{\cal N} h^{2}\frac{\partial h}{\partial\phi} \right] = 0\,.
\end{multline*}
%

After the change of variable $x=-\cos\theta$ and upon neglecting gravity for the time being (i.e., setting ${\cal G}=0$) the equation becomes
%
%
\begin{multline*}
\frac{\partial h}{\partial t}+\frac{\partial}{\partial x}\left\{ {\cal S} h^{3}(1-x^{2})\frac{\partial}{\partial x}\left(2h+\frac{\partial}{\partial x}\left((1-x^{2})\frac{\partial h}{\partial x}\right)+\frac{1}{1-x^{2}}\frac{\partial^{2}h}{\partial\phi^{2}}\right)+{\cal N}(1-x^{2})h^{2}\frac{\partial h}{\partial x}\right\} \\
+\frac{1}{1-x^{2}}\frac{\partial}{\partial\phi}\left\{ {\cal S} h^{3}\frac{\partial}{\partial\phi}\left[2h+\frac{\partial}{\partial x}\left((1-x^{2})\frac{\partial h}{\partial x}\right)+\frac{1}{1-x^{2}}\frac{\partial^{2}h}{\partial\phi^{2}}\right]+{\cal N} h^{2}\frac{\partial h}{\partial\phi}\right\} =0\,.
\end{multline*}

In the absence of gravity and with a radially symmetric temperature field, it is expected that a uniform film thickness (e.g., $h(x,\phi,t)=1$) should be a solution to the above equation, which is obviously the case. We are interested in the stability of this solution. In order to investigate this, consider a slight perturbation of that solution in the form:
\[
h(x,\phi,t)=1+\varepsilon\tilde{h}(x,\phi,t)\,.
\]
Substituting into the evolution equation and taking the limit as $\varepsilon\rightarrow0$, we obtain the linearized equation
(dropping the tilde for convenience):
\begin{multline*}
\frac{\partial h}{\partial t}+\frac{\partial}{\partial x}\left\{ {\cal S} (1-x^{2})\frac{\partial}{\partial x}\left(2h+\frac{\partial}{\partial x}\left((1-x^{2})\frac{\partial h}{\partial x}\right)+\frac{1}{1-x^{2}}\frac{\partial^{2}h}{\partial\phi^{2}}\right)+{\cal N} (1-x^{2})\frac{\partial h}{\partial x}\right\} \\
+\frac{1}{1-x^{2}}\frac{\partial}{\partial\phi}\left\{ {\cal S} \frac{\partial}{\partial\phi}\left[2h+\frac{\partial}{\partial x}\left((1-x^{2})\frac{\partial h}{\partial x}\right)+\frac{1}{1-x^{2}}\frac{\partial^{2}h}{\partial\phi^{2}}\right]+{\cal N} \frac{\partial h}{\partial\phi}\right\} =0.
\end{multline*}
Upon assuming a modal decomposition for the small perturbation $h$ in the form
\[
h(x,\phi,t)=e^{\lambda t+im\phi}P_{l}^{m}(x)
\]
where $P_{l}^{m}(x)$ is the associated Legendre function, which is a solution of 
\[
\frac{d}{dx}\left((1-x^{2})\frac{dP}{dx}\right)+\left(l(l+1)-\frac{m^{2}}{1-x^{2}}\right)P=0,
\]
the growth or decay rates $\lambda$ are found to be
\[
\lambda_l=l(l+1)\left[{\cal S}(2-l(l+1))+{\cal N}\right]\,.
\]
Interestingly, the rates are independent of the index $m$ of the azimuthal modes.

The stability of a mode with index $l$ depends upon the sign of ${\cal N}$. (Note that the dimensionless group $\cal S$ is always positive.) If ${\cal N}<0$, $\lambda$ is negative for all positive $l$, which mean that the uniform film thickness is a stable solution. If ${\cal N}>0$, $\lambda$ may be positive for some small values of $l$, but will become negative with increasing $l$; therefore a certain range of mode numbers may be unstable. An intuitive explanation of the instability is as follows. For ${\cal N}>0$, surface tension is higher where the film is thicker and lower where the film is thinner. As such, if an initially uniform film is perturbed, the Marangoni effect which drives the surface flow from low to high surface tension regions causes the bulk fluid to be ``pumped'' away from the thin regions (with low surface tension) towards the thick regions (with high surface tension), thereby causing the thin regions to become thinner and the thick regions to become thicker, amplifying the initial perturbation.

\subsection{Numerical simulations}

\begin{figure}
\begin{centering}
\includegraphics{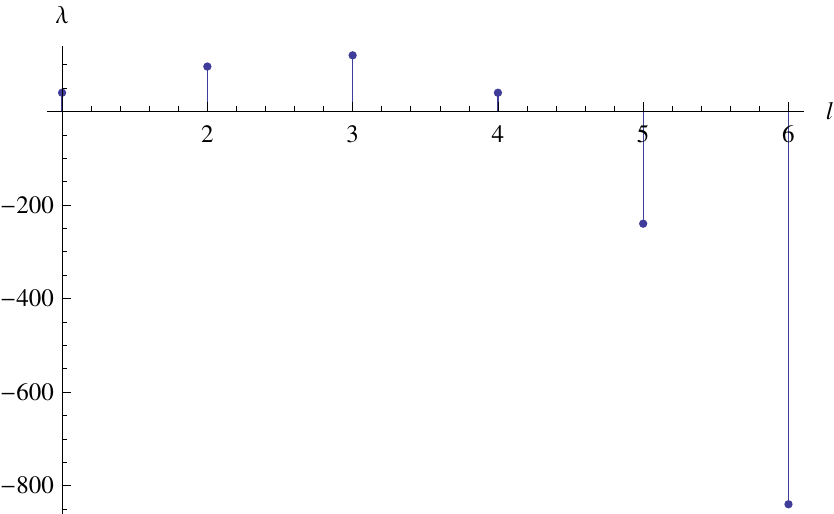}
\par\end{centering}
\protect\caption{Growth/decay rates $\lambda$ versus Legendre mode number $l$ 
for ${\cal S}=1$ and ${\cal N}=20$.\label{fig:lambda vs l} }
\end{figure}

To check the stability predictions numerically, we choose ${\cal S}=1$ and ${\cal N}=20$.
Also we choose the amplitude of the initial perturbation to be $\varepsilon=10^{-5}$. The
relationship between the eigenvalues $\lambda$ and modes $l$ is plotted 
in Figure~\ref{fig:lambda vs l} to illustrate that the higher modes are stabilized ($\lambda < 0$) 
due to surface tension $\cal S$. The most unstable mode for these parameter values corresponds to $l= 3$.

\begin{figure}
\begin{centering}
\includegraphics[scale=0.35]{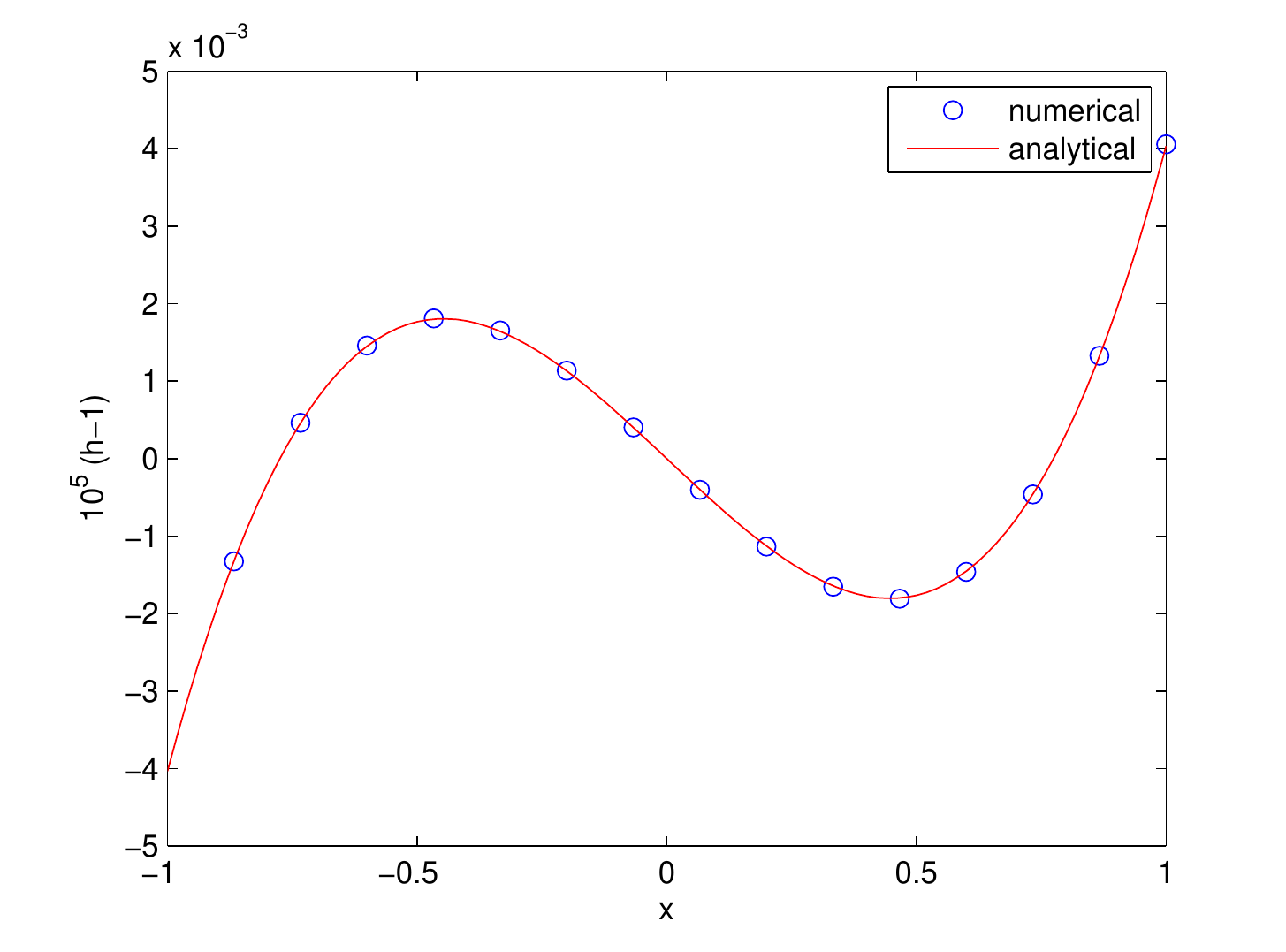}\includegraphics[scale=0.35]{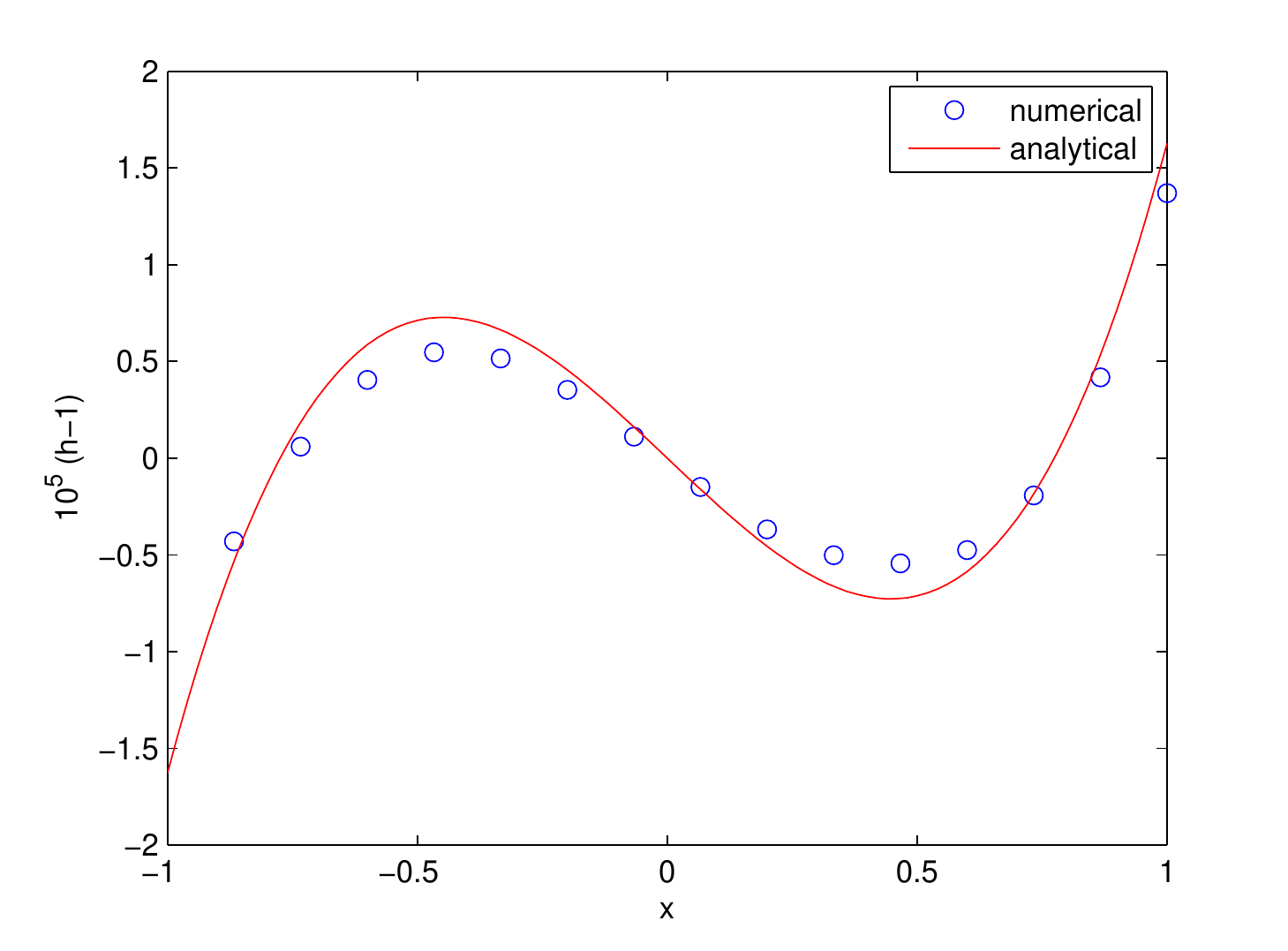}\includegraphics[scale=0.35]{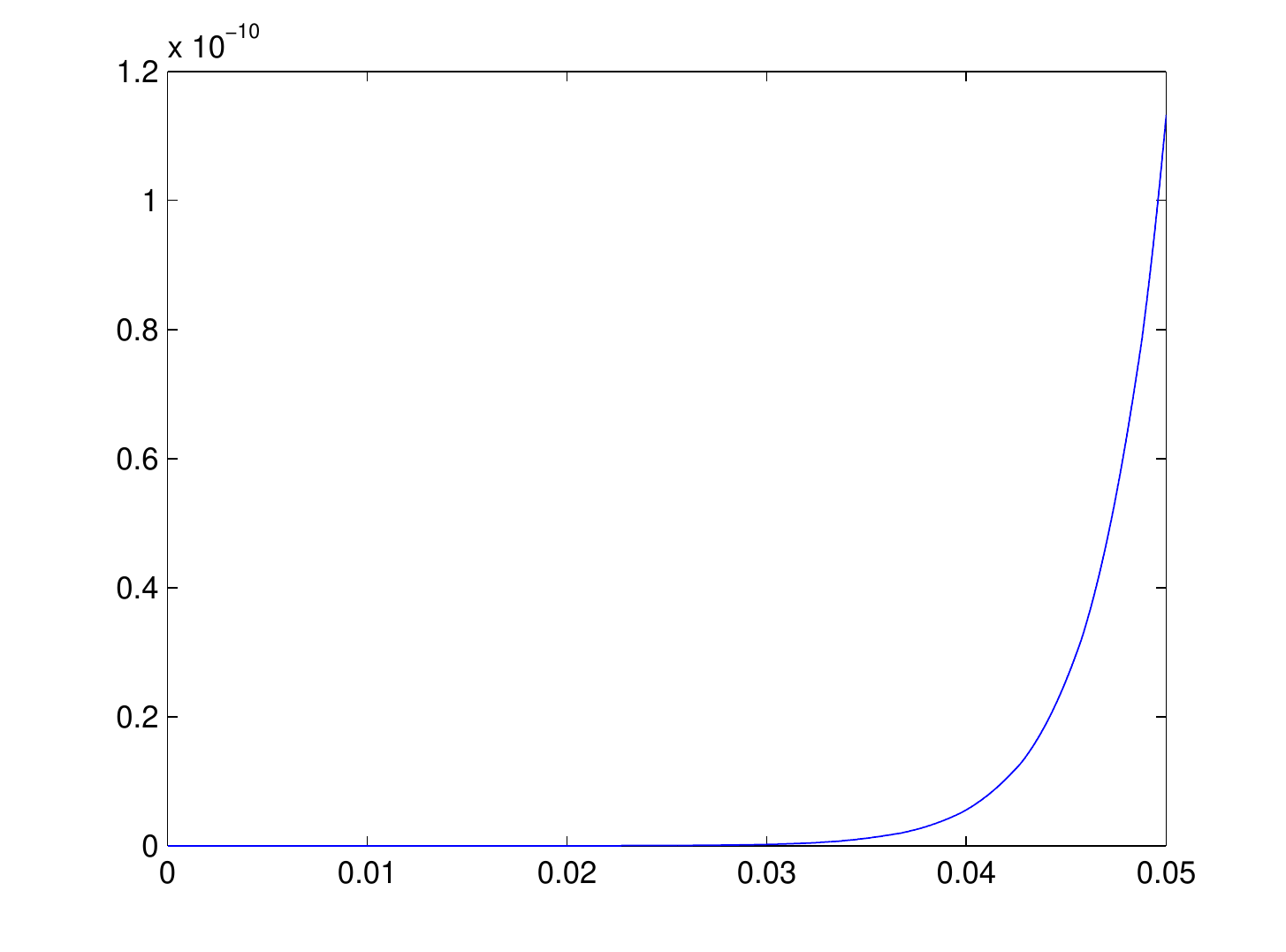}
\par\end{centering}
\protect\caption{Comparison between analytical and numerical results. Left panel
is at time $0.05$; middle panel is at time $0.1$. The $L^{2}$-distance between numerical and analytical results from
$t=0$ to $t=0.05$ is shown in the right panel. \label{fig:1d comparison result}}
\end{figure}

For an axisymmetric initial condition, we choose $l=3$ and $m=0$. We can simplify the evolution equation when the film thickness $h$ is independent of $\phi$. We carry out the numerical simulations of the full nonlinear equation in COMSOL Multiphysics. The initial condition is
\[
h(x,0)=h_{0}(x)=1+10^{-5} {(5x^{3}-3x)}/{2}\,.
\]
For $l=3$, the predicted growth rate is $\lambda=120$. We can estimate the growth rate from the numerical simulations using
\[
\lambda_{\mbox{\tiny num}}=\frac{1}{t}\ln\left(\frac{\max_{x}h(x,t)}{\max_{x}h(x,0)}\right)\,.
\]
The results are shown in Table \ref{tab:lambda symmetric case}. We find that the nonlinear time evolution of the film thickness is captured well by the linear approximation for small times. The analytically computed profile from the linearized model and the numerical solution of the full nonlinear equation are plotted together for comparison in Figure~\ref{fig:1d comparison result}, together with the $L^2$-distance between them as a function of time.
While for small times there is good agreement between the linear and nonlinear dynamics, as time goes by, the two solutions deviate significantly.  

\begin{table}
\begin{centering}
\begin{tabular}{|c|c|c|}
\hline
$t$ & $\max_{x}h(x,t)$ & $\lambda_{\mbox {\tiny num}}$\tabularnewline
\hline\hline
$10^{-3}$ & $1.1277$ & $120.18$\tabularnewline
\hline
$2\times10^{-3}$ & $1.2722$ & $120.37$\tabularnewline
\hline
$4\times10^{-3}$ & $1.6212$ & $121.06$\tabularnewline
\hline
\end{tabular}
\par\end{centering}
\protect\caption{Growth rate $\lambda_{\mbox{\tiny num}}$ calculated for the axisymmetric initial condition ($l=3$,
$m=0$) at different times with parameters ${\cal S}=1$ and ${\cal N}=20$. \label{tab:lambda symmetric case}}
\end{table}

If we consider both $\theta$ and $\phi$ dependence and carry out the simulations using the two-dimensional equation, with $l=3$ and different choices of $m$, we can verify the analytical stability result with varying degrees of accuracy depending on the value of $m$. Since
the three-dimensional results are difficult to compare visually, we also
provide a comparison of the height profiles along the a spiral line on the sphere: $\phi=\pi(1-\cos\theta)$, which corresponds to the diagonal line of the rectangular domain $-1<x<1$ and $0<\phi<2\pi$.

\begin{figure}
\begin{centering}
\includegraphics[scale=0.3]{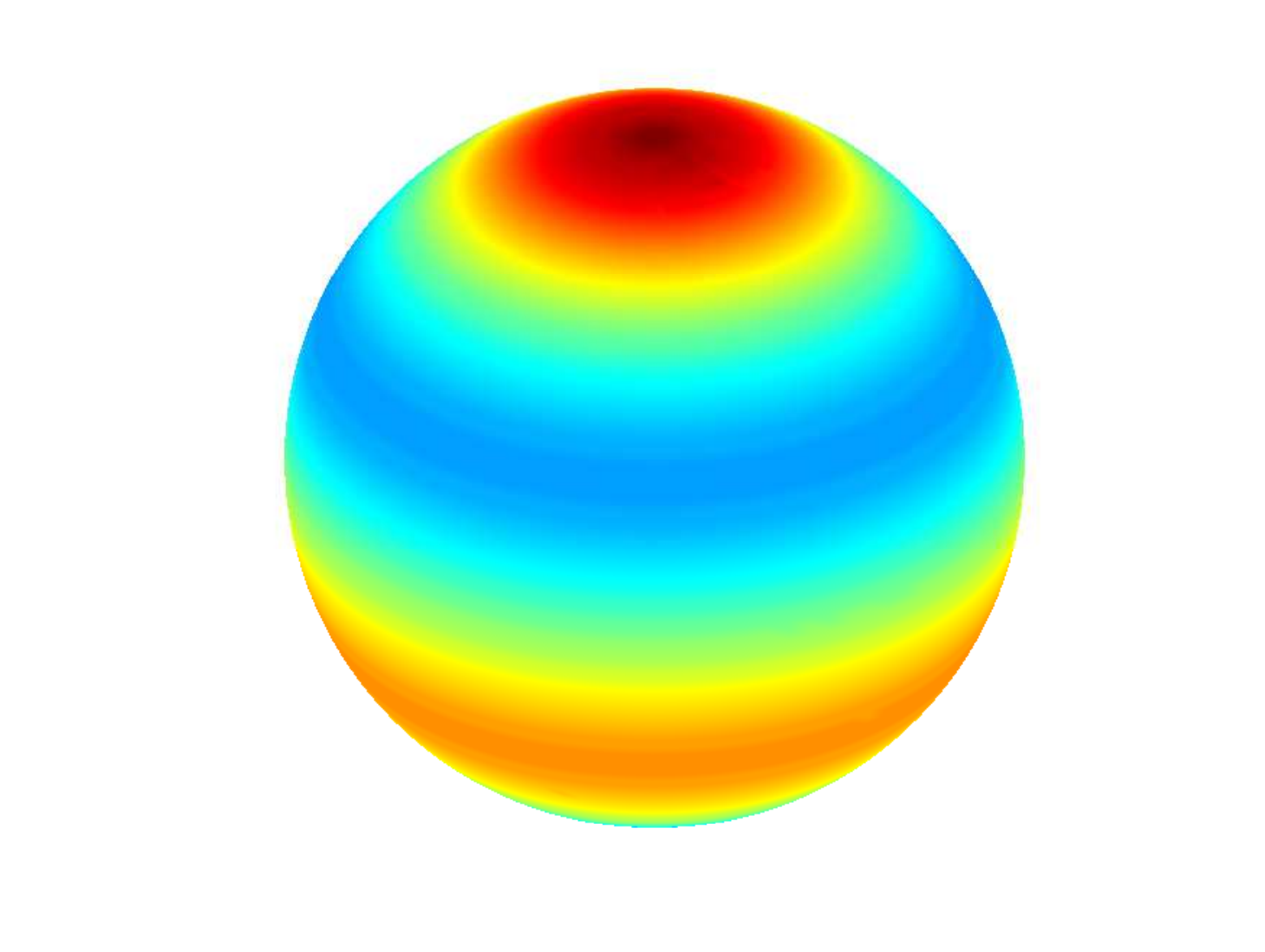}\includegraphics[scale=0.3]{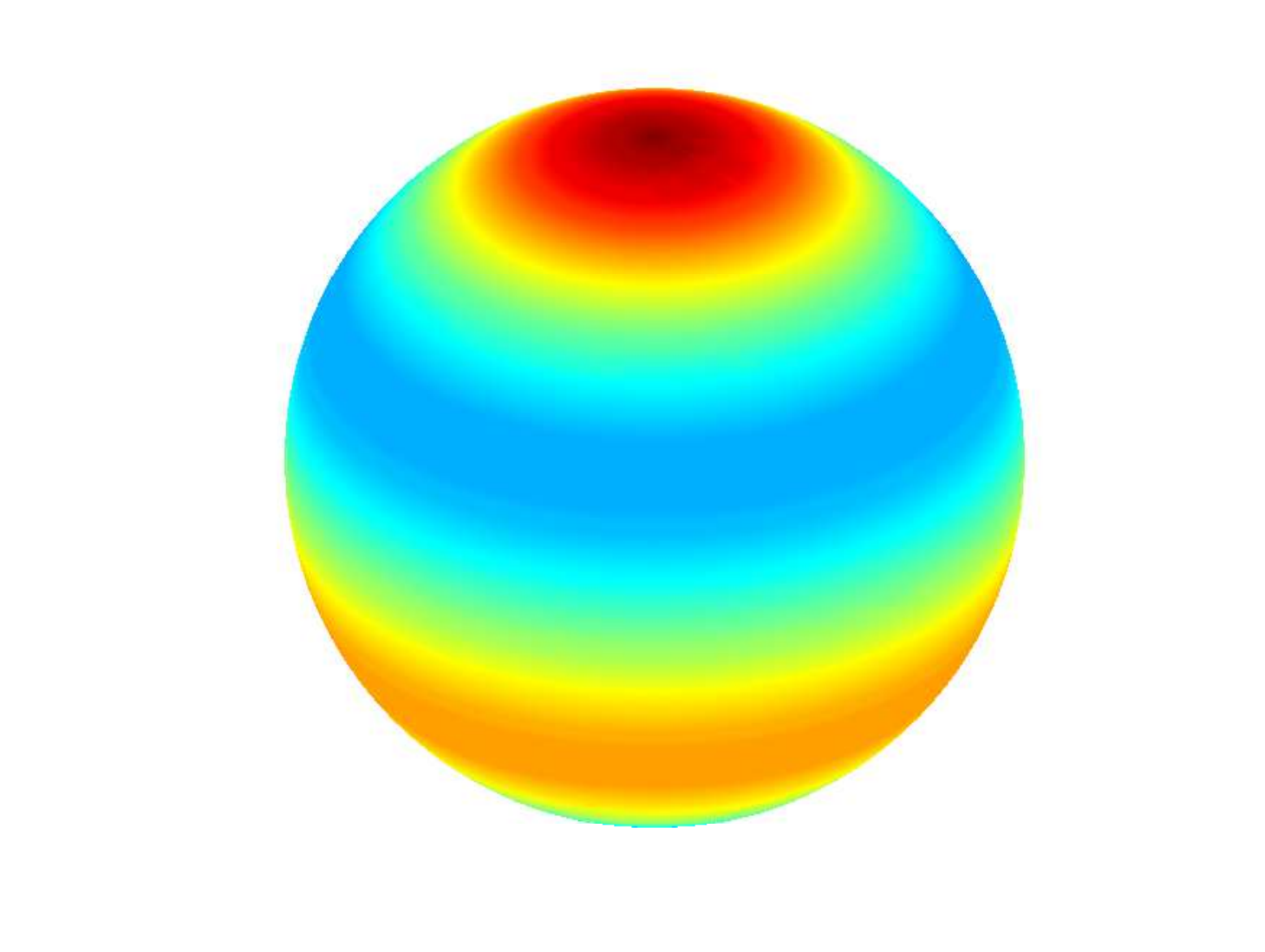}\includegraphics[scale=0.3]{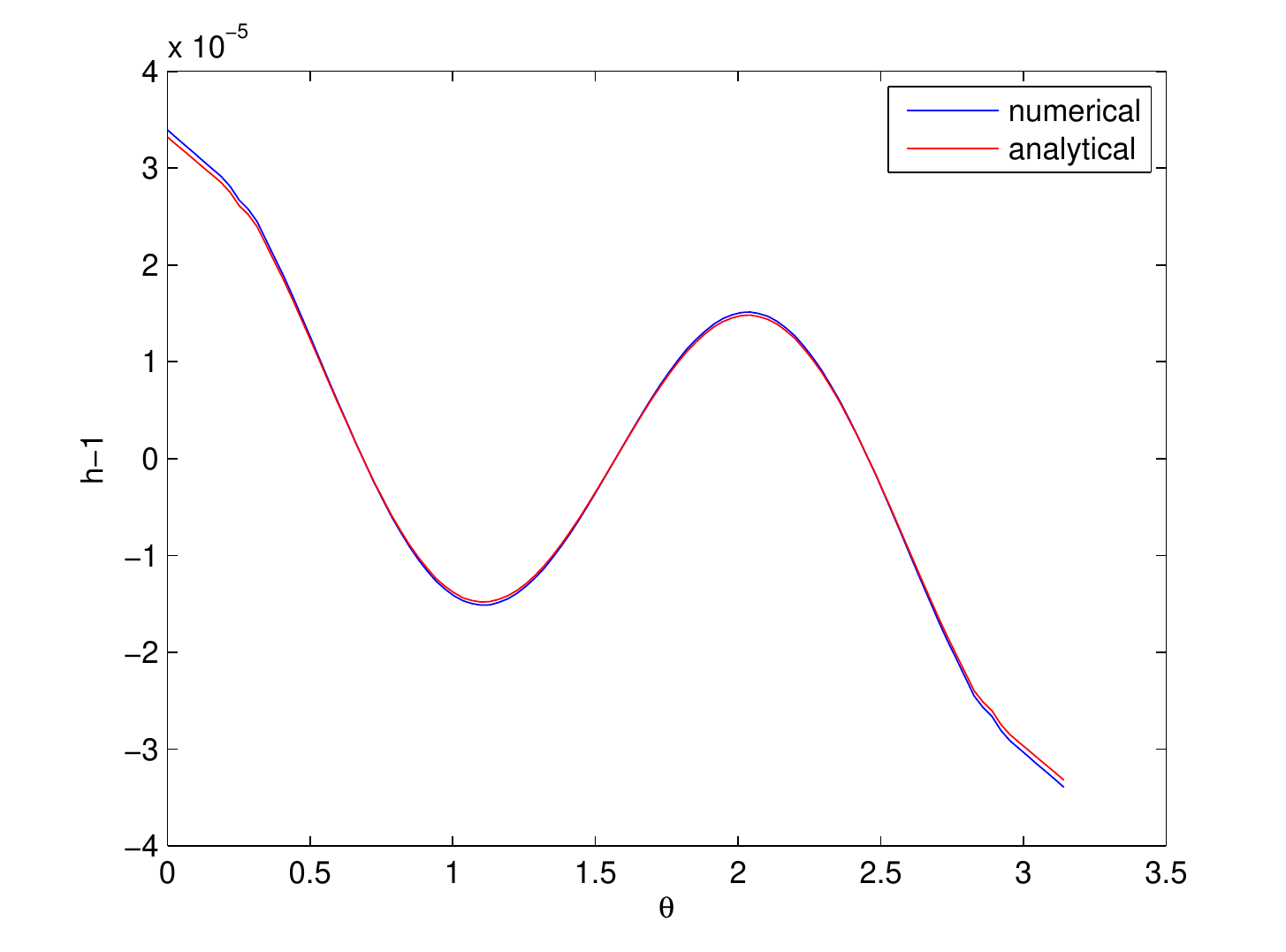}
\par\end{centering}
\protect\caption{Results for $m=0$ and $l=3$ at $t=0.01$. From left to right are the numerical result, the analytical result and the profile comparison on the spiral line $\phi=\pi(1-\cos\theta)$.
\label{fig:two dimensional result m=00003D0}}
\end{figure}
\begin{figure}
\begin{centering}
\includegraphics[scale=0.3]{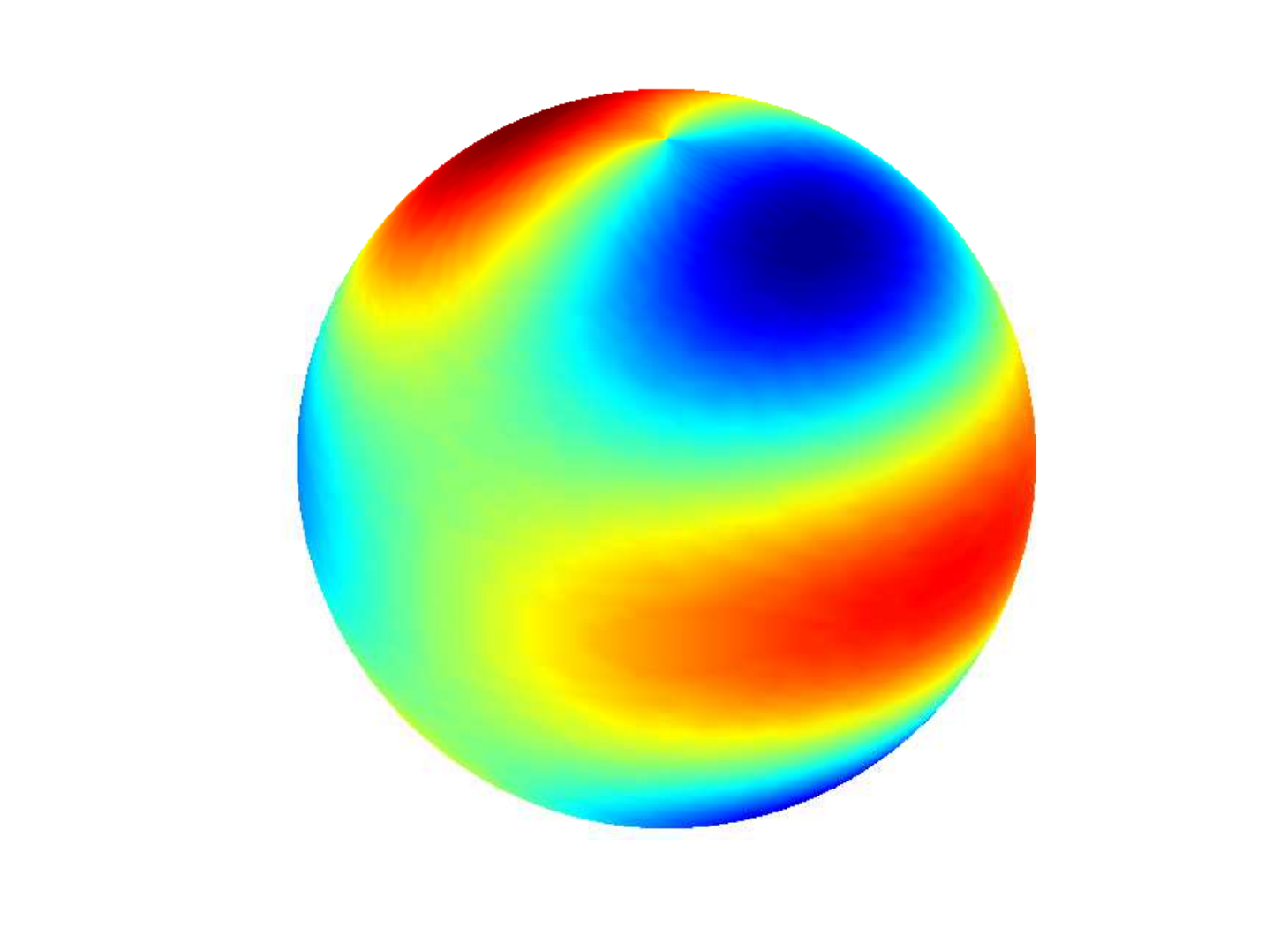}\includegraphics[scale=0.3]{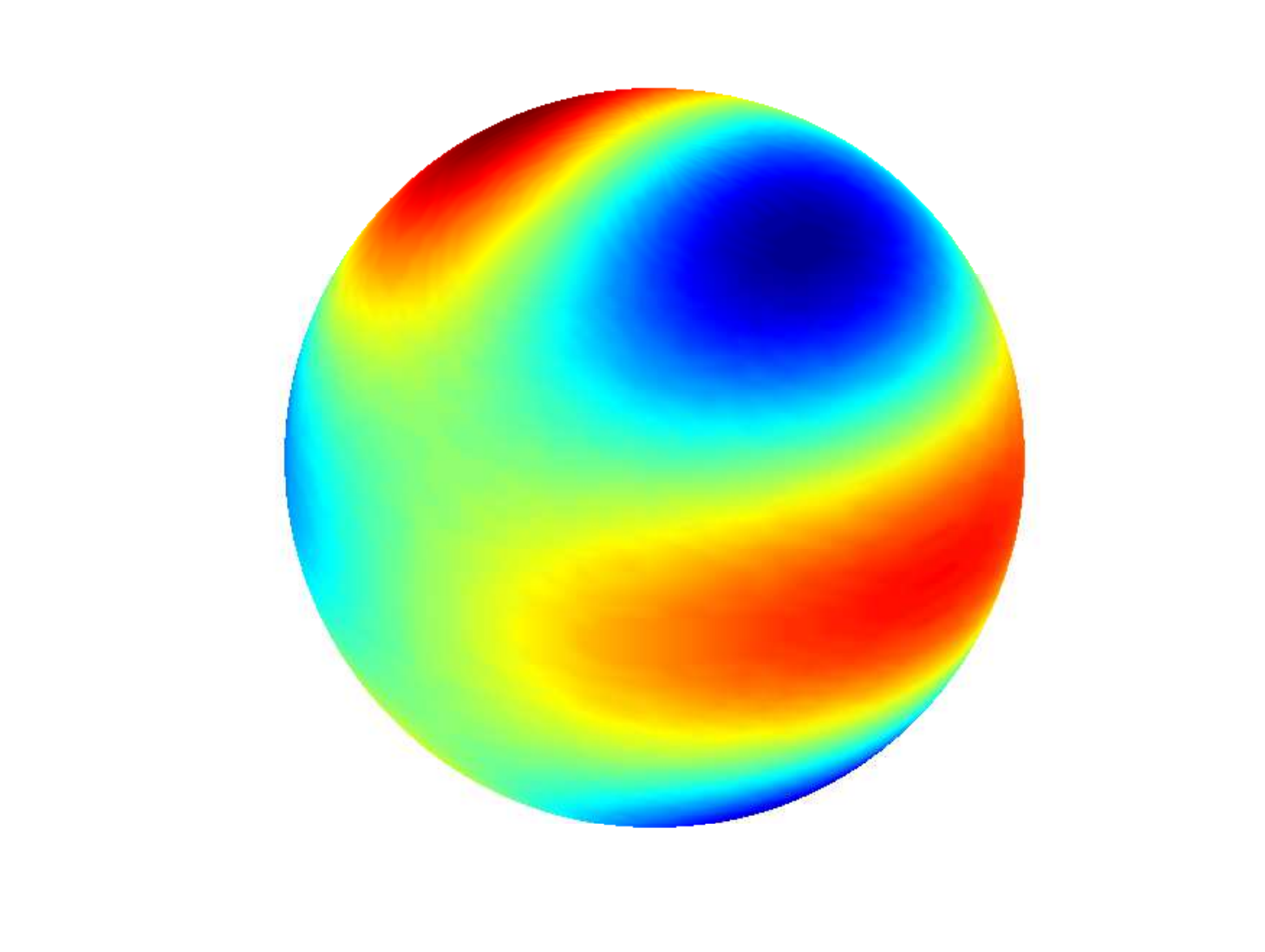}\includegraphics[scale=0.3]{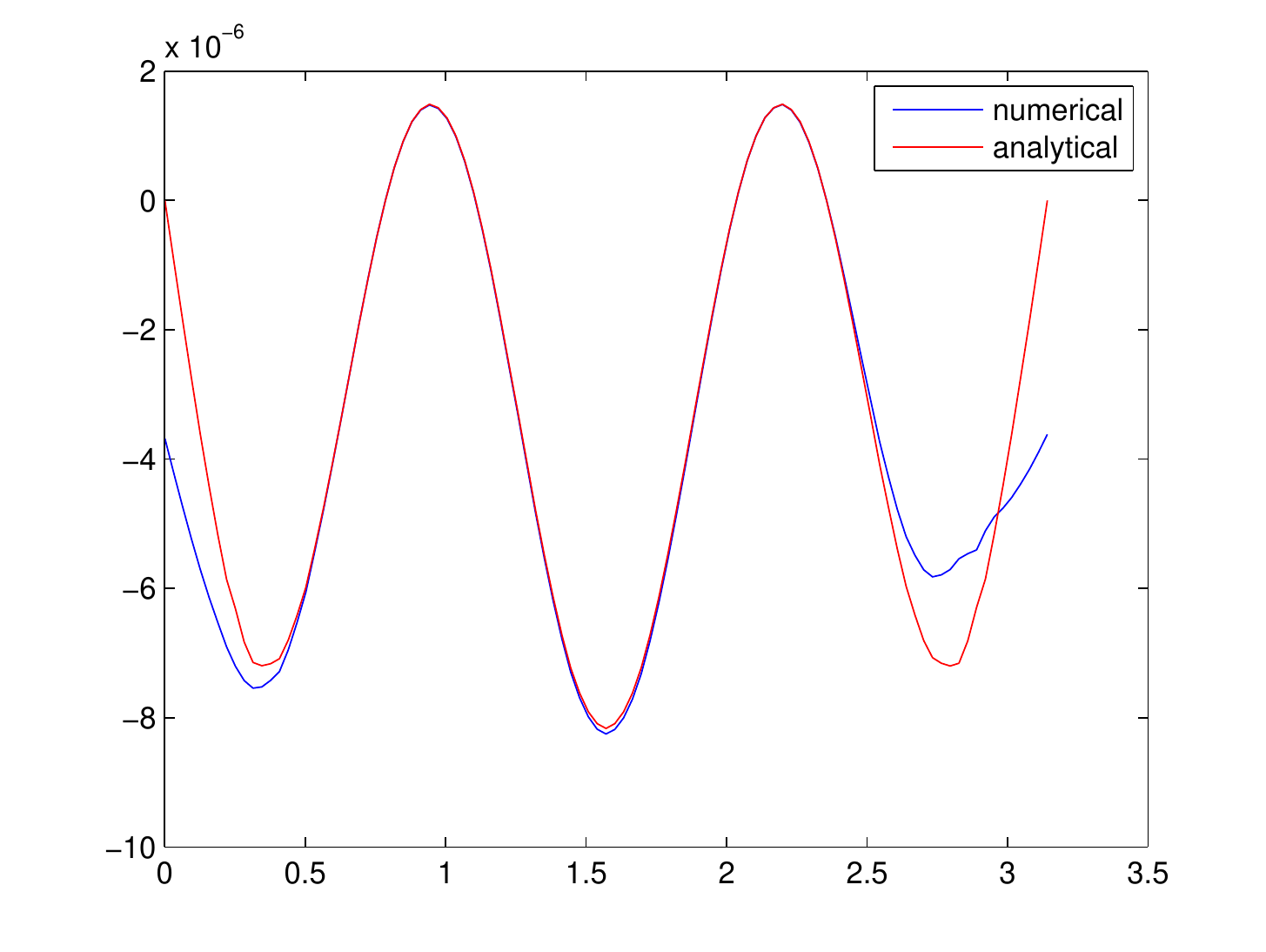}
\par\end{centering}
\protect\caption{Results for $m=1$ and $l=3$ at $t=0.001$. From left to right are the numerical result, the analytical result and the profile comparison on the spiral line $\phi=\pi(1-\cos\theta)$.}
\end{figure}
\begin{figure}
\begin{centering}
\includegraphics[scale=0.3]{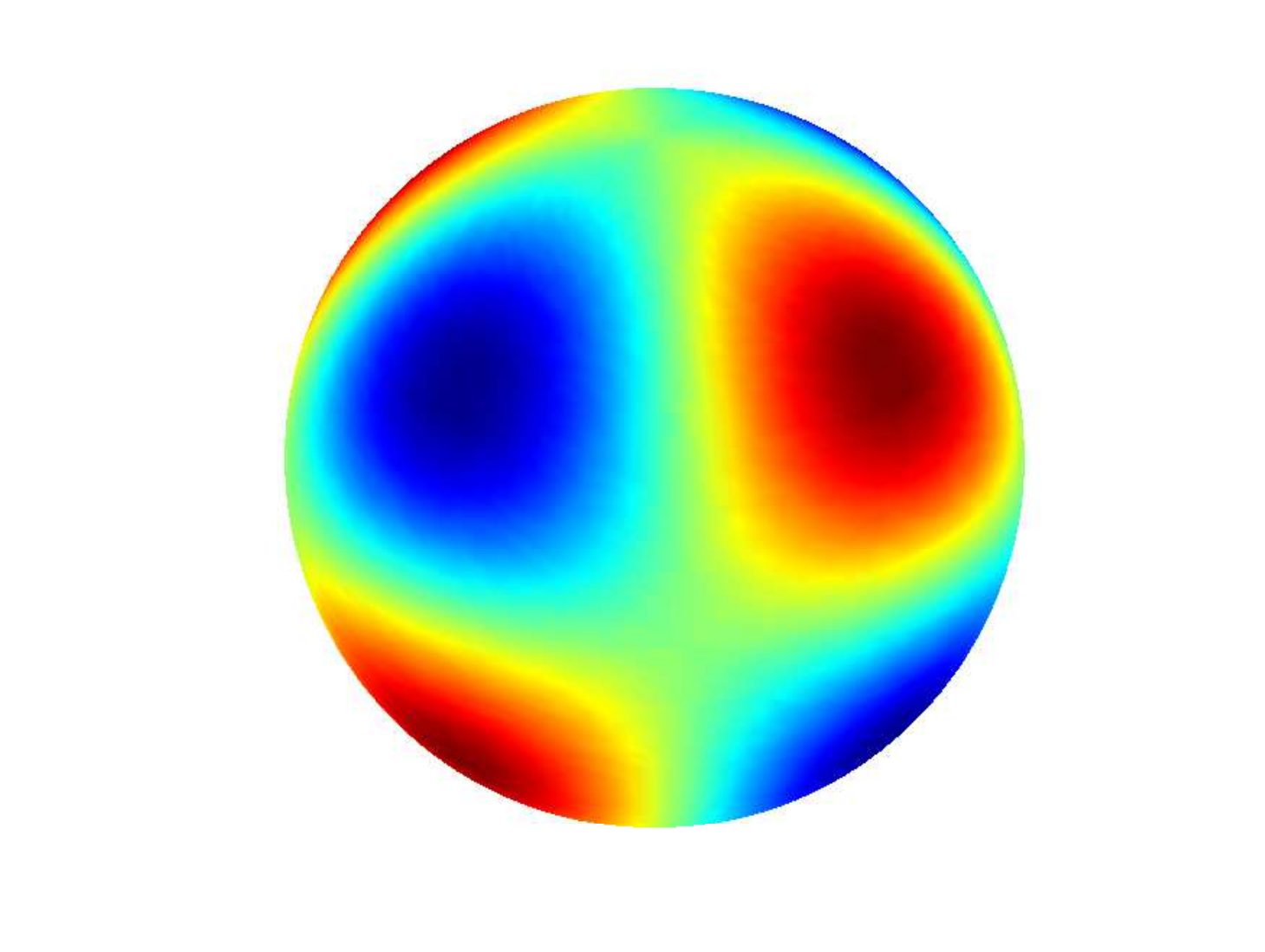}\includegraphics[scale=0.3]{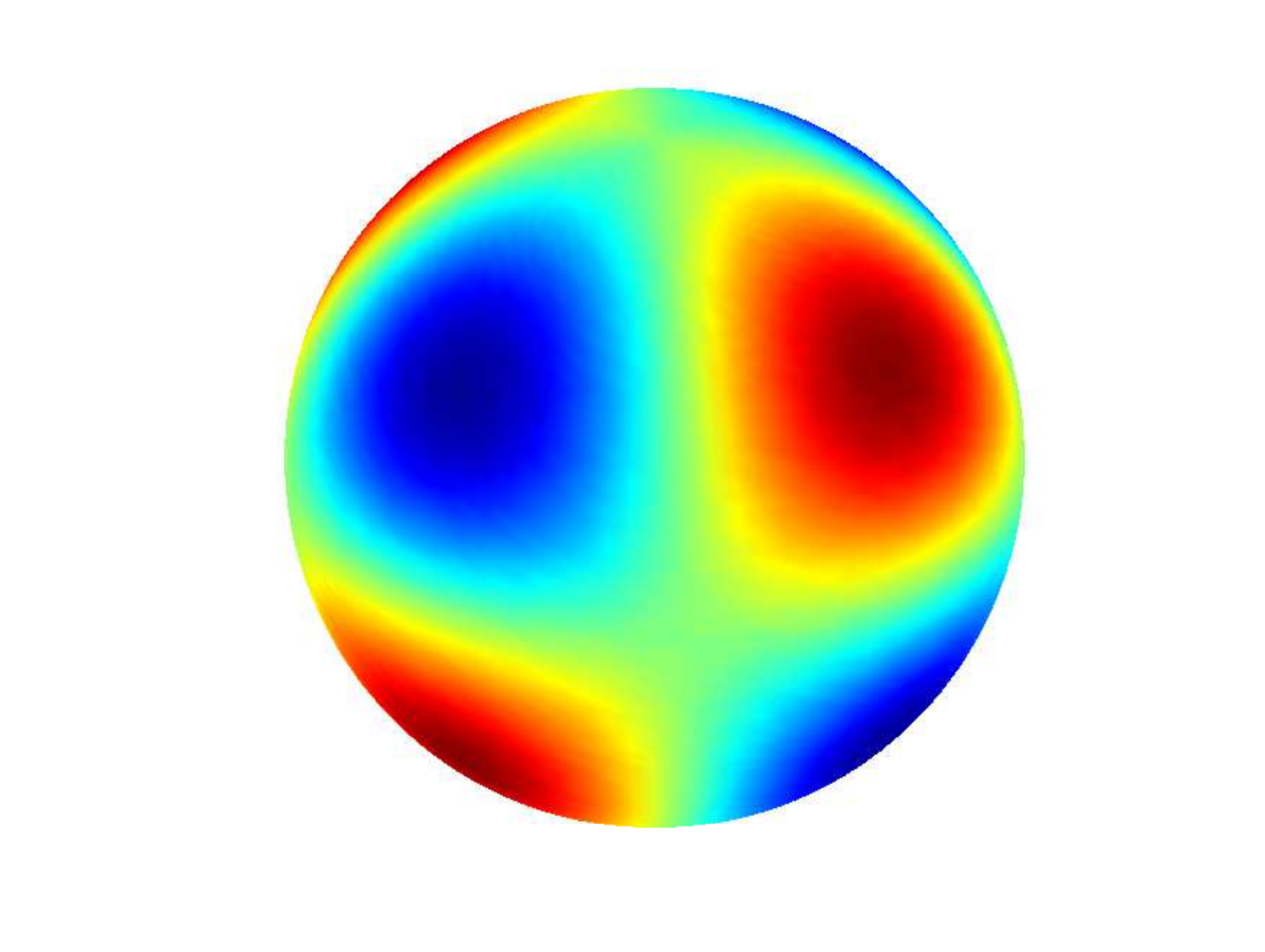}\includegraphics[scale=0.3]{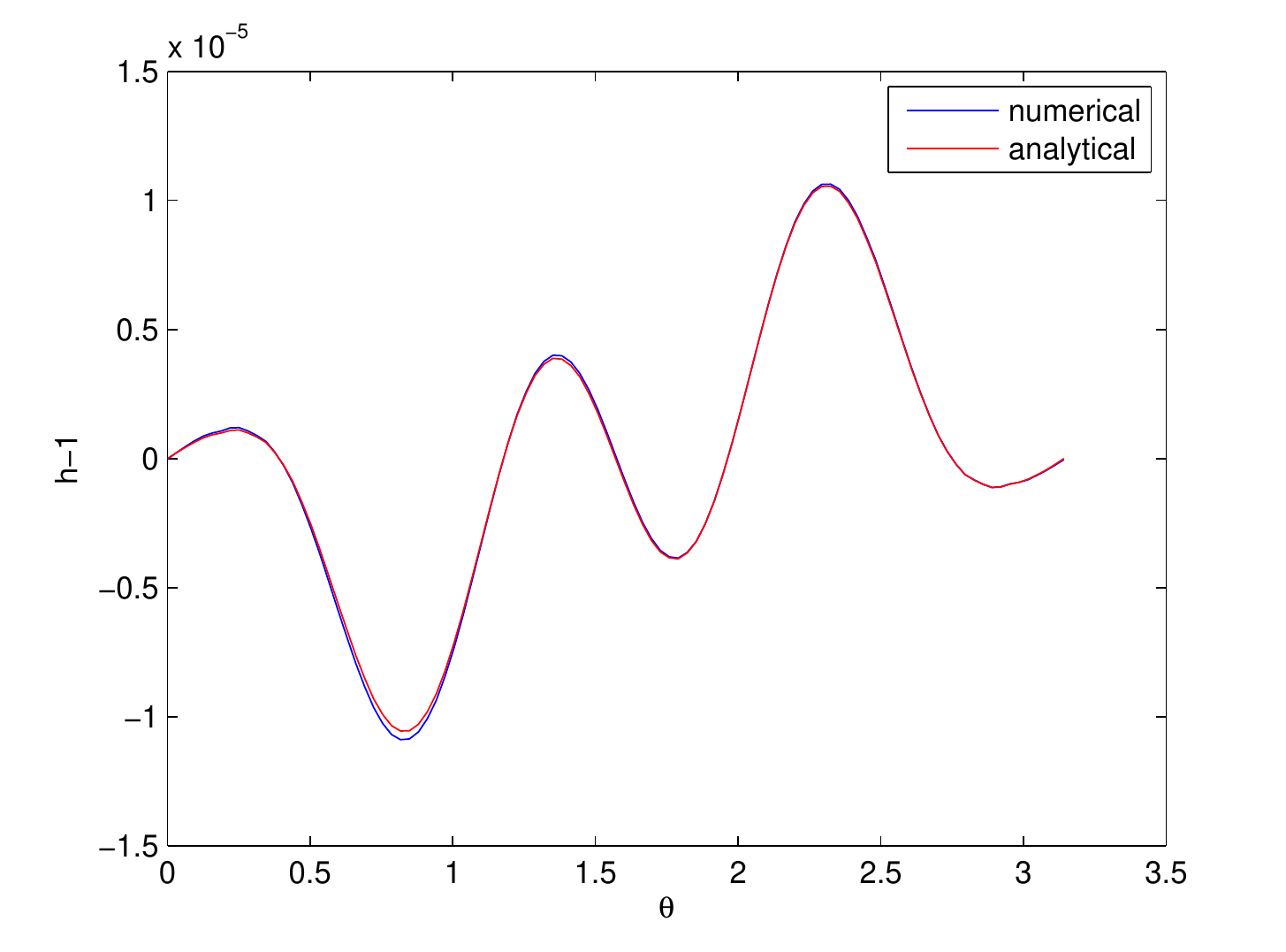}
\par\end{centering}
\protect\caption{Results for $m=2$ and $l=3$ at $t=0.001$. From left to right are the numerical result, the analytical result and the profile comparison on the spiral line $\phi=\pi(1-\cos\theta)$.
\label{fig:two dimensional result m=00003D2}}
\end{figure}
\begin{figure}
\begin{centering}
\includegraphics[scale=0.3]{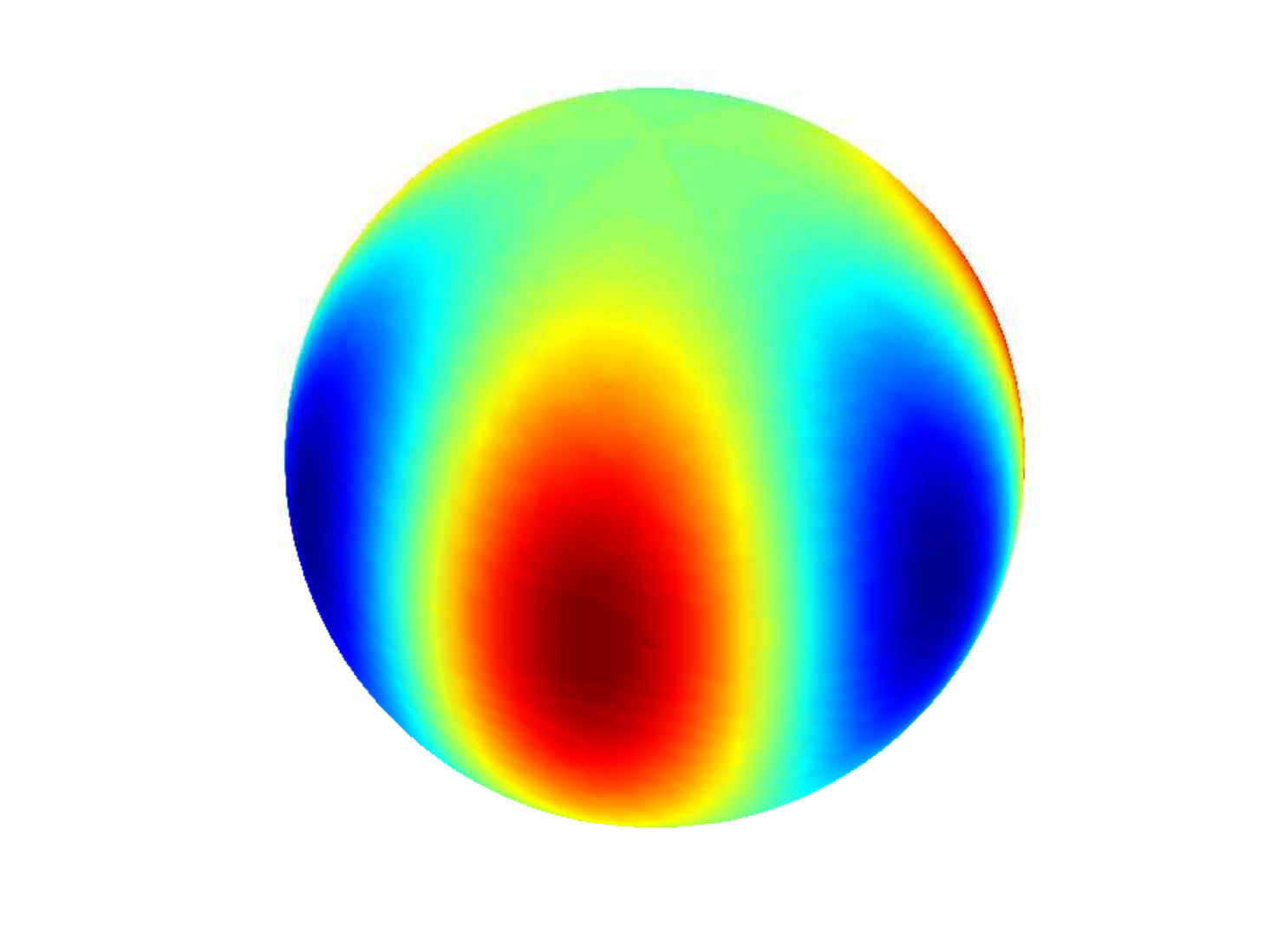}\includegraphics[scale=0.3]{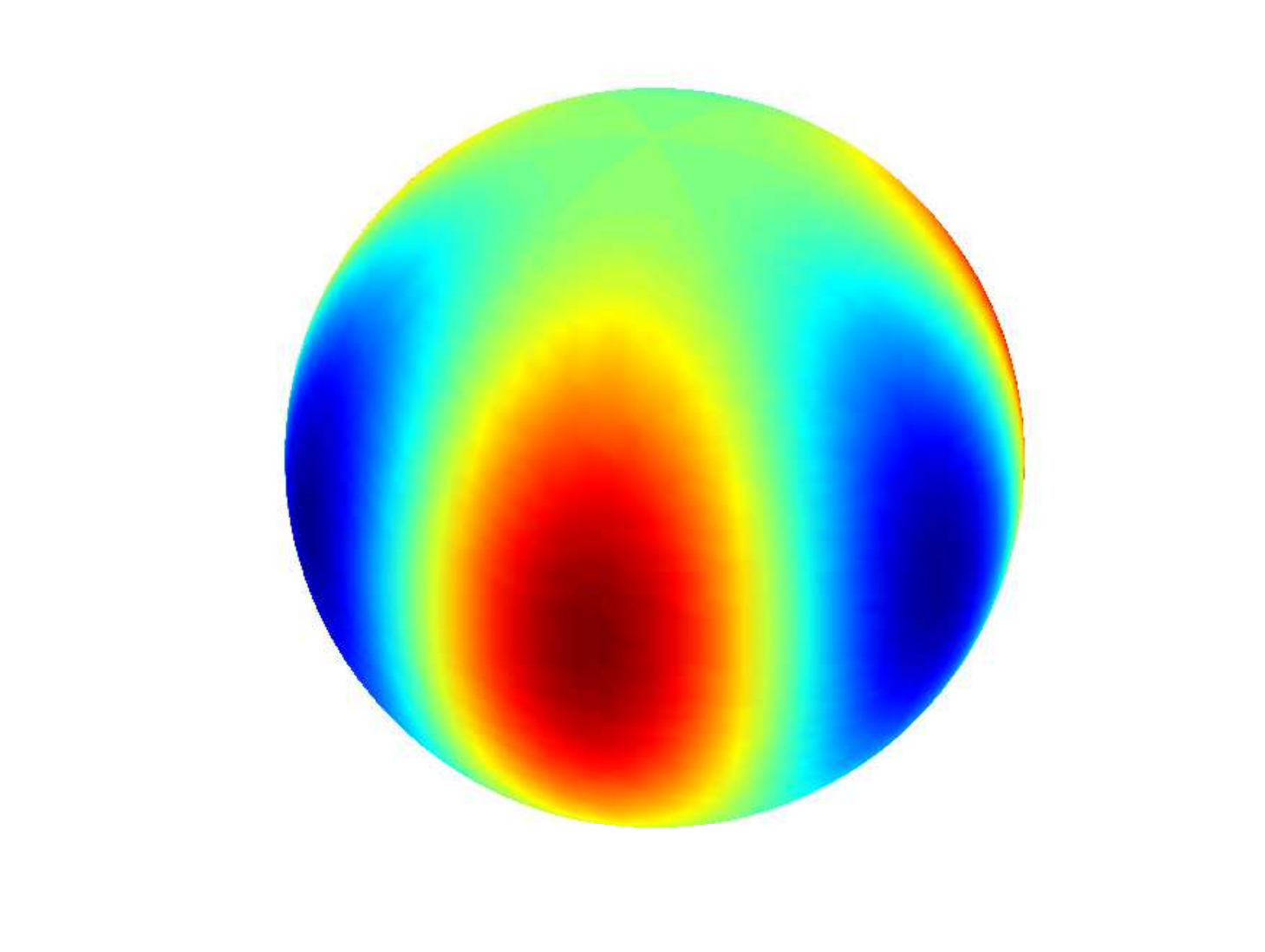}\includegraphics[scale=0.3]{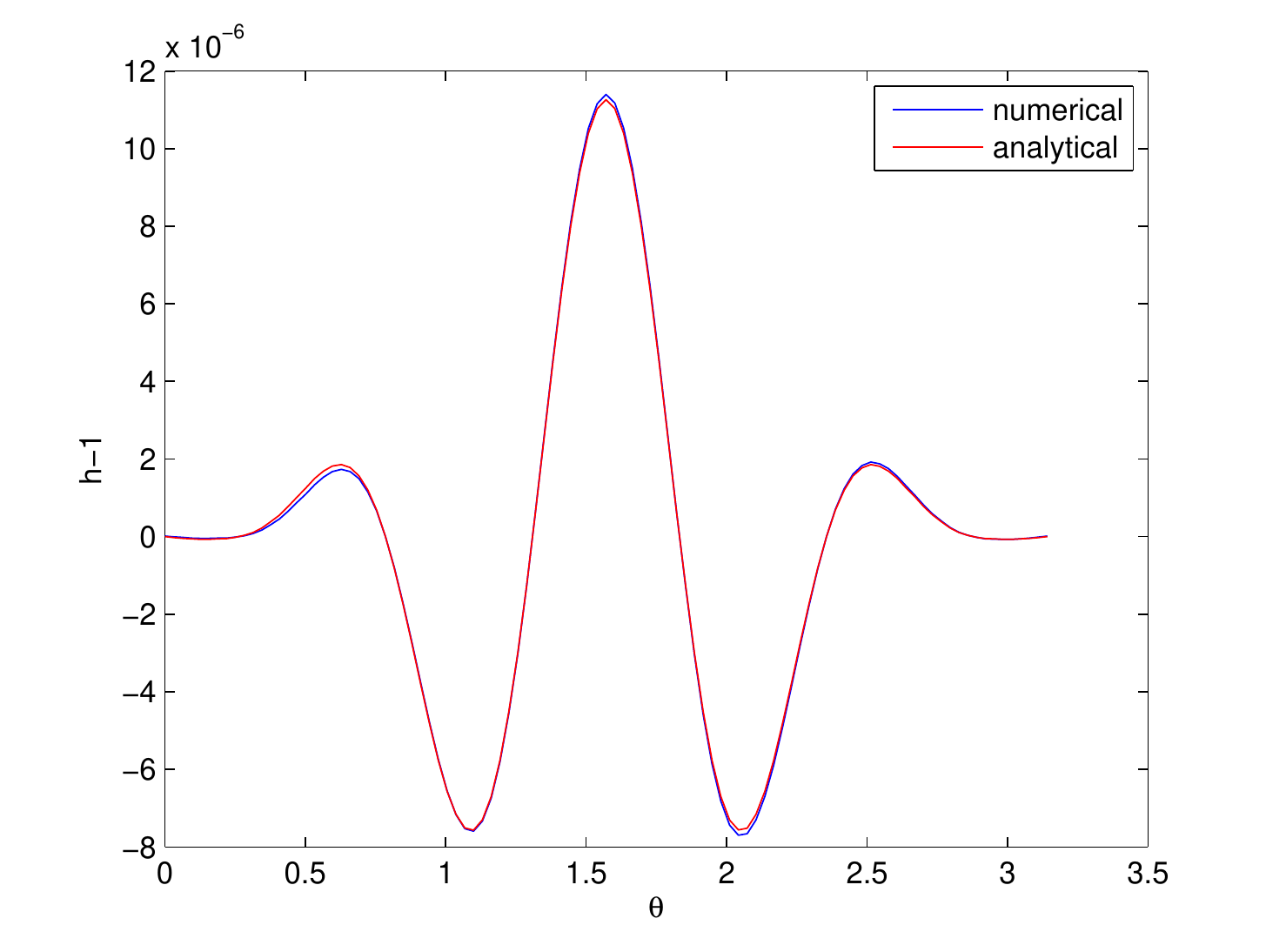}
\par\end{centering}
\protect\caption{Results for $m=3$ and $l=3$ at $t=0.001$. From left to right are the numerical result, the analytical result and the profile comparison on the spiral line $\phi=\pi(1-\cos\theta)$.
\label{fig:two dimensional result m=00003D3}}
\end{figure}

\begin{figure}
\hspace*{-2.4cm}
\includegraphics[scale=0.65]{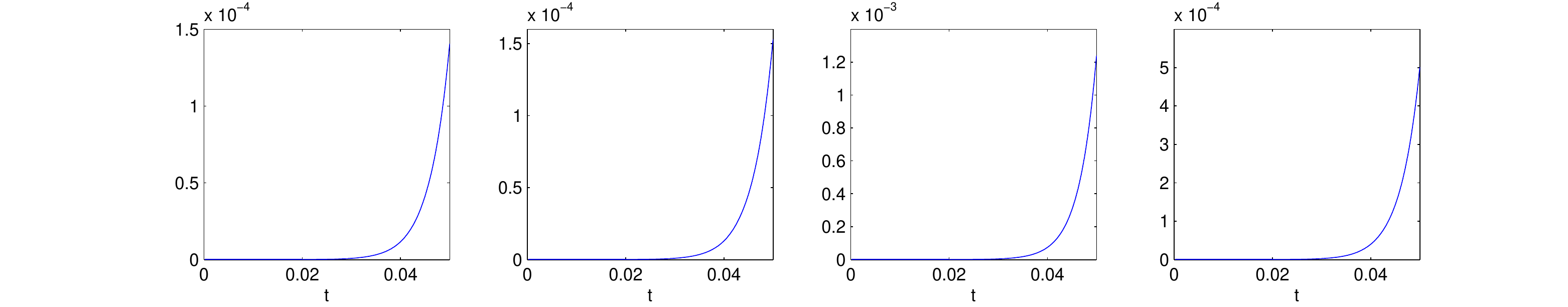}
\par
\protect\caption{$L^{2}$-difference between analytical and numerical results versus time. From left to right are $m=0$, $m=1$, $m=2$ and $m=3$.
\label{fig:L2 error two dimensional}}
\end{figure}

The result for $m=0$ through $m=3$ are shown in Figures \ref{fig:two dimensional result m=00003D0}--\ref{fig:two dimensional result m=00003D3} respectively. For the axisymmetric $m=0$ case, good agreement remains through time $t=0.01$ while for the $m=2$ and $m=3$ cases, similar agreement can only be obtained until a much shorter time of $t=0.001$. The case $m=1$ yields the worst agreement, which is associated with the Legendre function $P_3^1(x)$ being the least smooth among $P_3^m(x)$. The $L^{2}$-difference between computational and analytical results for different $m$ are shown in Figure \ref{fig:L2 error two dimensional}, illustrating that the linear model approximates the short time non-linear dynamics sufficiently well.

Using a similar procedure to the earlier axisymmetric case, the growth rate derived from the simulation is shown in Table \ref{tab:Eigenvalue two dimensional}. For each case, the growth rates are estimated using both $L^\infty$ and $L^1$ norms of the numerical results. Whereas the analytical growth rate is predicted to be 120 for all cases, the numerical results show varying degrees of agreement, with differences as high as 18\% in some cases.

\begin{table}
\begin{centering}
\begin{tabular}{|c|c|c|c|}
\hline
 & $t=10^{-5}$ & $t=10^{-4}$ & $t=5\times10^{-4}$\tabularnewline
\hline\hline
$m=0$ & $120.93$,~$120.16$ & $120.47$,~$120.36$ & $121.74$,~$121.73$\tabularnewline
\hline
$m=1$ & $121.81$,~$136,71$ & $133.53$,~$126.20$ & $123.20$,~$107.62$\tabularnewline
\hline
$m=2$ & $139.82$,~$136.07$ & $142.50$,~$134.93$ & $142.48$,~$134.67$\tabularnewline
\hline
$m=3$ & $129.90$,~$124.05$ & $129.35$,~$124.54$ & $132.51$,~$125.33$\tabularnewline
\hline
\end{tabular}
\par\end{centering}
\protect\caption{Growth rate $\lambda$ calculated from the numerical results. The pair of values for each case include the value based on the $L^{\infty}$ norm on the left and that based on the $L^{1}$ norm on the right. The numerical tolerance was set to $10^{-7}$ in COMSOL, and the maximum time step was taken to be $\Delta t=10^{-6}.$\label{tab:Eigenvalue two dimensional}}
\end{table}

\subsection{Energy functional for the axisymmetric case}

When the film profile is independent of $\phi$, we have been able to find a corresponding energy functional whose time-derivative is negative under the governing dynamics, showing that the system is dissipative. We still take the temperature field to to be purely in the radial direction and gravity to be zero. The corresponding evolution equation is given by
\[
\frac{\partial h}{\partial t}+\frac{\partial}{\partial x}\left\{ {\cal S} h^{3}(1-x^{2})\frac{\partial}{\partial x}\left[2h+\frac{\partial}{\partial x}\left((1-x^{2})\frac{\partial h}{\partial x}\right)\right]+{\cal N} h^{2}(1-x^{2})\frac{\partial h}{\partial x}\right\} =0
\]
where parameter $\cal S$ relates to surface tension and $\cal N$ to the Marangoni effect.

Multiply $h$ by the multiplier
\[
-\frac{{\cal S}}{2}\left(2h+\frac{\partial}{\partial x}\left((1-x^{2})\frac{\partial h}{\partial x}\right)\right)-{\cal N}(\ln h-1)
\]
and integrate over $x$ in $[-1,1]$ to obtain the functional $E(h)$:
\begin{align}
E(h)&=\int_{-1}^{1}-\frac{{\cal S}h}{2}\left(2h+\frac{\partial}{\partial x}\left((1-x^{2})\frac{\partial h}{\partial x}\right)\right)-{\cal N}h(\ln h-1)dx\\
\label{eq:EnergyFunctional} &=\int_{-1}^{1}-{\cal S}h^{2}+\frac{{\cal S}}{2}(1-x^{2})h_{x}^{2}-{\cal N}h(\ln h-1)dx\,.
\end{align}
Here, subscripts on $h$ refer to corresponding partial derivatives.

Take the derivative of $E$ with respect to time $t$ and simplify using integration-by-parts and by invoking the evolution equation to relate the time-derivative $h_t$ to the spatial derivative term, resulting in
\begin{align*}
\frac{d}{dt}E(h)&=\int_{-1}^{1}-2{\cal S}hh_{t}+{\cal S}(1-x^{2})h_{x}h_{xt}-{\cal N}h_{t}\ln h\,dx\\
&=-\int_{-1}^{1}\left[2{\cal S}h+{\cal S}((1-x^{2})h_{x})_{x}+{\cal N}\ln h\right]h_{t}\,dx \\
&=-\int_{-1}^{1}h^{3}(1-x^{2})\left[ 2{\cal S}h_{x}+{\cal S}((1-x^{2})h_{x})_{xx}+{\cal N}\frac{1}{h}h_{x}\right]^{2}\,dx \,.
\end{align*}
As $h$ and $(1-x^{2})$ are non-negative on $[-1,1]$, we conclude that
$\frac{d}{dt}E(h)\leq 0$.

Mathematically, the energy functional (\ref{eq:EnergyFunctional}) allows one to prove the existence of the minimizer for the axisymmetric case. However, in addition to the energy decay property, one must also show that this functional is bounded from below. To establish a lower bound for $E(h)$, we can use the result from \cite{kang2016weak} that $h(x,t)$ belongs to a Sobolev space $H^1$ with weight $1-x^2$. This establishes that the integral $\int_{-1}^{1}(1-x^{2})h_{x}^{2} dx$ is bounded from below. The other two terms contributing to $E(h)$ are easily seen to be bounded from below and therefore so is $E(h)$ itself.

\begin{figure}[h]
\begin{centering}
\includegraphics[scale=0.35]{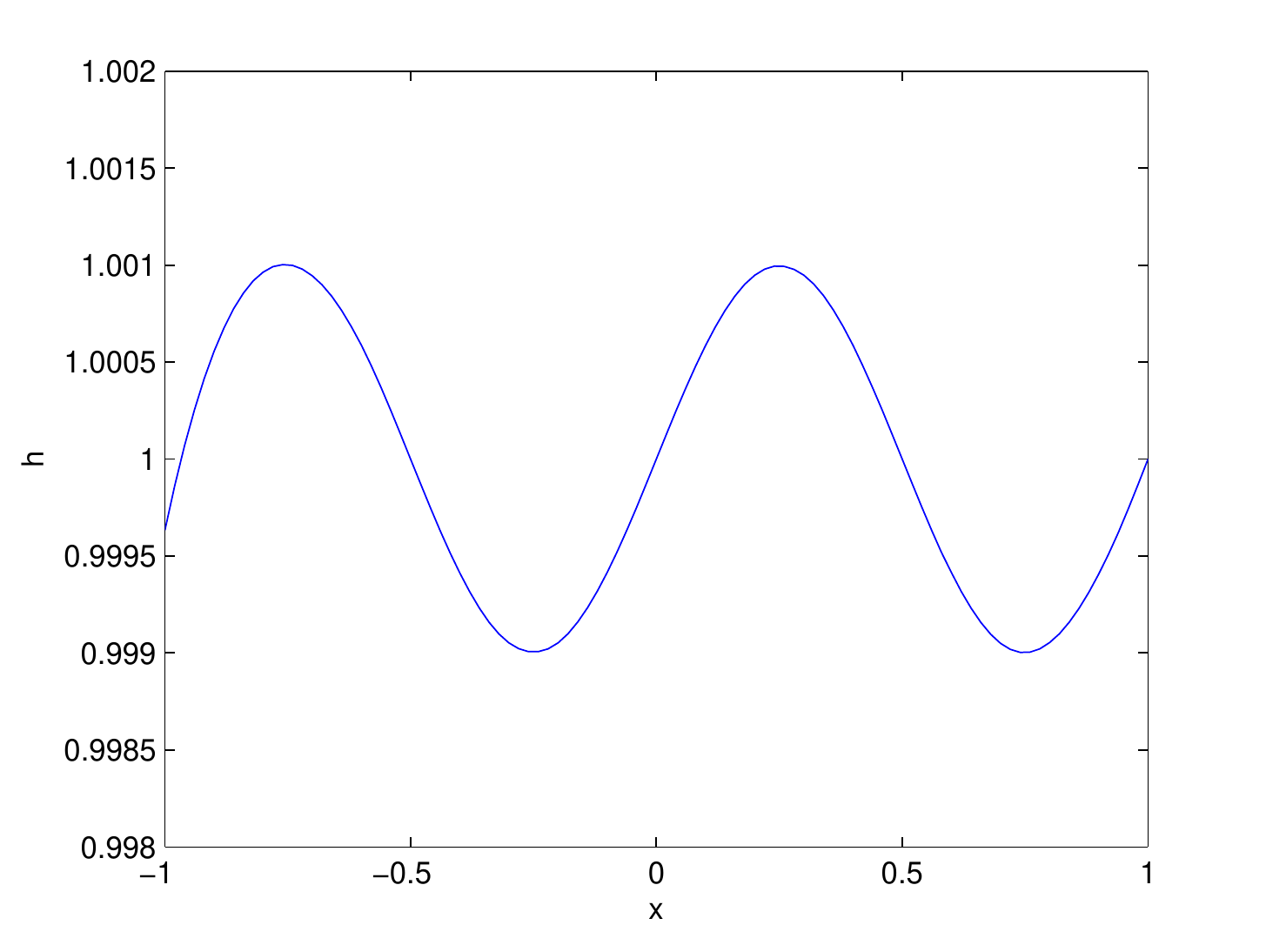}\includegraphics[scale=0.35]{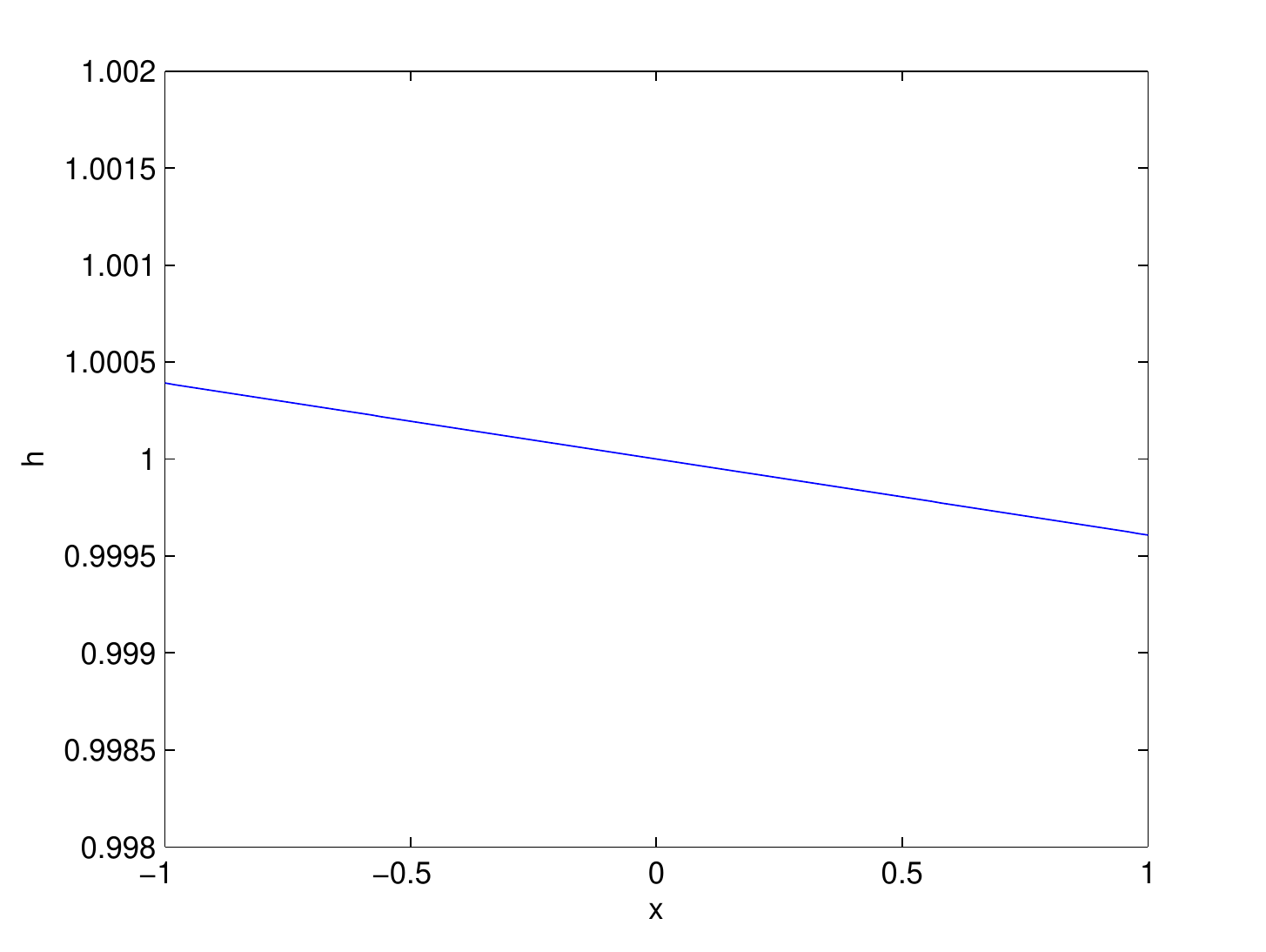}\includegraphics[scale=0.35]{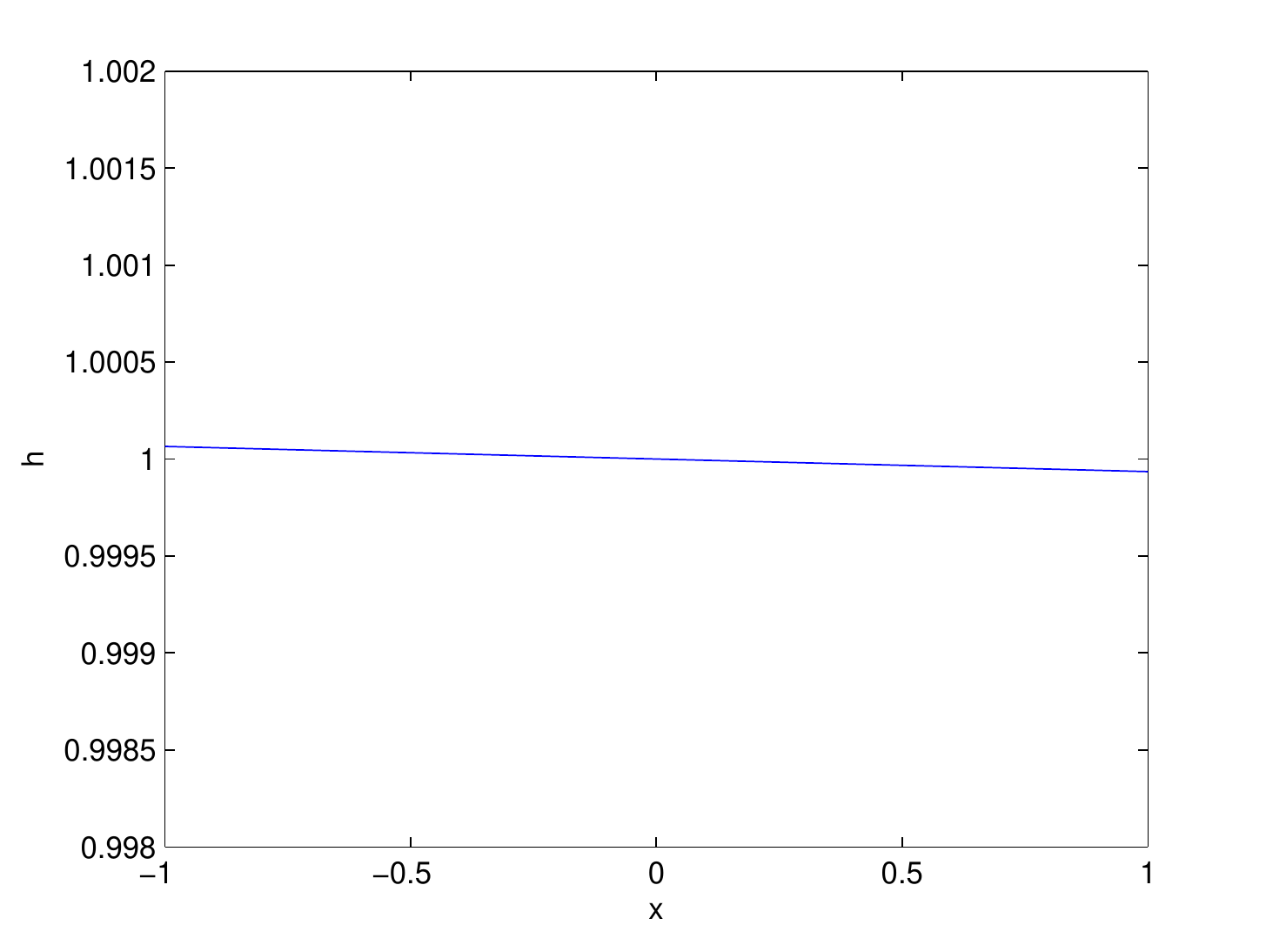}
\includegraphics[scale=0.6]{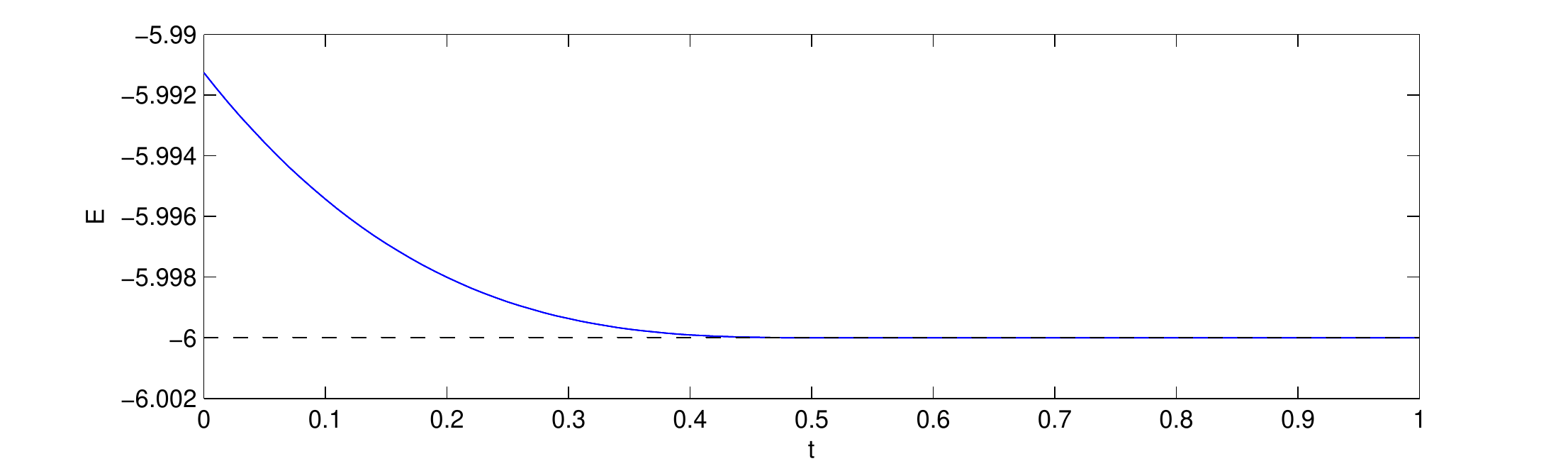}
\par\end{centering}
\protect\caption{Top: $h(x,t)$ at time $t=0,0.05,0.5$  with ${\cal G}=0$, ${\cal S}=1$ and ${\cal N}=-2$; bottom: energy approach to its equilibrium.  
\label{fig:energydecaynog}}
\end{figure}

Upon minimizing $E(h)$ subject to the mass conservation constraint
\[
\int_{-1}^{1}h\,dx=M.
\]
we obtain the Euler-Lagrange equation with a Lagrange multiplier $\lambda$ (note that $\lambda$ in this section is not the same as the growth rate from the linearized stability analysis):
\begin{equation}
\label{eq:EL}
{\cal S}\frac{d}{dx}\left((1-x^{2})\frac{dh}{dx}\right)+2{\cal S}h+{\cal N}\ln h=\lambda \,.
\end{equation}
Numerically we find that with a slightly perturbed initial condition $h_0(x)=1+0.001\sin(2\pi x)$ and parameters ${\cal G}=0$, ${\cal S}=1$ and ${\cal N}=-2$, the thin film will become uniform and during this process, energy will approach to an equilibrium corresponding to $\lambda=2$ (see Figure~\ref{fig:energydecaynog}).

\begin{figure}[h]
\begin{centering}
\includegraphics[scale=0.35]{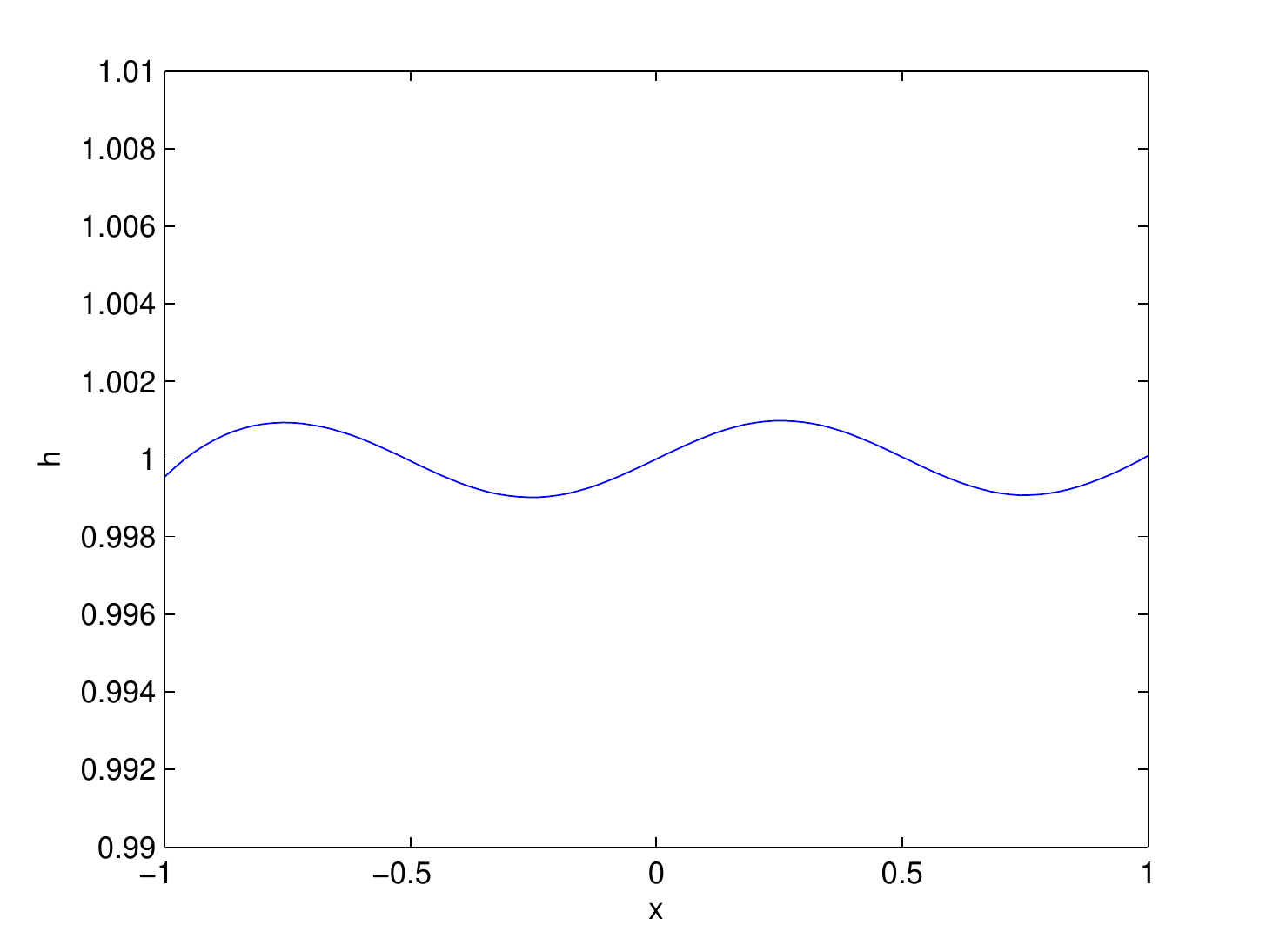}\includegraphics[scale=0.35]{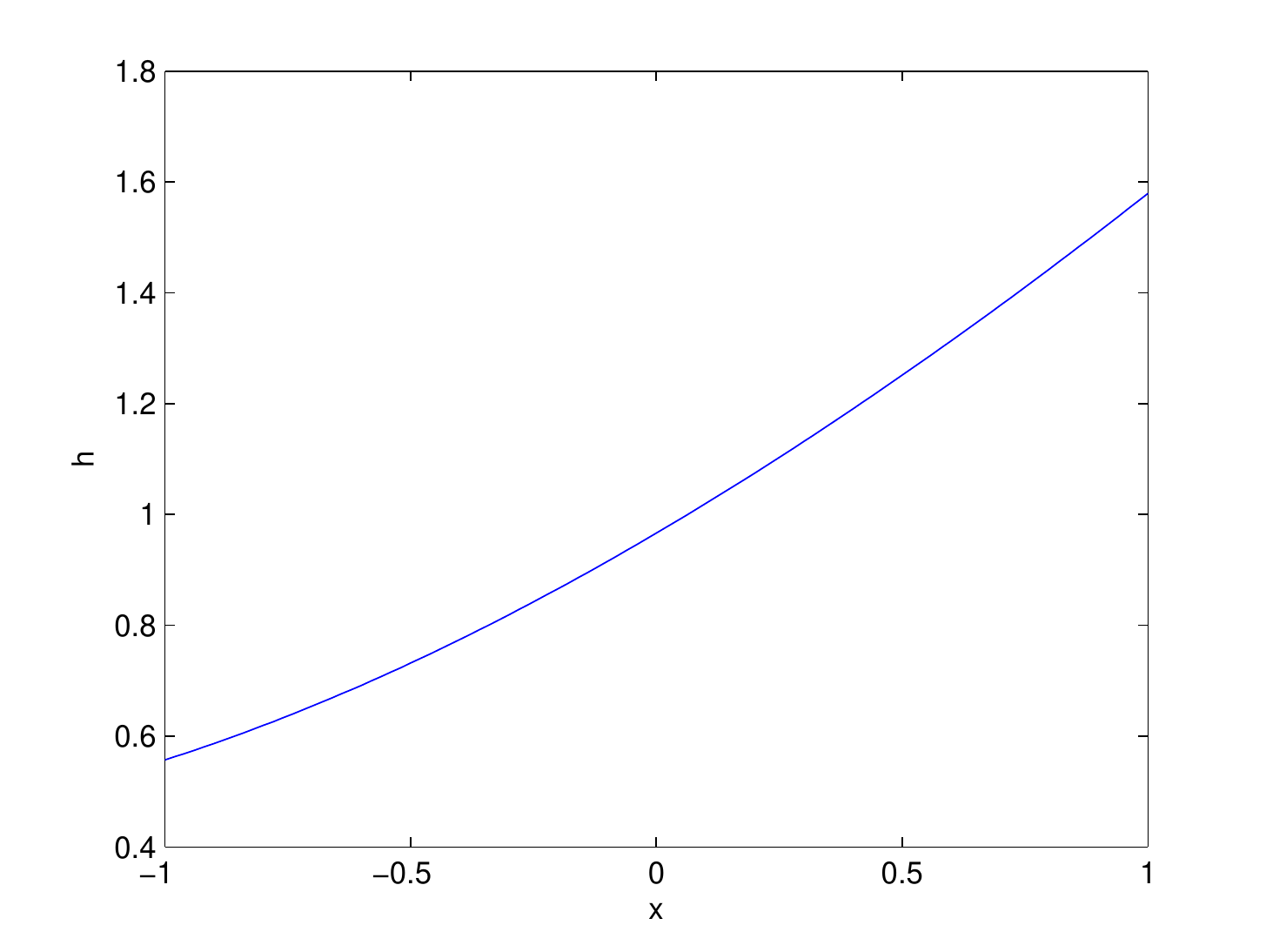}\includegraphics[scale=0.35]{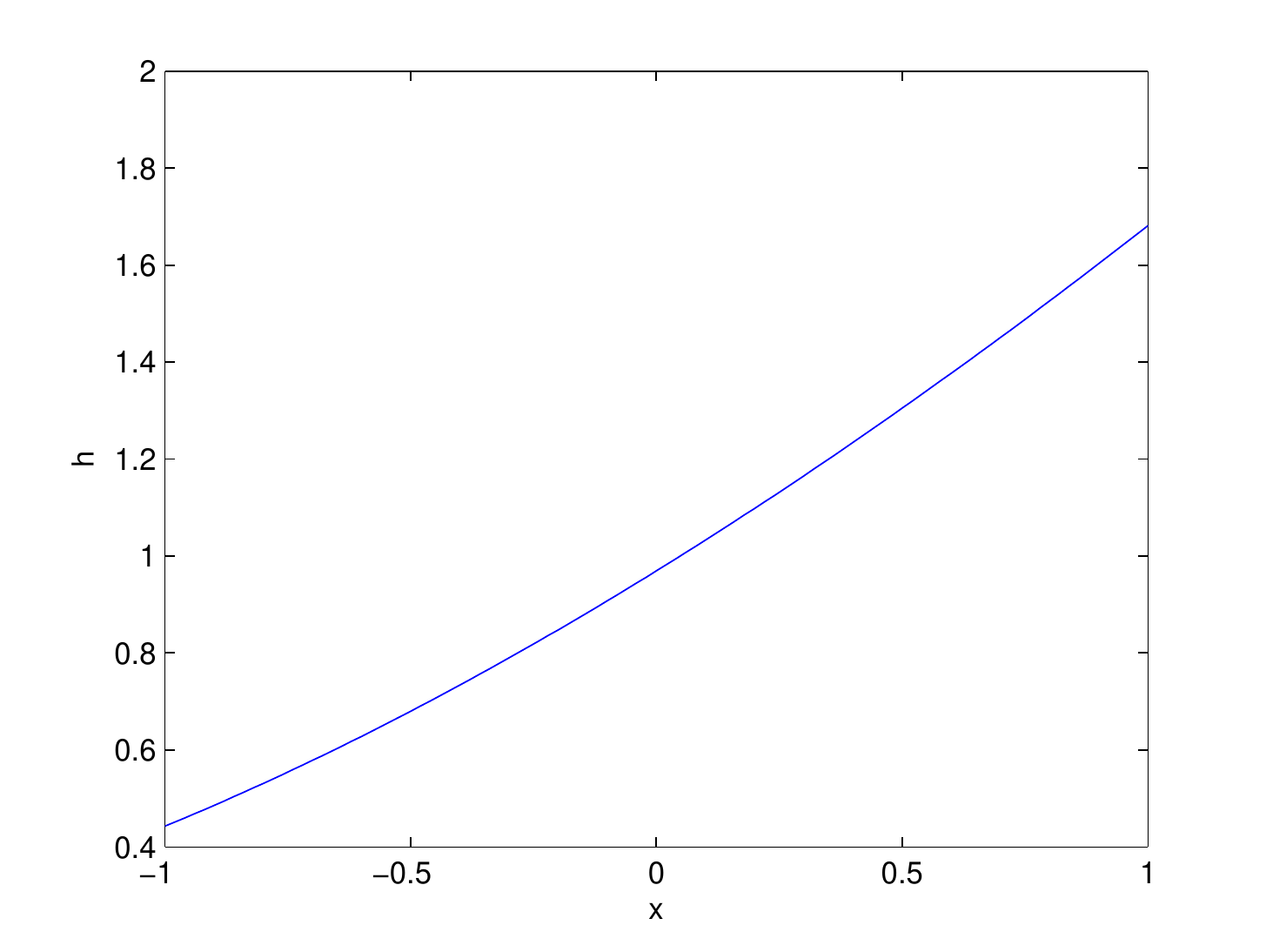}
\includegraphics[scale=0.6]{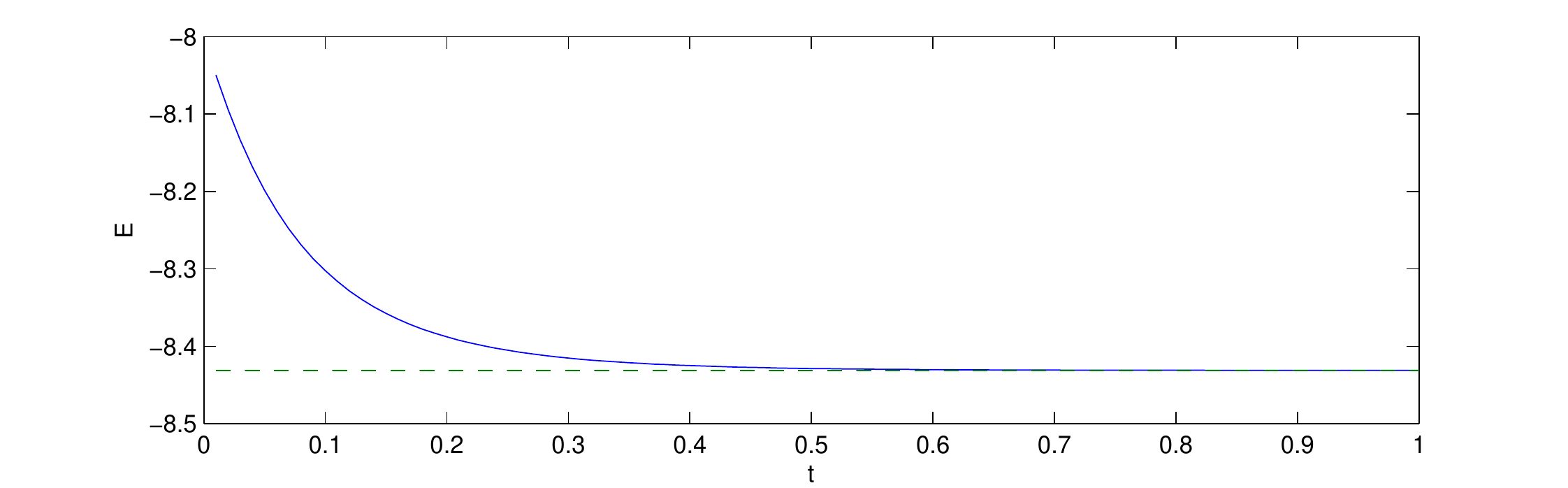}
\par\end{centering}
\protect\caption{Top: $h(x,t)$ at time $t=0,0.3,1$  with ${\cal G}=2$, ${\cal S}=1$ and ${\cal N}=-3$;
bottom: energy approach to its equilibrium.  
\label{fig:energydecaywithg}}
\end{figure}

It turns out to be possible to include the effects of gravity, characterized by parameter $\cal G$, in this derivation. In that case, one can show that the energy functional becomes:
%
%
\[
E(h)=\int_{-1}^{1}-{\cal G}xh-\frac{{\cal S}h}{2}\left(2h+\frac{\partial}{\partial x}\left((1-x^{2})\frac{\partial h}{\partial x}\right)\right)-{\cal N} h(\ln h-1)\,dx\,,
\]
with the corresponding Euler-Lagrange equation:
\[
{\cal S}\frac{d}{dx}\left((1-x^{2})\frac{dh}{dx}\right)+2{\cal S} h+{\cal G}x+{\cal N}\ln h=\lambda\,.
\]
These nonlinear differential equations that include the logarithmic term $\ln(h)$ can be solved to obtain the long-time, energy-minimizing, equilibrium shape of the axisymmetric film in the presence of a radial temperature gradient. Numerically we find that with a slightly perturbed initial condition $h_0(x)=1+0.001\sin(2\pi x)$ and parameters ${\cal G}=2$, ${\cal S}=1$ and ${\cal N}=-3$, gravity will drive the fluid to accumulate near the bottom of the sphere and during this process, the energy approaches an equilibrium corresponding to the value $\lambda=2.21$ (see Figure~\ref{fig:energydecaywithg}).

\section{Discussion}
\label{sec:5}

For the case of a thin viscous liquid film coating a sphere with an imposed axial (i.e. vertical) or radial temperature gradient, we have been able to derive a closed-form nonlinear fourth-order partial differential equation that governs the evolution of the film thickness accounting for gravitational, capillary and Marangoni effects. 

Assuming a linear variation of surface tension with temperature and a linearly varying axial temperature field, we found that there is a particular uniform film thickness (i.e., fluid volume) where, despite a continual internal flow, the film profile remains constant and steady. In that state, the downward draining flow due to gravity is balanced exactly by the upward flow due to Marangoni effects (provided that the surface tension near the north pole is higher than that near the south pole). Moreover, our analysis showed that if the fluid volume is slightly increased from that state, the fluid tends to accumulate near the bottom of the sphere, whereas if the volume is slightly decreased, the film thins out near the bottom and bulges out above, though not exactly at the north pole. This analysis suggests a potential experimental method for measuring, indirectly, the slope of the surface tension versus temperature curve. For instance, one can slowly add volume to a liquid coating to see at what point the film thickness become exactly uniform. Conversely, for a fluid that might slowly evaporate over time, one can start with excess fluid, which produces a bulge near the bottom of the sphere and observe, as the volume decreases slowly by evaporation, the film first become uniform in thickness and then thin out near the bottom. If the (dimensional) steady uniform film thickness $h_s$ can be measured in this manner, the parameter $\sigma_1$ can be obtained from our theoretical prediction: $\sigma_1=2 \rho g h_s / 3 k_1$. Typically, both $k_1$ and $\sigma_1$ would be negative numbers in this case.

For a radially varying temperature field, we were able to show that an initially uniform film may become unstable provided that surface tension becomes lower in regions where the film thins out and higher where it bulges out. So, if surface tension decreases with increasing temperature, the imposed temperature field must be higher at the sphere surface and decrease radially outward. This is consistent with the finding by \cite{wilson1994onset}. In the axisymmetric case, we could also find an energy functional with the dissipation property and bounded from below, so that the evolution of the film thickness takes the energy level toward a minimum. To show that the Euler-Lagrange equation that describes the minimum of the energy functional does have a proper mathematical structure, we can
apply local analysis to show that it admits zero-contact-angle non-negative solutions near the transition between dry and wet zones on the surface. Without loss of generality, let us assume that the touch-down point is at $x_0 = 0$. Hence, locally in the neighborhood of $x_0$ the differential equation (\ref{eq:EL}) simplifies to 
$$h'' + 2h + \alpha \ln{h} = \beta $$
where $\alpha = {\cal N}/{\cal S}$ and $\beta = \lambda/{\cal S}$.
By reduction of order, i.e., multiplying the equation by $h'$ and integrating we obtain
$$ \frac{(h')^2}{2} + h^2 + \alpha h (\ln{h}  - 1) - \beta h = c$$
where $c$ is the integration constant. A zero contact angle $h'(0) = 0$ implies that the integration constant is $c = 0$ resulting in
$$  h' = \pm \sqrt{2[(\alpha + \beta)h - h^2 - \alpha h \ln{h}]}\,.$$
To get a local non-negative solution we would choose the plus sign. Local existence follows from the  continuity of the right side of the equation at $h(0) = 0$ at $0^+$.

%

\begin{acknowledgments}
This work was partially supported by a grant from the Simons Foundation (\#275088 to MC). DK thanks the Institute of Mathematical Sciences at Claremont Graduate University for a Director's Fellowship that supported his doctoral studies.
\end{acknowledgments}


\end{document}